%% file: TCPD_20220725.tex
\documentclass[11pt]{article}
\usepackage{epsfig,color,amsmath,amssymb,euscript}
\usepackage{fancyhdr}
\usepackage[square,sort,comma,numbers,sectionbib,round]{natbib}
\usepackage{rotating}
\usepackage{amsmath}
\usepackage{amsfonts}
\usepackage{setspace}
\usepackage{fancyhdr}
\usepackage[top=0.9in,bottom=0.9in,left=0.8in,right=0.8in]{geometry}
\usepackage{lscape}
\usepackage{amsthm}
\usepackage{enumerate}
\usepackage{threeparttable}
\usepackage{diagbox}
\usepackage{multirow}
\usepackage{makecell}
\usepackage{xcolor}
\usepackage{amsmath}
\usepackage{graphicx}
\usepackage{enumerate}
\usepackage{mathrsfs}
\usepackage{authblk}
\usepackage[square,sort,comma,numbers,sectionbib,round]{natbib}
 \usepackage{graphicx}
\theoremstyle{definition}

\newtheorem{remark}{Remark}

\def\T{\scriptscriptstyle \top}
\def\H{\scriptscriptstyle \mathrm{H}}
\def\t{\scriptscriptstyle \top}

\renewcommand{\baselinestretch}{1.4}
\makeatletter

\newcommand{\bXi}{{\boldsymbol{\Xi}}}
\newcommand{\bJ}{{\bf{J}}}

\input{QYalias}

\newcommand{\Date}[1]{\def\@Date{#1}}
\def\today{\number\day~\ifcase\month\or
 January\or February\or March\or April\or May\or June\or
 July\or August\or September\or October\or November\or December\fi~\number\year}

\newtheorem{Condition}{Condition}

\allowdisplaybreaks[3]

\begin{document}

\bibliographystyle{asa}
\bibpunct{(}{)}{,}{a}{}{;}

\title{\bf Modelling Matrix Time Series via a Tensor CP-Decomposition}

%\author{Jinyuan Chang \qquad Jing He \qquad Qiwei Yao\\
%School of Statistics, Southwestern University of Finance and Economics\\
% Chengdu, Sichuan 611130, China\\
%changjinyuan@swufe.edu.cn \qquad  he\_jing@swufe.edu.cn\\[1ex]
%Qiwei Yao\\
% {Department of Statistics, The London School of Economics and Political Science,}\\
% { Houghton Street, London, WC2A 2AE, U.K.}\\
% q.yao@lse.ac.uk }

\author[a]{Jinyuan Chang}
\author[a]{Jing He}
\author[a]{Lin Yang}
\author[b]{Qiwei Yao}
\affil[a]{\it \small Joint Laboratory of Data Science and Business Intelligence, Southwestern University of Finance and Economics, Chengdu, Sichuan 611130, China}
\affil[b]{\it \small Department of Statistics, The London School of Economics and Political Science, London, WC2A 2AE, U.K.}

% \date{}
\maketitle

\begin{abstract}
We consider to model matrix time series based on a tensor CP-decomposition.
Instead of using an iterative algorithm which is the standard practice for estimating CP-decompositions, we propose a new and
one-pass estimation procedure based on a generalized eigenanalysis constructed
from the serial dependence structure of the underlying process.
To overcome the intricacy of solving a rank-reduced generalized eigenequation, we propose a further refined approach which projects it into a lower-dimensional full-ranked
eigenequation. This refined method improves significantly the finite-sample performance of the estimation.
% {\color{red}A key idea of the new procedure is to project a generalized eigenequation defined in terms of rank-reduced matrices to a lower-dimensional one with full-ranked matrices, to avoid the intricacy of the former of which the number of eigenvalues can be zero, finite and infinity.}
The asymptotic theory has been established under a general setting without the stationarity. It shows, for example, that all the component coefficient vectors in the
CP-decomposition are estimated consistently with certain convergence rates. 
%depending on the relative sizes between the dimensions of time series and
%the sample size. 
The proposed model and the estimation method are also illustrated with both simulated and real data;
showing effective dimension-reduction in modelling and forecasting matrix time series.
\end{abstract}


\newpage

\section{Introduction}

Let $\bY_t= (y_{i,j,t})$ be a $p\times q$ matrix time series, i.e. there are
$pq$ recorded values at each time $t$ from, for example, $p$ individuals
and over $q$ indices or variables, and $y_{i,j,t}$ is then the value of the
$j$-th variable on the $i$-th individual at time $t$.
% We denote by $\by_{1:t}, \cdots, \by_{p:t}$ and $\by_{:1t}, \cdots,
% \by_{:qt}$, respectively, the row vectors and the column vectors of $\bY_t$.
% Hence $\by_{i:t}$ consists of the $q$ values taken from the $i$-th individual
% (over the $q$ variables) at time $t$, and $\by_{:jt}$ consists of the $p$
% values on the $j$-th variable (across the $p$ individuals) at time $t$.
%
Given available observations $\bY_1, \ldots, \bY_n$,
the goal is to build a dynamic model for $\bY_t$ and to forecast the future values $\bY_{n+\ell}$ for $\ell \ge 1$. With moderately large $p$ and $q$,
any direct attempts based on the time series ARMA framework are unlikely to be successful
due to overparametrization. We seek a low-dimensional
structure via a tensor canonical polyadic (CP) decomposition. To this end,
we denote by $\eulbY$ the $p\times q\times n$ tensor with $\bY_1, \ldots, \bY_n$
as its $n$ frontal slices \citep{KoldaBader_2009}. Then $y_{i,j,t}$ is the $(i,j,t)$-th
element of $\eulbY$.
Conceptually we decompose $\eulbY$ into two parts:
\begin{equation} \label{a1}
\eulbY = \eulbX + \eulbE\,,
\end{equation}
where all the dynamic structure of $\eulbY$ is reflected by $\eulbX$, and the frontal
slices of $\eulbE \equiv (\ve_{i,j,t})$ are matrix white noise, i.e.
$
\cov(\ve_{i,j,t},  \ve_{k,\ell, s}) = 0$ for any $t \ne s$.
The key idea is to perform a CP-decomposition for $\eulbX$, i.e.
to express it as a sum of rank one tensors (see (\ref{b1}) below). This
effectively represents the dynamic structure of matrix process $\bY_t$
in terms of that of a vector process, and, hence, achieving an effective
dimension-reduction in modeling the dynamic behaviour of the process.
% Note that the CP-decomposition is unique for tensors of order three (or
% higher) under some mild conditions; see Theorems 1.5 and 1.7 of Domanov
% and De Lathauwer (2014).

The `workhorse' method for CP-decompositions is the
so-called alternative least squares (ALS) algorithm which is
easy to understand and implement.
See Section 3.4 of \cite{KoldaBader_2009} and the
references therein. However it has obvious drawbacks. For example, an ALS
algorithm takes many iterations to
converge. It is not guaranteed to converge to the global minimum even for
moderately large $p$, $q$ or $n$.
Furthermore it also depends sensitively on the selection of the initial values.
Substantial effort has been made to improve the convergence and the performance of the ALS algorithm, including, among others,
\cite{agj14}, \cite{lsfcc14}, \cite{cv16}, \cite{sllc17},
\cite{sv17}, \cite{ws17}, \cite{zx18}, and \cite{hz21}.

We propose a new and one-pass estimation procedure
in this paper.
The new method is inspired by \cite{SanchezKowalski_1990} which transforms a CP-decomposition into a generalized eigenanalysis problem.
While Sanchez and Kowalski's approach does not require iteration, it only works
for the noise-free cases with $\eulbE \equiv \bzero$ in (\ref{a1}). In contrast, our
new procedure eliminates the impact of the noise by incorporating the
serial dependence into the estimation.
Furthermore to overcome the intricacy
in solving a generalized eigenequation defined by rank-reduced matrices (see Section 7.7 of \cite{GolubVan_2013}), we propose a new refined approach which projects a rank-reduced generalized eigenequation to a full-ranked lower-dimensional one which
is, therefore, equivalent to a standard eigenequation.
The numerical results in simulation also demonstrate the
significant improvement in the finite sample performance
by this refined method.

Most existing literature on matrix time series is based on the factor modelling
via the Tucker decomposition; see \cite{ChenChen_2019}, \cite{WangLiuChen_2019} and \cite{ChenTsayChen_2019}.
%\cite{HanChenZhangYao_2021} proposed a contemporaneous bilinear transformation whichsegments a matrix time series into serially uncorrelated submatrices.
The key difference between our approach and the Tucker decomposition
based approaches is two fold. First, a Tucker decomposition
represents a matrix process as a linear combination of a smaller matrix process while
a CP-decomposition is more canonical in the sense that it represents a matrix process in
terms of a vector process; see also the real data example in Section 5.2 below.
Secondly,  a Tucker decomposition entails more conventional factor models, and,
therefore, we only need to identify and estimate the factor loading spaces, for which
the standard factor model methods (e.g. \cite{LamYao_2012}, and \cite{ChangGuoYao_2015}) are applicable.
However for a CP-decomposition, we need to identify and estimate
the component coefficient vectors
precisely. % (upto permutation and a scaling factor).
Therefore a radically new inference procedure is required.
The other approaches for modelling matrix time series include: the matrix-coefficient autoregressive models of
\cite{ChenXiaoYang_2021}, and the bilinear transformation segmentation method of
\cite{HanChenZhangYao_2021}. \cite{HanZhangChen_2021} models tensor time series also based on a CP-decomposition. But their approach is  radically different from ours, as they estimate the CP-decomposition based on an iterative simultaneous orthogonalization algorithm
with a warm-start initialization using the so-called composite
principal component analysis for tensors; see Section 3 of \cite{HanZhangChen_2021}. Note that our estimation is an one-pass procedure,
and no iterations are required.

The rest of the paper is organized as follows. The matrix time series model based on
a CP-decomposition is presented in Section \ref{sec:model}. Section \ref{sec:method} deals with the model identification and presents the newly proposed estimation procedures. %The key idea of our approach is first elucidated for weakly stationary processes. The formal identification result and the estimation procedure are presented under a general setting without the stationarity condition.
The asymptotic results, including the convergence rates for the estimated component vectors in the CP-decomposition, are presented in Section \ref{sec4}. Numerical illustration with both simulated and real data sets is given in Section \ref{sec:numerical}. All the technical proofs are relegated to the supplementary material.

{\it Notations.} For a positive integer $m$, write $[m]=\{1,\ldots,m\}$,  and denote by $\bI_m$ the $m\times m$ identity matrix. Let $I(\cdot)$ be the indicator function. For an $m_1\times m_2$ matrix $\bH=(h_{i,j})_{m_1\times m_2}$, let $\|\bH\|_2$, ${\rm rank}(\bH)$, $\sigma_{\min}(\bH)$ and ${\rm vec}(\bH)$ be, respectively, its spectral norm, its rank, its smallest singular value, and a vector obtained by stacking together the columns of $\bH$. Specifically, if $m_2=1$, we use $|\bH|_2=(\sum_{i=1}^{m_1}|h_{i,1}|^2)^{1/2}$ to denote the $\ell^2$-norm of the $m_1\times 1$ vector $\bH$. Also, denote by $\bH^{\t}$ and $\bH^{\H}$, respectively, the transpose and conjugate transpose of $\bH$. When ${\rm rank}(\bH)=m_2$, denote by $\bH^+$, an $m_2\times m_1$ matrix, the Moore-Penrose inverse of $\bH$ such that $\bH^+\bH=\bI_{m_2}$. When $m_1=m_2$, denote by ${\rm det}(\bH)$ and ${\rm tr}(\bH)$ the determinant and the trace of $\bH$, respectively. %For any $m\times r$ matrix $\bH$, denote by $\mathcal{M}(\bH)$ the linear space spanned by the $r$ columns of $\bH$.
 Let $\otimes$ and $\circ$ denote
 the Kronecker product and the vector outer product, respectively. For any vector $\bh=(h_1,\ldots,h_m)^{\t}$, we write %$\bar{\bh}={\rm Re}(\bh)-{\rm i}\,{\rm Im}(\bh)$ with ${\rm i}=\sqrt{-1}$,
 ${\rm Re}(\bh)=\{{\rm Re}(h_1),\ldots,{\rm Re}(h_m)\}^{\t}$ and ${\rm Im}(\bh)=\{{\rm Im}(h_1),\ldots,{\rm Im}(h_m)\}^{\t}$, where ${\rm Re}(h_i)$ and ${\rm Im}(h_i)$ denote, respectively, the real part and the imaginary part of $h_i$.

\section{Model}\label{sec:model}

We impose a low-dimensional dynamic structure in model (\ref{a1}) as follows:
\begin{equation} \label{b1}
\eulbY = \sum_{\ell=1}^d \ba_\ell \circ \bb_\ell \circ \bx_\ell + \eulbE\,,
\end{equation}
where $\ba_\ell=(a_{1,\ell},\ldots,a_{p,\ell})^{\t}$ and $\bb_\ell=(b_{1,\ell},\ldots,b_{q,\ell})^{\t}$ are, respectively, $p\times 1$ and $q\times 1$
constant vectors, $\bx_\ell=(x_{1,\ell},\ldots,x_{n,\ell})^{\t}$ is an $n\times 1$ random vector, and $1\leq d<\min(p,q)$ is an unknown integer.
Put
\begin{equation*}%\label{b2}
\bA\equiv (a_{i,\ell})_{p\times d} =(\ba_1, \ldots, \ba_d)~~\textrm{and}~~
\bB\equiv (b_{j,\ell})_{q\times d} =(\bb_1, \ldots, \bb_d)\,.%~\bX \equiv (x_{t,k})_{n\times d}=
%(\bx_1, \ldots, \bx_d)\,.
\end{equation*}
Then componentwisely (\ref{b1}) admits the representation
\begin{equation} \label{b0}
 y_{i,j,t} = \sum_{\ell=1}^d a_{i,\ell} b_{j,\ell} x_{t,\ell} + \ve_{i,j,t}\,.
\end{equation}
Hence the dynamic structure in $\eulbY$ is entirely determined by
that of the $d$ time series $\bx_1, \ldots, \bx_d$. There is a clearly scaling
indeterminacy in (\ref{b1}), as the triple $(\ba_\ell, \bb_\ell, \bx_\ell)$ can be replaced
by $(\alpha_\ell\ba_\ell, \beta_\ell\bb_\ell, \ga_\ell\bx_\ell)$ as long as $\alpha_\ell \beta_\ell \gamma_\ell=1$.
We assume that all $\ba_\ell$ and $\bb_\ell$ are unit vectors (i.e. $|\ba_\ell|_2=
|\bb_\ell|_2=1$). Once $\ba_\ell$ and $\bb_\ell$ are specified, $|\bx_\ell|_2$ will be determined
by (\ref{b1}) accordingly. Note that $\ba_1, \ldots, \ba_d$ (or $\bb_1, \ldots, \bb_d$) are not required to be orthogonal with each other.

Model (\ref{b1}) is resulted from  applying the CP-decomposition to $\eulbX$ in (\ref{a1}),
where $d$ is the order of the CP-decomposition. Note that this decomposition is
unique upto the scaling and permutation indeterminacy if
$
\calR(\bA) + \calR(\bB) + \calR(\mathscr{X}) \ge 2 d + 2$,
where $\mathscr{X}=(\bx_1,\ldots,\bx_d)$ and $\calR(\cdot) = \max\{ k: {\rm any}\;k \; {\rm columns\; of \; the\; matrix} \cdot
{\rm are \; linear \; independent} \}$. Such requirement provides a sufficient
condition for the uniqueness \cite[p.467]{KoldaBader_2009}. See also Theorems 1.5 and
1.7 of \cite{DomanovDeLathauwer_2014} for more refined results on the uniqueness
of the CP-decomposition.
% Unfortunately the property of the CP decomposition for tensors is complex,
% though formally it is a natural extension of the sigular value
% decomposition for matrices. See Section 3 of Kolda and Bader (2009) and
% the reference within. To clarify the goal of our endeavor,
% we assume model (\ref{b1}) in the sense that $(\ba_k, \bb_k, \bx_k), k=1, \cdots, d$,
% exist and unique upto the scaling and permutation indeterminacy, and
% $d<<(pq)$.
%The setting adopted by
%\cite{DunlavyKoldaAcar_2011} is equivalent to (\ref{b1}) with $\eulbE \equiv \bzero$.

Though $y_{i,j,t}$ is a linear combination of $x_{t,1}, \ldots, x_{t,d}$
under (\ref{b1}), the factor representation of the model admits some
special structure, i.e. the elements of the factor loading matrix are of the
form of $a_{i,\ell}b_{j,\ell}$; see (\ref{b0}).
In fact, we need to identify and estimate all the vectors in the
first term on the RHS of (\ref{b1}) precisely (upto the permutation and
scaling indeterminacy). Therefore the conventional factor model estimation
methods such as \cite{LamYao_2012} and \cite{ChangGuoYao_2015} do not apply.

The frontal slice equation of (\ref{b1}) admits the form
\begin{equation} \label{b3}
\bY_t = \sum_{\ell=1}^d  \ba_\ell \circ \bb_\ell \,x_{t,\ell} + \bve_t
= \sum_{\ell=1}^d x_{t,\ell} \, \ba_\ell \bb_\ell^{\t}  + \bve_t = \bA \bX_t \bB^{\t} + \bve_t\,,
\end{equation}
where $\bX_t = \diag(x_{t,1}, \ldots, x_{t,d})$ and $\bve_t$ denotes the $p \times q$ matrix with $\ve_{i,j,t}$ as its $(i,j)$-th element. We impose the following regularity condition on the model. %error sequence $\{\bve_t\}_{t=1}^n$:
\begin{Condition}\label{as:1}
It holds that ${\rm rank}(\bA)=d={\rm rank}(\bB)$. Furthermore,
 $
\mathbb{E}(\bve_{t})=\bzero$ for any $t$,
$\mathbb{E}(\bve_{t} \otimes \bve_{s}) =\bzero$ for all $t \ne s$, and
$\mathbb{E}(x_{t,\ell} \bve_{s})=\bzero$ for any $\ell\in[d]$ and $t\leq s$.
\end{Condition}

\begin{remark}
Write ${\bf f}_t=(x_{t,1},\ldots,x_{t,d})^{\t}$. Model \eqref{b3} is then equivalent to
\begin{equation}\label{eq:m2}
{\rm vec}(\bY_t)=(\bb_1\otimes\ba_1,\ldots,\bb_d\otimes \ba_d){\bf f}_t+{\rm vec}(\bve_t)\,.
\end{equation}
%Another idea for modelling $\bY_t$ is to stack it as a $(pq)$-dimensional vector time series ${\rm vec}(\bY_t)$ and then consider the classic factor model:
This may entice to consider a factor model for the vector process
${\rm vec}(\bY_t)$ directly:
\begin{equation}\label{eq:m3}
{\rm vec}(\bY_t)=\bC\tilde{\bf f}_t+\tilde{\bve}_t\,,
\end{equation}
where $\bC$ is a $(pq)\times d$ loading matrix, $\tilde{\bf f}_t$ is a $d\times 1$ factor, and $\tilde{\bve}_t$ is an error term. In comparison to %the classical factor model
\eqref{eq:m3}, our model \eqref{b3} has the following advantages:
(i) The number of parameters to be estimated in \eqref{b3} is $(p+q)d$ which is smaller than $pqd$, i.e. the number of parameters in \eqref{eq:m3}, and %There are $(p+q)d$ unknown parameters in model \eqref{eq:m2}, while there are $pqd$ unknown parameters in model \eqref{eq:m3}.
(ii) model \eqref{b3} preserves the original column and row structures of the data while model \eqref{eq:m3} does not.
More precisely, the row and column variables of $\bY_t$
are typically
of different nature. For example, the rows stand for
$p$ individuals and the columns stand for $q$ indices.
Note that \eqref{b3} implies
%in In real matrix-valued time series,  the observations in the same column or in the same row usually have stronger inter-relationship. Recall $\bB=(b_{j,\ell})_{q\times d}$. If we focus on $\bY_{\cdot,j,t}$, the $j$-th column of $\bY_t$, it follows from \eqref{eq:m2} that
$
\bY_{\cdot,j,t}=\sum_{\ell=1}^db_{j,\ell}\ba_\ell x_{t,\ell}+\bve_{\cdot,j,t}$,
i.e. the dynamic part of the $j$-th column $\bY_{\cdot,j,t}$ of
$\bY_t$ is a randomly weighted linear combination of
$\ba_1, \ldots, \ba_d$. By the symmetry, the dynamic part of
any row of $\bY_t$ is a randomly weighted linear combination of $\bb_1, \ldots, \bb_d$.
%where $\bve_{\cdot,j,t}$ is the $j$-th column of $\bve_t$. %Hence, the inter-relationship among the observations in the same column of $\bY_t$ can be reflected by $\ba_1,\ldots,\ba_d$. Analogously, the inter-relationship among the observations in the same row of $\bY_t$ can be reflected by $\bb_1,\ldots,\bb_d$. Since the classical factor model does not impose any structural assumption between different rows (or columns) of the loading matrix $\bC$, model \eqref{eq:m3} does not maintain the matrix structure of $\bY_t$ and will be difficult to interpret.
In contrast model \eqref{eq:m3} treats the rows and
the columns of $\bY_t$ on an equal footing; losing
the original meaning and interpretation of the matrix process.
\end{remark}

\section{Methodology}\label{sec:method}

\subsection{Direct estimation for $\bA$, $\bB$ and $d$} \label{sec31}

%To present our basic idea for estimating $\bA, \bB$ and $d$, we first assume ${\rm rank}(\bA)={\rm rank}(\bB)=d$. As we will discuss in Section \ref{sec32}, such requirement is not necessary for our refined estimation procedure.
Without loss of generality, we assume $q\leq p$ in this section, as $\bA$ and $\bB$ are on the equal footing in model
(\ref{b1}); see also (\ref{b0}). Then both the identification and the estimation
of $\bA$, $\bB$ and $d$ essentially reduce to solving a generalized eigenequation defined by two rank-reduced $q\times q$ matrices.

\subsubsection{Identification}
%Recall $\bA$ and $\bB$ are, respectively, $p\times d$ and $q\times d$ matrices with $d<\min(p,q)$. We assume rank$(\bA)=d={\rm rank}(\bB)$.
Let $\bB^+\equiv (\bb^1,\ldots,\bb^d)^{\t}$ be the Moore-Penrose inverse of
$\bB$, i.e. $\bb_k^{\t} \bb^\ell=I(k=\ell)$ for $k, \ell \in [d]$.  Hence it follows from (\ref{b3}) that
\begin{equation} \label{b4}
\bY_t \bb^\ell = x_{t,\ell} \ba_\ell + \bve_t\bb^\ell\,, \quad \ell \in[d]\,.
\end{equation}
When $\bve_t \equiv \bzero$, this leads to $\bY_t \bb^\ell = \la \bY_{t+1} \bb^\ell$ with $\lambda=x_{t,\ell}/x_{t+1, \ell}$.
Thus, $\bb^\ell$ can be obtained from solving this generalized eigenequation. This is essentially the idea of \cite{SanchezKowalski_1990}.
We proceed differently from this point onwards  in order (i) to eliminate the impact
of non-zero $\bve_t$, (ii) to increase the estimation efficiency by
augmenting the information over time $t$, and (iii) to improve the estimation performance in solving a generalized eigenequation with rank-reduced matrices.
%, as the number of the eigenvalues of such an equation may be 0, finite  or infinite; see Section 7.7 of \cite{GolubVan_2013}.
%To highlight the key idea of our new approach, we proceed within this subsection with the assumption that $\{\bY_t\}$ and $\{\bve_t\}$ are both weakly stationary.
%This stationarity condition will be relaxed hereafter when
%we formally present the identification result,  the estimation method,
%and the associated asymptotic theory.
%Condition C1 below assumes that $y_{ijt}$ is weakly stationary in $t$
%and $\bve_t$ is white noise.
% Thus
%\[
%{\rm Cov}(y_{ij, {t-k}}, \bY_t) \bb^\ell = {\rm Cov}(y_{ij, {t-k}}, x_{t\ell}) \ba_\ell, \quad k\ne 0.
%\]
%Note that ${\rm Cov}(y_{ij, {t-k} } x_{t\ell})$ does not depend on $t$.

Let
$\xi_{t}$ be a linear combination of $\bY_t$. For any $k\geq 1$ and $t\geq k+1$, we define
$
\bXi_{t,k}=\mathbb{E}[\{\bY_t-\mathbb{E}(\bar{\bY})\}\{\xi_{t-k}-\mathbb{E}(\bar{\xi})\}]
$
with $\bar{\bY}=n^{-1}\sum_{t=1}^n\bY_t$ and $\bar{\xi}=n^{-1}\sum_{t=1}^n\xi_t$. Let
\begin{align} \label{SigmaYxi}
\bSigma_{\bY,\xi}(k)=\frac{1}{n-k}\sum_{t=k+1}^n\bXi_{t,k}
\end{align}
for any $k\geq1$. Furthermore, write $\lambda_{t,k,\ell}=\mathbb{E}[\{\xi_{t-k}-\mathbb{E}(\bar{\xi})\}\{x_{t,\ell}-\mathbb{E}(\bar{x}_{\cdot,\ell})\}]$ with $\bar{x}_{\cdot,\ell}=n^{-1}\sum_{t=1}^nx_{t,\ell}$ for any $k\geq1$, $t\geq k+1$ and $\ell\in[d]$. By \eqref{b4} and Condition \ref{as:1}, it holds that $\bXi_{t,k}\bb^\ell=\lambda_{t,k,\ell}\ba_\ell$, which implies
\begin{align}\label{eq:identA}
\bSigma_{\bY,\xi}(k)\bb^\ell=\bigg(\frac{1}{n-k}\sum_{t=k+1}^n\lambda_{t,k,\ell}\bigg)\ba_\ell\,,\quad \ell \in[d]\,.
\end{align}
Then, we have
\begin{equation}\label{eq:bl}
 \bSigma_{\bY,\xi}(2)\bb^\ell =\tilde{\lambda}_\ell\bSigma_{\bY,\xi}(1)\bb^\ell~~\textrm{with}~~\tilde{\lambda}_\ell=\frac{(n-1)\sum_{t=3}^n\lambda_{t,2,\ell}}{(n-2)\sum_{t=2}^n\lambda_{t,1,\ell}}\,.
 \end{equation}
 Write
 \begin{align*}
\bK_{1,q}={\bSigma}_{\bY,\xi}(1)^{\t}{\bSigma}_{\bY,\xi}(1)~~\textrm{and}~~\bK_{2,q}={\bSigma}_{\bY,\xi}(1)^{\t}{\bSigma}_{\bY,\xi}(2)\,.\notag%\\
%\bK_1^*={\bSigma}_{\bY,\xi}(1){\bSigma}_{\bY,\xi}(1)^{\T}\,,~~~\bK_2^*={\bSigma}_{\bY,\xi}(1){\bSigma}_{\bY,\xi}(2)^{\T}\,.
\end{align*}
 Hence the rows of $\bB^+=(\bb^1,\ldots,\bb^d)^{\t}$ are the eigenvectors of the generalized eigenequation
\begin{equation} \label{qeigen1}
\bK_{2,q}\bb = \la \bK_{1,q}\bb\,.
\end{equation}
This is a generalized eigenequation defined by rank-reduced matrices $\bK_{1,q}$ and $\bK_{2,q}$. In general, the number of eigenvalues of a generalized eigenequation defined by rank-reduced matrices may be 0, finite  or infinite; see Section 7.7 of \cite{GolubVan_2013}. However, since $\bK_{1,q}$ is positive
definite with rank $d$, (\ref{qeigen1}) admits exactly
$d$ eigenvalues. To verify this statement,
recall $\min(p,q)>d$ and $\bSigma_{\bY,\xi}(1)=\bA\bLambda\bB^{\t}$ for some $d\times d$ diagonal matrix $\bLambda$. If the elements in the main-diagonal of $\bLambda$ are nonzero, together with Condition \ref{as:1}, we know ${\rm rank}(\bK_{1,q})=d$.
Hence $\bK_{1,q}=\bGamma \bC\bGamma^{\t}$, where $\bGamma$
is a $q\times q$ orthogonal matrix, and $\bC={\rm diag}(c_1,\ldots,c_d,0,\ldots,0)$ with $c_1\geq\cdots\geq c_d>0$.
%Performing the spectral decomposition to $\bK_{1,q}$, there exists a $q\times q$ orthogonal matrix $\bGamma$ such that $\bK_{1,q}=\bGamma \bC\bGamma^{\t}$, where $\bC={\rm diag}(c_1,\ldots,c_d,0,\ldots,0)$ with $c_1\geq\cdots\geq c_d>0$.
Then the characteristic equation of the generalized eigenequation \eqref{qeigen1} is
$$
0={\rm det}(\bK_{2,q}-\lambda\bK_{1,q})={\rm det}^2(\bGamma){\rm det}(\bGamma^{\t}\bK_{2,q}\bGamma-\lambda\bC).$$
The RHS of the above equation is
a polynomial in $\lambda$ of order $d$, which, therefore,
has $d$ roots.

Let $\tilde{\lambda}_1,\ldots,\tilde{\lambda}_d$ specified in \eqref{eq:bl} be distinct. Then the rows of $\bB^+$ can be identified by \eqref{qeigen1} upto the scaling and permutation indeterminacy. However, to specify
$\bB^+$ completely, both the length and direction of each row need to be
determined precisely, which is beyond what can be learned
from \eqref{qeigen1}. Nevertheless the eigenvectors of \eqref{qeigen1} can identify the columns of
$\bA=(\ba_1,\ldots,\ba_d)$ based on the following identity:
%Notice that $|\ba_\ell|_2=1$. By \eqref{eq:identA}, the columns of $\bA=(\ba_1,\ldots,\ba_d)$ can be uniquely identified by
\begin{equation}\label{eq:ideaell}
\ba_\ell=\frac{\bSigma_{\bY,\xi}(1)\bb^\ell}{|\bSigma_{\bY,\xi}(1)\bb^\ell|_2}\,,\quad \ell\in[d]\,,
\end{equation}
which is implied by \eqref{eq:identA}. For $\bA$ specified above, let $\bA^+=(\ba^1,\ldots,\ba^d)^{\t}$ be its Moore-Penrose inverse. By the symmetry,  the columns of $\bB=(\bb_1,\ldots,\bb_d)$ are uniquely identified by
\begin{align}\label{eq:idenB}
\bb_\ell=\frac{\bSigma_{\bY,\xi}(1)^{\t}\ba^\ell}{|\bSigma_{\bY,\xi}(1)^{\t}\ba^\ell|_2}\,,\quad \ell\in[d]\,.
\end{align}

\subsubsection{Estimation} \label{sec312}
With the available observations $\bY_1, \ldots,\bY_n$, we define
\begin{align}\label{eq:hatSigmaYxi}
\wh \bSigma_{\bY,\xi}(k)  = \frac{1}{n-k}\sum_{t=k+1}^n
(\bY_t - \bar \bY)(\xi_{t-k} - \bar \xi)
\end{align}
for $k\geq 1$. When $pq\gg n$, $\wh \bSigma_{\bY,\xi}(k)$ is no longer a consistent estimator for $\bSigma_{\bY,\xi}(k)$ under the spectral norm $\|\cdot\|_2$. In the spirit of \cite{BickelandLevina_2008}, we select $\widehat{\bSigma}_k$ defined as follows for the estimate of $\bSigma_{\bY,\xi}(k)$:
\begin{equation}\label{eq:thresholdop}
\widehat{\bSigma}_{k}=T_{\de_1}\{ \widehat{\bSigma}_{\bY,\xi}(k) \}\,,%\quad k\in\{1,2\}\,,
\end{equation}
where $T_{\de_1}(\cdot)$ is a threshold operator
$
T_{\de_1}(\bW) = \{ w_{i,j} I(|w_{i,j}|\ge \de_1) \}_{m_1 \times m_2}
$
for any matrix $\bW=(w_{i,j})_{m_1 \times m_2}$ with the threshold level $\delta_1\geq0$. We choose $\de_1>0$ when
$pq \gg n$. When $\delta_1=0$, we have $\widehat{\bSigma}_k=\widehat{\bSigma}_{\bY,\xi}(k)$, which is appropriate when, for example, $p$ and $q$ are fixed constants. Then
 %\begin{align*}
$\widehat{\bK}_{1,q}=\widehat{\bSigma}_1^{\t}\widehat{\bSigma}_1$ and $\widehat{\bK}_{2,q}=\widehat{\bSigma}^{\t}_1\widehat{\bSigma}_2$
%\widehat{\bK}_1^*=\widehat{\bSigma}_1\widehat{\bSigma}_1^{\T}\,,~~~\widehat{\bK}_2^*=\widehat{\bSigma}_1\widehat{\bSigma}_2^{\T}\,,
%\end{align*}
provide the estimates of $\bK_{1,q}$ and $\bK_{2,q}$, respectively. %$\bK_1^*$ and $\bK_2^*$, respectively.

Let $\hat{\lambda}_1(\widehat{\bK}_{1,q})\geq \cdots\geq \hat{\lambda}_{q}(\widehat{\bK}_{1,q})\geq0$ be the eigenvalues of $\widehat{\bK}_{1,q}$. Since ${\rm rank}(\bK_{1,q})=d$, following \cite{ChangGuoYao_2015}, we can estimate $d$ as
\begin{align}\label{eq:ratio1}
\hat{d}=\arg\min_{j\in[R]}\frac{\hat{\lambda}_{j+1}(\widehat{\bK}_{1,q})+c_n}{\hat{\lambda}_{j}(\widehat{\bK}_{1,q})+c_n}\,,
\end{align}
where $R=\lfloor \alpha \min(p,q)\rfloor$ for a prescribed constant $\alpha\in(0,1)$ and some $c_n\rightarrow0^+$ as $n\rightarrow\infty$. In practice, we may set $\alpha=0.5$. Note that the true eigenvalues of $\bK_{1,q}$ satisfy the condition
$\la_1(\bK_{1,q}) \ge \cdots \ge \la_d(\bK_{1,q}) >0 = \la_{d+1}(\bK_{1,q}) = \cdots = \la_{q}(\bK_{1,q})$.
Adding a small constant $c_n>0$ in \eqref{eq:ratio1} is to avoid the ratio `$0/0$'. Under some regularity conditions, $\hat{d}$ defined in \eqref{eq:ratio1} is a consistent estimate for $d$ in the sense that $\mathbb{P}(\hat{d}=d)\rightarrow1$ as $n\rightarrow\infty$.

Applying the spectral decomposition to $\widehat{\bK}_{1,q}$, we have
$
\widehat{\bK}_{1,q}=\widehat{\bGamma}\widehat{\bC}\widehat{\bGamma}^{\t}$, where $\widehat{\bGamma}=(\hat{\bgamma}_1,\ldots,\hat{\bgamma}_q)$ is a $q\times q$ orthogonal matrix, and $\widehat{\bC}={\rm diag}(\hat{c}_1,\ldots,\hat{c}_q)$ with $\hat{c}_1\geq\cdots\geq\hat{c}_q\geq0$. For $\hat{d}$ specified in \eqref{eq:ratio1}, we define
\begin{equation}\label{eq:tildeK}
\widetilde{\bK}_{1,q}=\sum_{j=1}^{\hat{d}}\hat{c}_j\hat{\bgamma}_j\hat{\bgamma}_j^{\t}\,,
\end{equation}
which is a truncated version of $\widehat{\bK}_{1,q}$. Then ${\rm rank}(\widetilde{\bK}_{1,q})=\hat{d}$. Let $\hat{\bb}^1,\ldots,\hat{\bb}^{\hat{d}}$ be the eigenvectors of the generalized eigenequation
\begin{align}\label{eq:samplegenequ}
\widehat{\bK}_{2,q}\bb=\lambda\widetilde{\bK}_{1,q}\bb\,,
\end{align}
which is a sample version of \eqref{qeigen1}. We can use the function {\tt geigen} in the R package {\tt geigen} to solve \eqref{eq:samplegenequ}. Then the columns of $\bA$ can be estimated as
\begin{equation}\label{eq:esta}
\hat{\ba}_\ell=\frac{\widehat{\bSigma}_1\hat{\bb}^\ell}{|\widehat{\bSigma}_1\hat{\bb}^\ell|_2}\,,~~~~\ell\in[\hat{d}]\,.
\end{equation}
%Analogously, applying the spectral decomposition to $\widehat{\bK}_1^*$, we have $\widehat{\bK}_1^*=\widehat{\bGamma}^*\widehat{\bC}^*(\widehat{\bGamma}^*)^{\T}$, where $\widehat{\bGamma}^*=(\hat{\bgamma}_1^*,\ldots,\hat{\bgamma}_p^*)$ is a $p\times p$ orthogonal matrix, and $\widehat{\bC}^*={\rm diag}(\hat{c}_1^*,\ldots,\hat{c}_p^*)$ with $\hat{c}_1^*\geq\cdots\geq\hat{c}_p^*\geq0$. Define
%\begin{equation}\label{eq:tildeK*}
%\widetilde{\bK}_1^*=\sum_{j=1}^{\hat{d}}\hat{c}_j^*\hat{\bgamma}_j^*(\hat{\bgamma}_j^*)^{\T}
%\end{equation}
%and let $\hat{\ba}^1,\ldots,\hat{\ba}^{\hat{d}}$ be the eigenvectors of the generalized eigenequation $\widehat{\bK}_2^*\ba=\lambda\widetilde{\bK}_1^*\ba$.
Let $\widehat{\bA}^+=(\hat{\ba}^1,\ldots,\hat{\ba}^{\hat{d}})^{\t}$ be the Moore-Penrose inverse of $\widehat{\bA}=(\hat{\ba}_1,\ldots,\hat{\ba}_{\hat{d}})$. Then the columns of $\bB$ can be estimated as
\begin{equation}\label{eq:estb}
\hat{\bb}_\ell=\frac{\widehat{\bSigma}_1^{\t}\hat{\ba}^\ell}{|\widehat{\bSigma}_1^{\t}\hat{\ba}^\ell|_2}\,,~~~~\ell\in[\hat{d}]\,.
\end{equation}
The truncation of $\widehat{\bK}_{1,q}$ given in \eqref{eq:tildeK} is necessary here for estimating the rows of $\bB^+$. Note that $\widehat{\bK}_{1,q}$ is a $q\times q$ matrix with $q>d$. Since ${\rm rank}(\widehat{\bK}_{1,q})$ may be larger than $\hat{d}$ in finite samples, the generalized eigenequation $\widehat{\bK}_{2,q}\bb=\lambda\widehat{\bK}_{1,q}\bb$ may have more than $\hat{d}$ eigenvectors. Since we do not know which eigenvalues %of  $\widehat{\bK}_{2,q}\bb=\lambda\widehat{\bK}_{1,q}\bb$
are associated with our required eigenvectors, it will be extremely difficult (if not impossible) for us to pick out $\hat{\bb}^1,\ldots,\hat{\bb}^{\hat{d}}$ from all the eigenvectors of $\widehat{\bK}_{2,q}\bb=\lambda\widehat{\bK}_{1,q}\bb$.

Based on $\hat{\ba}_\ell$ and $\hat{\bb}_\ell$ specified in \eqref{eq:esta} and \eqref{eq:estb}, we define
$$
\widehat{\bH}=(\hat{\bb}_1\otimes\hat{\ba}_1,\ldots,\hat{\bb}_{\hat{d}}\otimes\hat{\ba}_{\hat{d}})\,.
$$
By \eqref{eq:m2}, we can recover $\bX_t$ by $\widehat{\bX}_t={\rm diag}(\hat{x}_{t,1},\ldots,\hat{x}_{t,\hat{d}})$ with
$$(\hat{x}_{t,1},\ldots,\hat{x}_{t,\hat{d}})^{\t}= \wh\bH^{+}\mathrm{vec}(\bY_t)\,.$$
We need to point out that the eigenvalues of the generalized eigenequation \eqref{eq:samplegenequ} are not necessary to be real. Proposition \ref{conj} shows that  its complex eigenvalues always occur in complex conjugate pairs. %Let $\lambda$ and $\bar{\lambda}$ be two complex conjugate eigenvalues of \eqref{eq:samplegenequ}. The next proposition summarizes the relationship between their associated $(\hat{\ba}_\ell,\hat{\bb}_\ell,\hat{x}_{t,\ell})$'s.

\begin{proposition}\label{conj}
Assume the eigenvalues of the generalized eigenequation \eqref{eq:samplegenequ} are distinct. If $\lambda_*\in\mathbb{C}$ is a complex eigenvalue of
\eqref{eq:samplegenequ} such that $\widehat{\bK}_{2,q}\hat{\bb}^\ell=\lambda_*\widetilde{\bK}_{1,q}\hat{\bb}^\ell$ for some $\ell\in[\hat{d}]$, then $\overline{\lambda_*}$, the complex conjugate of $\lambda_*$, is also a complex eigenvalue of \eqref{eq:samplegenequ}. More specifically, there exists some $\tilde{\ell}\in[\hat{d}]$ and a constant $\kappa\in\{-1,1\}$ satisfying $\widehat{\bK}_{2,q}\hat{\bb}^{\tilde{\ell}}=\overline{\lambda_*}\widetilde{\bK}_{1,q}\hat{\bb}^{\tilde{\ell}}$, $\hat{\ba}_{\tilde{\ell}}=\kappa\overline{\hat{\ba}_{\ell}}$, $\hat{\bb}_{\tilde{\ell}}=\kappa\overline{\hat{\bb}_{\ell}}$, and $\hat{x}_{t,\tilde{\ell}}=\overline{\hat{x}_{t,\ell}}$, where $\overline{\hat{\ba}_\ell}$, $\overline{\hat{\bb}_\ell}$ and $\overline{\hat{x}_{t,\ell}}$ are the complex conjugate of $\hat{\ba}_\ell$, $\hat{\bb}_\ell$ and $\hat{x}_{t,\ell}$, respectively.
\end{proposition}

Assume the generalized eigenequation \eqref{eq:samplegenequ} has $s$ real eigenvalues and $\hat{d}-s$ complex eigenvalues. Since the complex eigenvalues always occur in complex conjugate pairs, $\hat{d}-s$ is an even integer. Write $\hat{d}-s=2m$. Let $\lambda_1,\overline{\lambda_1},\ldots,\lambda_m,\overline{\lambda_m}$ be the $\hat{d}-s$ complex eigenvalues, where $\overline{\lambda_1},\ldots,\overline{\lambda_m}$ are the complex conjugate of $\lambda_1,\ldots,\lambda_m$, respectively. For each $j\in[m]$, there exist $\ell_j, \tilde{\ell}_j\in[\hat{d}]$ such that the eigenvectors associated with $\lambda_j$ and $\overline{\lambda_j}$ are, respectively, $\hat{\bb}^{\ell_j}$ and $\hat{\bb}^{\tilde{\ell}_j}$. By Proposition \ref{conj}, there exists $(\kappa_1,\ldots,\kappa_m)\in\{-1,1\}^m$ such that $\hat{\ba}_{\tilde{\ell}_j}=\kappa_j\overline{\hat{\ba}_{\ell_j}}$, $\hat{\bb}_{\tilde{\ell}_j}=\kappa_j\overline{\hat{\bb}_{\ell_j}}$, and $\hat{x}_{t,\tilde{\ell}_j}=\overline{\hat{x}_{t,\ell_j}}$ for each $j\in[m]$. Then
\[
\sum_{j=1}^m(\hat{x}_{t,\ell_j}\hat{\ba}_{\ell_j}\hat{\bb}_{\ell_j}^{\t}+\hat{x}_{t,\tilde{\ell}_j}\hat{\ba}_{\tilde{\ell}_j}\hat{\bb}_{\tilde{\ell}_j}^{\t})\in\mathbb{R}^{p\times q}\,.
\]
Write $\{k_1,\ldots,k_s\}=[\hat{d}]\setminus\{\ell_1,\tilde{\ell}_1,\ldots,\ell_m,\tilde{\ell}_m\}$. Then $\hat{\bb}^{k_1},\ldots,\hat{\bb}^{k_s}$ are the eigenvectors of the generalized eigenequation \eqref{eq:samplegenequ} associated with the $s$ real eigenvalues. Hence, $\hat{\ba}_{k_j}\in\mathbb{R}^p$, $\hat{\bb}_{k_j}\in\mathbb{R}^q$ and $\hat{x}_{t,k_j}\in\mathbb{R}$ for each $j\in[s]$. To do prediction of $\bY_t$ based on \eqref{b3}, we only need to model $\hat{d}$ univariate time series $\{\hat{x}_{t,k_1}\},\ldots,\{\hat{x}_{t,k_s}\}, \{{\rm Re}(\hat{x}_{t,\ell_1})\}, \{{\rm Im}(\hat{x}_{t,\ell_1})\},\ldots, \{{\rm Re}(\hat{x}_{t,\ell_m})\}, \{{\rm Im}(\hat{x}_{t,\ell_m})\}$.

\begin{remark}\label{re:2}

Solving a generalized eigenequation defined by rank-reduced matrices could be a complex computational task. See Section 7.7 of \cite{GolubVan_2013}. In principle we can also estimate $\bA^+$ first; leading to the estimate for $\bB$ and then that for $\bA$. Technically this
boils down to solving a generalized eigenequation defined by two $p\times p$ rank-reduced matrices, which is computationally more expensive and less stable by using the R function {\tt geigen} when $p>q$; often leading to, for example, more than $\hat d$ eigenvalues/vectors.

\end{remark}

\subsection{A refined estimation procedure} \label{sec32}
To overcome the complication in solving a rank-reduced generalized eigenequation, which plays the key role in
the method proposed in Section  \ref{sec31},
we propose a refinement which reduces the $q$-dimensional
rank-reduced generalized eigenequation to a $d$-dimensional
full-ranked one. Therefore, effectively the new refined method only requires to solve a $d$-dimensional eigenequation.
Simulation results in Section \ref{sec:numerical} indicate that this new procedure
%procedure based on the refined idea
outperforms the direct estimation, proposed in
Section \ref{sec312}, uniformly over various settings. %Proposition \ref{conj} shows that the basic estimation procedure may estimate some $\ba_\ell$ and $\bb_\ell$ as complex vectors. Although the ref

\subsubsection{Identification}
For a prescribed integer $K\ge 1$, define
\begin{align}\label{eq:m1m2stat}
\bM_1 = \sum_{k=1}^K \bSigma_{\bY,\xi}(k) \bSigma_{\bY,\xi}(k)^{\t}~~\textrm{and}~~
\bM_2 = \sum_{k=1}^K \bSigma_{\bY,\xi}(k)^{\t} \bSigma_{\bY,\xi}(k)
\end{align}
with $\bSigma_{\bY,\xi}(k)$ defined as \eqref{SigmaYxi}. Recall $
\bSigma_{\bY,\xi}(k)=\bA\bG_k\bB^{\t}$, where $\bG_k= (n-k)^{-1}\sum_{t=k+1}^n\mathbb{E}[\{\bX_t-\mathbb{E}(\bar{\bX})\}\{\xi_{t-k}-\mathbb{E}(\bar{\xi})\}]$ is a $d\times d$ diagonal matrix with $\bar{\bX}=n^{-1}\sum_{t=1}^n\bX_t$ and $\bar{\xi}=n^{-1}\sum_{t=1}^n\xi_t$; see \eqref{b3}. Then
\begin{align}\label{eq:M1M2}
\bM_1=\bA\bigg(\sum_{k=1}^K\bG_k\bB^{\t}\bB\bG_k\bigg)\bA^{\t}~~\textrm{and}~~\bM_2=\bB\bigg(\sum_{k=1}^K\bG_k\bA^{\t}\bA\bG_k\bigg)\bB^{\t}\,.
\end{align}
Since both $p$ and $q$ are much greater than $d$ in practice, it is reasonable to impose the following assumption.
\begin{Condition}\label{as:2}
It holds that ${\rm rank}(\bM_1)=d={\rm rank}(\bM_2)$. Furthermore the nonzero eigenvalues of $\bM_1$ and $\bM_2$ are uniformly bounded away from zero.
\end{Condition}

\begin{remark}
Write $\bG=(\bG_1^{\t},\ldots,\bG_K^{\t})^{\t}$. Then $\bM_1=\bA\bG^{\t}(\bI_K\otimes \bB)^{\t}(\bI_K\otimes \bB)\bG\bA^{\t}$, which implies that ${\rm rank}(\bM_1)={\rm rank}\{(\bI_K\otimes\bB)\bG\bA^{\t}\}$. Notice that each $\bG_k$ is a $d\times d$ diagonal matrix and $(\bI_K\otimes\bB)\bG\bA^{\t}=(\bA\bG_1\bB^{\t},\ldots,\bA\bG_K\bB^{\t})^{\t}$. It holds that $d\geq{\rm rank}(\bG)\geq{\rm rank}\{(\bI_K\otimes\bB)\bG\bA^{\t}\}\geq \max_{k\in[K]}{\rm rank}(\bA\bG_k\bB^{\t})$. Since $\bSigma_{\bY,\xi}(k)=\bA\bG_k\bB^{\t}$, ${\rm rank}(\bM_1)=d$ provided that there exists some $k\in[K]$ such that ${\rm rank}\{\bSigma_{\bY,\xi}(k)\}=d$. By the same argument, ${\rm rank}(\bM_2)=d$ provided that there exists some $k\in[K]$ such that ${\rm rank}\{\bSigma_{\bY,\xi}(k)\}=d$. Since ${\rm rank}(\bA)=d={\rm rank}(\bB)$ (see Condition \ref{as:1}),  ${\rm rank}\{\bSigma_{\bY,\xi}(k)\}={\rm rank}(\bG_k)$. Consequently, ${\rm rank}(\bM_1)=d={\rm rank}(\bM_2)$ if the elements in the main diagonal of some $\bG_k$ are nonzero.
\end{remark}

   Perform the spectral decomposition:
\begin{align*}%\label{eq:svd1}
\bM_1=\bP\bLambda_{1}\bP^{\t}~~\textrm{and}~~\bM_2=\bQ\bLambda_2\bQ^{\t}\,,
\end{align*}
where % $\bP$ and $\bQ$ are, respectively, $p\times d$ and $q\times d$ matrices,
the columns of $\bP$ and $\bQ$ are, respectively, the $d$ orthonormal eigenvectors corresponding to the $d$ non-zero eigenvalues of $\bM_1$ and $\bM_2$,  $\bLambda_{1}$ and $\bLambda_2$ are the
diagonal matrices with the corresponding eigenvalues as the diagonal
elements. This, together with \eqref{eq:M1M2}, implies that
\begin{align}\label{eq:AB}
\bA=\bP\bU~~\textrm{and}~~\bB=\bQ\bV\,,
\end{align}
where $\bU$ and $\bV$ are two $d\times d$ invertible matrices. Furthermore
all the columns of $\bU$ and $\bV$ are unit vectors, which is implied by
the assumption that all $\ba_\ell$ and $\bb_\ell$ are unit vectors.

To identify $\bA$ and $\bB$, we only need to identify $\bU$ and $\bV$, which can be solved from a generalized eigenequation with two full-ranked matrices. To this end, define $d\times d$ matrix process $\bZ_t = \bP^{\t} \bY_t \bQ$. It follows from (\ref{b3}) and (\ref{eq:AB}) that
\begin{equation*} %\label{eq:Z}
\bZ_t = \bU \bX_t \bV^{\t} + \bDelta_t = \sum_{\ell=1}^d x_{t,\ell} \bu_\ell \bv_\ell^{\t} + \bDelta_t\,,
\end{equation*}
where $\bDelta_t= \bP^{\t} \bve_t \bQ$ is uncorrelated with $\{\bX_s\}_{s\leq t}$, %and
$\bu_\ell$ and $\bv_\ell$ are, respectively, the $\ell$-th column of $\bU$ and $\bV$. Choose
$\eta_t$ to be a linear combination of $\bZ_t$ such that
\begin{align}\label{eq:sigmazeta}
\bSigma_{\bZ,\eta}(k)=\frac{1}{n-k}\sum_{t=k+1}^n\mathbb{E}[\{\bZ_t-\mathbb{E}(\bar{\bZ})\}\{\eta_{t-k}-\mathbb{E}(\bar{\eta})\}]
%=&~\bU\bigg(\frac{1}{n-k}\sum_{t=k+1}^n\mathbb{E}[\{\bX_t-\mathbb{E}(\bar{\bX})\}\{\eta_{t-k}-\mathbb{E}(\bar{\eta})\}]\bigg)\bV^{\T}\,,
\end{align}
is full-ranked for $k\in\{1,2\}$, where $\bar{\bZ}=n^{-1}\sum_{t=1}^n\bZ_t$, and $\bar{\eta}=n^{-1}\sum_{t=1}^n\eta_t$. Then the same argument towards
(\ref{qeigen1}) implies that the rows of the $d\times d$ inverse matrix $\bV^{-1}=(\bv^1, \ldots, \bv^d)^{\t}$ are the eigenvectors of the generalized eigenequation
\begin{equation} \label{qeigen2}
\bSigma_{\bZ,\eta}(1)^{\t} \bSigma_{\bZ,\eta}(2)\bv = \la \bSigma_{\bZ,\eta}(1)^{\t} \bSigma_{\bZ,\eta}(1)\bv\,,
\end{equation}
which has exactly $d$ eigenvectors. Furthermore
those $d$ eigenvectors are unique
upto the scaling indeterminacy if the $d$ eigenvalues associated with \eqref{qeigen2} are distinct. %Let $$ be the $d$ eigenvectors of (\ref{qeigen2}).
Parallel to \eqref{eq:ideaell} and \eqref{eq:idenB}, the columns of $\bU=(\bu_1,\ldots,\bu_d)$ and $\bV=(\bv_1,\ldots,\bv_d)$ can be
identified as follows:
\begin{equation}\label{eq:ideuell}
\bu_\ell = \frac{\bSigma_{\bZ,\eta}(1) \bv^\ell}{|\bSigma_{\bZ,\eta}(1) \bv^\ell|_2}~~\textrm{and}~~\bv_\ell = \frac{\bSigma_{\bZ,\eta}(1)^{\t} \bu^\ell}{ |\bSigma_{\bZ,\eta}(1)^{\t} \bu^\ell|_2}
\end{equation}
for each $\ell\in[d]$, where $(\bu^1,\ldots,\bu^d)^{\t}$ is the inverse of $\bU$. %By the symmetry, the columns of  are obtained as follows:
%\begin{equation}\label{eq:idevell}
%\bv_\ell = \frac{\bSigma_{\bZ,\eta}(1)^{\T} \bu^\ell}{ |\bSigma_{\bZ,\eta}(1)^{\T} \bu^\ell|_2}\,, \quad \ell \in [d]\,.
%\end{equation}
With $\bU$ and $\bV$ specified above, $\bA $ and $\bB$ can be  determined by (\ref{eq:AB}). Write
\begin{equation}\label{eq:J}
\bJ_1=\{\bSigma_{\bZ,\eta}(1)^{\t} \bSigma_{\bZ,\eta}(1)\}^{-1}
   \bSigma_{\bZ,\eta}(1)^{\t}\bSigma_{\bZ,\eta}(2)\,.
\end{equation}

%Write $\bU^{-1}=(\bu^1,\ldots,\bu^d)^{\T}$ and $\bV^{-1}=(\bv^1,\ldots,\bv^d)^{\T}$. Hence, for any $\ell\in[d]$, \eqref{eq:sigmazeta} implies that
%\begin{align}\label{b9}
%\bSigma_{\bZ,\eta}(k)\bv^\ell={c}_{k,\ell} \bu_\ell\,,
%\end{align}
%where ${c}_{k,\ell}=(n-k)^{-1}\sum_{t=k+1}^n\mathbb{E}[\{x_{t,\ell}-\mathbb{E}(\bar{x}_\ell)\}\{\eta_{t-k}-\mathbb{E}(\bar{\eta})\}]$. Consequently, for any $\ell\in[d]$,
%\begin{equation}\label{b6}
%\bSigma_{\bZ,\eta}(2)\bv^\ell = \mu_\ell \bSigma_{\bZ,\eta}(1)\bv^\ell\,,~~~
%\mu_{\ell} = {c}_{2,\ell}{c}_{1,\ell}^{-1}\,.
%\end{equation}
\begin{proposition}
     \label{prop_ID}
   Let Conditions \ref{as:1} and \ref{as:2} hold, and the eigenvalues of the $d\times d$
   matrix $\bJ_1$ specified in \eqref{eq:J} %$$\{\bSigma_{\bZ,\eta}(1)^{\T} \bSigma_{\bZ,\eta}(1)\}^{-1}
   %\bSigma_{\bZ,\eta}(1)^{\T}\bSigma_{\bZ,\eta}(2)$$
   be distinct. Then $\bA$
   and $\bB$ are uniquely defined as in (\ref{eq:AB}) upto the reflection
   and permutation indeterminacy, where the columns of
   $\bU=(\bu_1, \ldots, \bu_d)$ and $\bV=(\bv_1, \ldots, \bv_d)$ are defined as
   \eqref{eq:ideuell}. % with $\bv^\ell$ and $\bu^\ell$, $\ell \in [d]$, being, respectively, the eigenvectors of the generalized eigenequations
   %\begin{equation}
    %   \label{2qeigen}
   %\bSigma_{\bZ,\eta}(1)^{\T} \bSigma_{\bZ,\eta}(2) \bv = \la
   %\bSigma_{\bZ,\eta}(1)^{\T} \bSigma_{\bZ,\eta}(1) \bv
   %~~{\rm and} ~~
   %\bSigma_{\bZ,\eta}(1) \bSigma_{\bZ,\eta}(2)^{\T} \bu = \la
   %\bSigma_{\bZ,\eta}(1)\bSigma_{\bZ,\eta}(1)^{\T} \bu\,.
   %\end{equation}
   %Furthermore those two generalized eigenequations share the same $d$ eigenvalues.
\end{proposition}

\begin{remark} \label{rm1}
By the symmetry, we also know that $\bu^1,\ldots,\bu^d$ are the $d$ eigenvectors of the generalized eigenequation $\bSigma_{\bZ,\eta}(1) \bSigma_{\bZ,\eta}(2)^{\t} \bu = \la\bSigma_{\bZ,\eta}(1)\bSigma_{\bZ,\eta}(1)^{\t} \bu$.  Write
%\begin{equation}\label{eq:tildeJ}
$$\bJ_2=\{\bSigma_{\bZ,\eta}(1)\bSigma_{\bZ,\eta}(1)^{\t}\}^{-1}
\bSigma_{\bZ,\eta}(1) \bSigma_{\bZ,\eta}(2)^{\t}\,.$$
%\end{equation}
 It holds that $\bv^\ell$ and $\bu^\ell$ are, respectively, the eigenvectors of the $d\times d$ matrices $\bJ_1$
and
$\bJ_2$ associated with the same eigenvalue.
\end{remark}

%, and the last step is based on Condition \ref{as:1}.

%Put
%\begin{align}\label{b7}
%\bS_1 = \bSigma_{\bZ,\eta}(1)^{\T} \bSigma_{\bZ,\eta}(1)~~\textrm{and}~~\bS_2 = \bSigma_{\bZ,\eta}(1)^{\T} \bSigma_{\bZ,\eta}(2)\,.
%\end{align}
%By \eqref{b6}, we have
%\begin{equation} \label{b8}
%$\bS_2 \bv^\ell = \mu_\ell \bS_1 \bv^\ell$ for any $\ell\in[d]$, which implies the rows of $\bV^{-1}=(\bv^1,\ldots,\bv^d)^{\T}$ are the eigenvectors of the generalized eigenequation
%\begin{equation*}
%\bS_2 \bdelta = \mu  \bS_1 \bdelta\,.
%\end{equation*}
%Put
%\begin{align}\label{eq:s1*s2*}
%\bS_1^*=\bSigma_{\bZ,\eta}(1)\bSigma_{\bZ,\eta}(1)^{\T}~~\textrm{and}~~\bS_2^* = \bSigma_{\bZ,\eta}(1)\bSigma_{\bZ,\eta}(2)^{\T}\,.
%\end{align}
%By the symmetry of $\bU$ and $\bV$
%in (\ref{eq:Z}), we also know $\bu^\ell$ is the eigenvector of the generalized eigenequation
%$\bS_2^*\bdelta=\mu_\ell\bS_1^*\bdelta$ for $\mu_\ell$ defined in \eqref{b6}. Hence, we can identify $\bV=(\bv_1,\ldots,\bv_d)$ by

%\subsection{Estimation of $\bA$, $\bB$ and $d$ based on the refined idea}
\subsubsection{Estimation}\label{sec:estAB}

 Let $\eta_t=\bw^{\T}{\rm vec}(\bZ_t)$ be a linear combination of $\bZ_t$ for some constant vector $\bw\in\mathbb{R}^{d^2}$.
 Any $\bw\in\mathbb{R}^{d^2}$ such that the associated $d\times d$ matrix $\bJ_1$ specified in \eqref{eq:J} has $d$ distinct eigenvalues is valid for the identification of $\bU$ and $\bV$. See Proposition \ref{prop_ID} for details. Write  $\bTheta=\bI_p\otimes\{(\bQ\otimes \bP)\bw\}$ and
$
\bSigma_{\mathring{\bY}}(k)=(n-k)^{-1}\sum_{t=k+1}^n\mathbb{E}[\{\bY_t-\mathbb{E}(\bar{\bY})\}\otimes {\rm vec}\{\bY_{t-k}-\mathbb{E}(\bar{\bY})\}]
$. Then $\bSigma_{\bZ,\eta}(k)$ defined as \eqref{eq:sigmazeta} can be reformulated as
\begin{align}\label{eq:sigmaz}
\bSigma_{\bZ,\eta}(k)=\bP^{\T}\bTheta^{\T}\bSigma_{\mathring{\bY}}(k)\bQ\,.
\end{align}

For $\widehat{\bSigma}_{\bY,\xi}(k)$ defined as \eqref{eq:hatSigmaYxi}, %with the available observations $\bY_1, \ldots,\bY_n$,
we
define the threshold estimators for $\bM_1$ and $\bM_2$ given in \eqref{eq:m1m2stat} as follows:
\begin{align}\label{eq:mesthd}
\widehat{\bM}_{1}=\sum_{k=1}^K\widehat{\bSigma}_{k}\widehat{\bSigma}_{k}^{\T}~~\textrm{and}~~\widehat{\bM}_{2}=\sum_{k=1}^K\widehat{\bSigma}_{k}^{\T} \widehat{\bSigma}_{k}\,,
\end{align}
where $\widehat{\bSigma}_{k}$ is defined as \eqref{eq:thresholdop}. %See the arguments below \eqref{eq:thresholdop} for the discussion of $T_{\de_1}(\cdot)$. %We choose $\de_1>0$ when
%$pq \gg n$. When $\de_1=0$, $\wh\bM_1$ and $\wh\bM_2$ are defined directly based on
%the sample covariance matrices $\wh \bSigma_{\bY,\xi}(k)$ without
%truncation, which is appropriate when, for example, $p$ and $q$ are fixed.
Let $\hat{\lambda}_{1}(\widehat{\bM}_1)\geq \cdots\geq \hat{\lambda}_{p}(\widehat{\bM}_1)\ge 0$ be the eigenvalues of the $p\times p$ matrix $\wh\bM_1$. Recall ${\rm rank}(\bM_1)=d$.
%Given $(\widehat{\bM}_1,\widehat{\bM}_2)$, the estimate of $(\bM_1,\bM_2)$,
%we denote by $\hat{\lambda}_{1,1}\geq \cdots\geq \hat{\lambda}_{1,p}$ and $\hat{\lambda}_{2,1}\geq\cdots\geq\hat{\lambda}_{2,q}$, respectively, the eigenvalues of $\widehat{\bM}_1$ and $\widehat{\bM}_2$. We will specify in Section \ref{sec4} for the selections of $(\widehat{\bM}_1,\widehat{\bM}_2)$. Due to ${\rm rank}(\bM_1)=d$,
Analogous to \eqref{eq:ratio1}, we can also estimate $d$ as
\begin{align}\label{eq:ratio}
\hat{d}=\arg\min_{j\in[R]}\frac{\hat{\lambda}_{j+1}(\widehat{\bM}_1)+c_n}{\hat{\lambda}_{j}(\widehat{\bM}_1)+c_n}\,,
\end{align}
where $R$ and $c_n$ are same as those in \eqref{eq:ratio1}. The convergence rate of $c_n$ will be specified in Theorem \ref{hatd.exp} and Remark \ref{remark2} in Section \ref{sec4}. Theorem \ref{hatd.exp} shows that $\hat{d}$ is consistent, i.e. $\mathbb{P}(\hat{d}\neq d)\rightarrow0$ as $n\rightarrow\infty$.

\begin{remark} \label{rm4}
Analogously, we can also estimate $d$ by replacing $\wh\bM_1$ in \eqref{eq:ratio} by $\wh\bM_2$. Recall $\widehat{\bM}_1$ and $\widehat{\bM}_2$ are, respectively, $p\times p$ and $q\times q$ matrices. For $K=1$, since the nonzero eigenvalues of $\wh\bM_1$ and $\wh\bM_2$ are identical, such replacement will lead to a same estimate for $d$ as that by \eqref{eq:ratio}. For $K>1$, although the estimates based on $\widehat{\bM}_1$ and $\widehat{\bM}_2$ are both consistent, their finite sample performance is a little bit different.  More specifically, simulation results show that (i) the estimate based on $\widehat{\bM}_1$ has higher probability of correctly estimating $d$ when $p>q$, (ii) the estimate based on $\widehat{\bM}_2$ has higher probability of correctly estimating $d$ when $q>p$, and (iii) the estimates based on $\widehat{\bM}_1$ and $\widehat{\bM}_2$ are almost identical when $p=q$. See Tables \ref{SM:Dhat_d1}--\ref{SM:Dhat_d6} in the supplementary material for details. We suggest to estimate $d$ based on $\widehat{\bM}_1$ when $p\ge q$, and based on $\widehat{\bM}_2$ when $p < q$.
\end{remark}

Now let $\wh \bP$ be the $p\times \hat d$ matrix of which the columns are the $\hat d$ orthonormal eigenvectors of $\wh \bM_1$ corresponding to its $\hat d$ largest eigenvalues, and $\wh \bQ$ be the $q\times \hat d$ matrix of which the columns are
the $\hat d$ orthonormal eigenvectors of $\wh \bM_2$ corresponding to its
$\hat d$ largest eigenvalues.
Define
\begin{align}\label{eq:hateta}
\widehat{\bZ}_t=\widehat{\bP}^{\T}\bY_t\widehat{\bQ}~~\textrm{and}~~\hat{\eta}_t=\bw^{\T}{\rm vec}(\widehat{\bZ}_t)
\end{align}
for some constant vector $\bw\in\mathbb{R}^{\hat{d}^2}$ with bounded $\ell^2$-norm. Based on \eqref{eq:sigmaz}, we put
 \begin{align}\label{eq:estsigmazhd}
\wh \bSigma_{\bZ,\eta}(k)  = \widehat{\bP}^{\T}\widehat{\bTheta}^{\T}T_{\de_2}\{\widehat{\bSigma}_{\check{\bY}}(k)\}\widehat{\bQ}\,, %\quad k=1, 2\,,
\end{align}
where $T_{\de_2}(\cdot)$ is a threshold operator with the threshold level $\de_2\ge 0$,
$\wh\bTheta=\bI_p\otimes\{(\wh\bQ\otimes \wh\bP)\bw\}$, and
 \begin{align}\label{eq:estsigmay}
\widehat{\bSigma}_{\check{\bY}}(k)=\frac{1}{n-k}\sum_{t=k+1}^n
(\bY_t-\bar{\bY})\otimes {\rm vec}(\bY_{t-k}-\bar{\bY})\,.
\end{align}
Write $
\widehat{\bJ}_1=\{\widehat{\bSigma}_{\bZ,\eta}(1)^{\T}\widehat{\bSigma}_{\bZ,\eta}(1)\}^{-1}
\widehat{\bSigma}_{\bZ,\eta}(1)^{\T}\widehat{\bSigma}_{\bZ,\eta}(2)$ and let $\hat \bv^1, \ldots, \hat \bv^{\hat{d}}$ be the $\hat{d}$ eigenvectors of the $\hat{d}\times \hat{d}$ matrix $\widehat{\bJ}_1$.
Now the estimators for $\bA$ and $\bB$ are defined as
\begin{align}\label{eq:hatAhatB}
\widehat{\bA}=\widehat{\bP}\widehat{\bU}~~\textrm{and}~~\widehat{\bB}=\widehat{\bQ}\widehat{\bV}\,,
\end{align}
where $\widehat{\bU}=(\hat{\bu}_1,\ldots,\hat{\bu}_{\hat{d}})$ and $\widehat{\bV}=(\hat{\bv}_1,\ldots,\hat{\bv}_{\hat{d}})$ with
\begin{equation*}%\label{eq:hatuellhatvell}
\hat{\bu}_\ell=
\frac{\widehat{\bSigma}_{\bZ,\eta}(1) \hat\bv^\ell}{
|\widehat{\bSigma}_{\bZ,\eta}(1) \hat\bv^\ell|_2}~~\textrm{and}~~
\hat{\bv}_\ell=
\frac{\widehat{\bSigma}_{\bZ,\eta}(1)^{\T} \hat\bu^\ell}{
|\widehat{\bSigma}_{\bZ,\eta}(1)^{\T} \hat\bu^\ell|_2}\,. % ~\quad
%\ell \in [\hat{d}]\,.
\end{equation*}
In the above expression, $(\hat \bu^1, \ldots, \hat \bu^{\hat{d}})^{\T}$ is the inverse of $\widehat{\bU}$.

\begin{remark}
Our above presented estimation procedure essentially estimates $\bV^{-1}, \bU, \bU^{-1}$ and $\bV$ sequentially. Parallel to Remark \ref{re:2} in Section \ref{sec31}, we can also consider estimating $\bU^{-1}$ first. Remark \ref{rm1} indicates that the $d$ rows of $\bU^{-1}$ are the $d$ eigenvectors of $\bJ_2$. Since $
\widehat{\bJ}_1$ %=\{\widehat{\bSigma}_{\bZ,\eta}(1)^{\T}\widehat{\bSigma}_{\bZ,\eta}(1)\}^{-1}
%\widehat{\bSigma}_{\bZ,\eta}(1)^{\T}\widehat{\bSigma}_{\bZ,\eta}(2)$ 
and $\widehat{\bJ}_2=\{\widehat{\bSigma}_{\bZ,\eta}(1)\widehat{\bSigma}_{\bZ,\eta}(1)^{\t}\}^{-1}
\widehat{\bSigma}_{\bZ,\eta}(1)\widehat{\bSigma}_{\bZ,\eta}(2)^{\t}$ are full-ranked, the difference between these two solutions are negligible, which is confirmned by the simulation not reported here.

% and let $\hat \bu^1, \ldots, \hat \bu^{\hat{d}}$ be its $\hat{d}$ eigenvectors. Then we can obtain $\widehat{\bV}=(\hat{\bv}_1,\ldots,\hat{\bv}_{\hat{d}})$ by \eqref{eq:hatuellhatvell}. Write $\widehat{\bV}^{-1}=(\hat{\bv}^1,\ldots,\hat{\bv}^{\hat{d}})^{\t}$. We can also further obtain $\widehat{\bU}=(\hat{\bu}_1,\ldots,\hat{\bu}_{\hat{d}})$ by \eqref{eq:hatuellhatvell}. Different from the undesirable phenomenon mentioned in Remark \ref{re:2} for the direct estimation procedures, respectively, based on estimating $\bA^{+}$ first and estimating $\bB^+$ first, numerical study shows that the performance of our refined idea with estimating $\bV^{-1}$ first is exactly equivalent to that with estimating $\bU^{-1}$ first since we do not need to solve a generalized eigenequation by the R function {\tt geigen} and transform the problem to solving a more feasible eigenequation problem.
\end{remark}

\section{Asymptotic properties} \label{sec4}
As we do not impose the stationarity on $\{\bY_t\}$, we use the concept of `$\alpha$-mixing' to characterize the serial dependence of $\{\bY_t\}$ with the $\alpha$-mixing coefficients defined as
\begin{align}\label{alpha.def}
\alpha(k)=\sup_{r}\sup_{A\in\mathcal{F}_{-\infty}^r,B\in\mathcal{F}_{r+k}^{\infty}}|\mathbb{P}(A\cap B)-\mathbb{P}(A)\mathbb{P}(B)|\,,~~~k\geq1\,,
\end{align}
where $\mathcal{F}_{r}^s$ is the $\sigma$-field generated by $\{ \bY_t: r\leq t\leq s  \}$. To simplify our presentation, we first present the theoretical results for the most challenging scenario with $p,q\gg n$ in Theorems \ref{hatd.exp} and \ref{a.exp}, and then give the associated results  in Remark \ref{remark2} for the cases with fixed $(p,q)$ or $(p,q)$ diverging at some polynomial rate of $n$. We need the following regularity conditions.

\begin{Condition}\label{b.x}
 (i) There exists a universal constant $C_1>0$ such that $\max_{k\in[K]}\|\bSigma_{\bY,\xi}(k)\|_2\leq C_1$. (ii) Write $\bSigma_{\bY,\xi}(k)=\{\sigma_{y,\xi,i,j}^{(k)}\}_{p\times q}$. It holds that $\max_{i\in[p]}\sum_{j=1}^{q}|\sigma_{y,\xi,i,j}^{(k)}|^{\iota}\leq s_1$ and $\max_{j\in[q]}\sum_{i=1}^{p}|\sigma_{y,\xi,i,j}^{(k)}|^{\iota}\leq s_2$ for some universal constant $\iota\in[0,1)$, where $s_1$ and $s_2$ may, respectively, diverge together with $p$ and $q$.
\end{Condition}

\begin{Condition}\label{exp.tail}
	(i) There exist some universal constants $C_2>0$, $C_3>0$ and  $r_1\in(0,2]$ such that $\max_{i\in[p]}\max_{j\in[q]}\max_{t\in[n]}\mathbb{P}( |y_{i,j,t}| >x)\leq C_2\exp(-C_3x^{r_1})$ and $\max_{t\in[n]}\mathbb{P}(|\xi_{t}| >x)\leq C_2\exp(-C_3x^{r_1})$ for any $x>0$. (ii) There exist some universal constants $C_4>0$, $C_5>0$ and $r_2\in(0,1]$ such that
	the mixing coefficients $\alpha(k)$ given in \eqref{alpha.def} satisfy $\alpha(k)\leq C_{4}\exp(-C_{5}k^{r_2})$ for all $k\geq 1$.
\end{Condition}

Recall $\bSigma_{\bY,\xi}(k)$ is a $p\times q$ matrix. Condition \ref{b.x}(i) requires the singular values of $\bSigma_{\bY,\xi}(k)$ to be uniformly bounded away from infinity for any $k\in[K]$. Our technical proofs indeed allow $\max_{k\in[K]}\|\bSigma_{\bY,\xi}(k)\|_2$ to diverge with $n$. We impose Condition \ref{b.x}(i) just for simplifying the presentation. Condition \ref{b.x}(ii) imposes some sparsity on $\bSigma_{\bY,\xi}(k)$. Notice that $
\bSigma_{\bY,\xi}(k)=\bA\bG_k\bB^{\T}$ for some $d\times d$ diagonal matrix $\bG_k$. Under some sparsity condition  on  $\bA$ and $\bB$, applying the technique used to derive Lemma 5 of \cite{ChangGuoYao_2018}, we can show that Condition \ref{b.x}(ii) holds for certain $(s_1,s_2)$. Condition \ref{exp.tail} is a common assumption in the literature on ultrahigh-dimensional data analysis, which ensures exponential-type upper bounds for the tail probabilities of the statistics concerned when $p, q\gg n$. See \cite{ChangChenWu2021} and reference therein. The $\alpha$-mixing assumption in Condition \ref{exp.tail}(ii) is mild. See the discussion below Equation (3) and Assumption 1 in \cite{ChangHuLiuTang_2022} for the widely used time series models which satisfy Condition \ref{exp.tail}(ii).  If we only require $\max_{i\in[p]}\max_{j\in[q]}\max_{t\in[n]}\mathbb{P}(|y_{i,j,t}|>x)=O\{x^{-2(l+\tau)}\}$ for any $x>0$, $\max_{t\in[n]}\mathbb{P}(|\xi_{t}|>x)=O\{x^{-2(l+\tau)}\}$ for any $x>0$ and $\alpha(k)=O\{k^{-(l-1)(l+\tau)/\tau}\}$ as $k\rightarrow\infty$ with two constants $l>2$ and $\tau>0$, we can apply Fuk-Nagaev-type inequalities to construct the upper bounds for the tail probabilities of the statistics concerned for which our procedure still works when $p$ and $q$ diverge at some polynomial rate of $n$. See Remark \ref{remark2}(ii) below. Let
\begin{align*}%\label{pi1n}
	\Pi_{1,n}=(s_1s_2)^{1/2}\{ {n}^{-1}{\log(pq)} \}^{(1-\iota)/2}\,.
\end{align*}
Theorem \ref{hatd.exp} shows that the ratio-based estimator $\hat{d}$ defined in \eqref{eq:ratio} is consistent.
% to the unknown integer $d$ involved in the model \eqref{b1}.

\begin{theorem}\label{hatd.exp}
	Let Conditions \ref{as:1}--\ref{exp.tail} hold and the threshold level $\de_1 =C_*\{n^{-1}\log(pq)\}^{1/2}$ for some sufficiently large constant $C_*>0$. For any $c_n$ in \eqref{eq:ratio} satisfying  $\Pi_{1,n}\ll c_n\ll 1$, it holds that $\mathbb{P}(\hat{d}= d)\rightarrow 1$ as $n\rightarrow\infty$, provided that $\Pi_{1,n}=o(1)$ and $\log(pq)=o(n^{c})$ for some constant $c\in(0,1)$ depending only on $r_1$ and $r_2$ specified in Condition \ref{exp.tail}.
\end{theorem}

To investigate the asymptotic properties of the estimator $(\widehat{\bA},\widehat{\bB})$ given in \eqref{eq:hatAhatB}, we first assume $\hat{d}=d$. Due to the consistency of $\hat{d}$ presented in Theorem \ref{hatd.exp}, we can prove, using the same arguments below Theorem 2.4 of \cite{ChangGuoYao_2015}, that the same results still hold without the assumption $\hat{d}=d$. See our discussion below Theorem \ref{a.exp}.

\begin{proposition}\label{hatp.exp}
 	Let Conditions \ref{as:1}--\ref{exp.tail} hold and the threshold level $\de_1 =C_*\{n^{-1}\log(pq)\}^{1/2}$ for some sufficiently large constant $C_*>0$.  If $\hat{d}=d$, there exist some orthogonal matrices $\bE_1$ and $\bE_2$ such that $\|\widehat{\bP}\bE_1-\bP\|_2=O_{\rm p}(\Pi_{1,n})=\|\widehat{\bQ}\bE_2-\bQ\|_2$, provided that $\Pi_{1,n}=o(1)$ and $\log(pq)=o(n^{c})$ for some constant $c\in(0,1)$ depending only on $r_1$ and $r_2$ specified in Condition \ref{exp.tail}.
\end{proposition}

Recall the columns of $\bP$ and $\bQ$ are, respectively, the $d$ orthonormal eigenvectors corresponding to the $d$ non-zero eigenvalues of $\bM_1$ and $\bM_2$. The presence of $\bE_1$ and $\bE_2$ accounts for the indeterminacy of
those eigenvectors due to reflections and/or possible tied (non-zero) eigenvalues.
%Involving some orthogonal matrices $\bE_1$ and $\bE_2$ here is a common phenomenon in the literature for estimating eigenvectors of certain matrices since (i) the orders of the eigenvalues associated with the columns of $\bP$ and $\bQ$ are unknown, (ii) there may exist several identical eigenvalues, and (iii) the eigenvectors have scaling indeterminacy. Let $\hat{d}=d$. For $(\bE_1,\bE_2)$ specified in Proposition \ref{hatp.exp},
Let $\tilde{\bw}=(\bE_2\otimes\bE_1)^{\T}\bw$,
with $\bw\in\mathbb{R}^{d^2}$ involved in \eqref{eq:hateta} for the definition of $\hat{\eta}_t=\bw^{\T}{\rm vec}(\widehat{\bZ}_t)$,
%Based on such defined $\tilde{\bw}$, we define
and define
\begin{align*}%\label{eq:sigmaztildeeta}
\bSigma_{\bZ,\tilde{\eta}}(k)=\bP^{\T}\widetilde{\bTheta}^{\T}\bSigma_{\mathring{\bY}}(k)\bQ\,,%~~~k\geq1\,,
\end{align*}
where $\widetilde{\bTheta}=\bI_p\otimes\{(\bQ\otimes \bP)\tilde{\bw}\}$, and $\bSigma_{\mathring{\bY}}(k)$ is specified in \eqref{eq:sigmaz}. As indicated in Lemma \ref{hatz.exp} in the supplementary material, $\bE_1^{\T}\widehat{\bSigma}_{\bZ,\eta}(k)\bE_2$ is consistent to $\bSigma_{\bZ,\tilde{\eta}}(k)$ under the spectral norm $\|\cdot\|_2$ rather than $\bSigma_{\bZ,\eta}(k)$ given in \eqref{eq:sigmaz}. In comparison to $\bSigma_{\bZ,\eta}(k)$, % in \eqref{eq:sigmaz},
we replace $\bw$ by $\tilde{\bw}$ in defining $\bSigma_{\bZ,\tilde{\eta}}(k)$. As we discussed in the beginning of Section \ref{sec:estAB}, the selection of $\bw$ for the identification of $\bU$ and $\bV$ is not unique.
 Define
\begin{align*}
\widetilde{\bS}_1={\bSigma}_{\bZ,\tilde{\eta}}(1)^{\T}{\bSigma}_{\bZ,\tilde{\eta}}(1)\,,~~\widetilde{\bS}_2={\bSigma}_{\bZ,\tilde{\eta}}(1)^{\T}{\bSigma}_{\bZ,\tilde{\eta}}(2)\,,\notag\\
\widetilde{\bS}_1^*={\bSigma}_{\bZ,\tilde{\eta}}(1){\bSigma}_{\bZ,\tilde{\eta}}(1)^{\T}\,,~~\widetilde{\bS}_2^*={\bSigma}_{\bZ,\tilde{\eta}}(1){\bSigma}_{\bZ,\tilde{\eta}}(2)^{\T}\,.
\end{align*}
Let $\tilde{\mu}_{\ell} = \tilde{c}_{2,\ell}\tilde{c}_{1,\ell}^{-1}$ with $\tilde{c}_{k,\ell} =(n-k)^{-1}\sum_{t=k+1}^n\tilde{\bw}^{\T}\mathbb{E}[{\rm vec}\{\bZ_{t-k}-\mathbb{E}(\bar{\bZ})\}\{x_{t,\ell}-\mathbb{E}(\bar{x}_{\cdot,\ell})\}]$. Under Condition \ref{as:3} below, parallel to Proposition \ref{prop_ID} in Section \ref{sec32}, we have that
the columns of
   $\bU=(\bu_1, \ldots, \bu_d)$ and $\bV=(\bv_1, \ldots, \bv_d)$ can be also defined, respectively, as
   \[
   \bu_\ell = \frac{\bSigma_{\bZ,\tilde{\eta}}(1) \bv^\ell}{|\bSigma_{\bZ,\tilde{\eta}}(1) \bv^\ell|_2}~~ {\rm and}~~
   \bv_\ell = \frac{\bSigma_{\bZ,\tilde{\eta}}(1)^{\T} \bu^\ell}{|\bSigma_{\bZ,\tilde{\eta}}(1)^{\T} \bu^\ell|_2}\,,
   \]
   with $\bv^\ell$ and $\bu^\ell$ being, respectively, the eigenvectors of the generalized eigenequations
\begin{align} \label{eq:finaleq}
\widetilde{\bS}_2 \bdelta = \tilde{\mu}_\ell \widetilde{\bS}_1 \bdelta~~\textrm{and}~~\widetilde{\bS}_2^*\bdelta=\tilde{\mu}_\ell\widetilde{\bS}_1^*\bdelta\,.
\end{align}
The following conditions are needed in our theoretical analysis.

\begin{Condition}\label{as:3}
(i) All the values $\tilde{\mu}_1,\ldots,\tilde{\mu}_d$ are finite and distinct. (ii) The eigenvalues of $\widetilde{\bS}_1$ are uniformly bounded away from zero.
\end{Condition}

\begin{Condition}\label{sp2}
(i) There exists a universal constant $C_6>0$ such that $\max_{k\in\{1,2\}}\|\bSigma_{\mathring{\bY}}(k)\|_2\leq C_6$. (ii) Write $\bSigma_{\mathring{\bY}}(k)=\{\sigma_{\mathring{y},r,s}^{(k)}\}_{(p^2q)\times q}$. It holds that $\max_{r\in[p^2q]}\sum_{s=1}^{q}|\sigma_{\mathring{y},r,s}^{(k)}|^{\iota}\leq s_3$ and $\max_{s\in[q]}\sum_{r=1}^{p^2q}|\sigma_{\mathring{y},r,s}^{(k)}|^{\iota}\leq s_4$ for some universal constant $\iota$ specified in Condition \ref{b.x}(ii), where $s_3$ and $s_4$ may, respectively, diverge together with $p$ and $q$.
\end{Condition}

Under Condition \ref{as:3}, $\bv^\ell$ and $\bu^\ell$ can be uniquely identified by the generalized eigenequations \eqref{eq:finaleq} upto the scaling and permutation indeterminacy. Recall $\bSigma_{\mathring{\bY}}(k)$ is a $(p^2q)\times q$ matrix. Condition \ref{sp2}(i) requires the largest singular value of $\bSigma_{\mathring{\bY}}(k)$ is uniformly bounded away from infinity. Our technical proofs indeed allow $\max_{k\in\{1,2\}}\|\bSigma_{\mathring{\bY}}(k)\|_2$ to diverge with $n$. We impose Condition \ref{sp2}(i) just for simplifying the presentation. Condition \ref{sp2}(ii) imposes some sparsity requirement on $\bSigma_{\mathring{\bY}}(k)$. Same as our discussion above for the validity of Condition \ref{b.x}(ii) imposed on the sparsity of $\bSigma_{\bY,\xi}(k)$, Condition \ref{sp2}(ii) holds automatically for certain $(s_3,s_4)$ under some sparsity condition imposed on the loading matrices $\bA$ and $\bB$.

Let $\bbeta_{\bv,\ell}$ and $\bbeta_{\bu,\ell}$ be the eigenvectors with unit $\ell^2$-norm of the generalized eigenequations \eqref{eq:finaleq} associated with $\tilde{\mu}_\ell$, i.e.,  $\widetilde{\bS}_2\bbeta_{\bv,\ell}=\tilde{\mu}_\ell\widetilde{\bS}_1\bbeta_{\bv,\ell}$ and $\widetilde{\bS}_2^*\bbeta_{\bu,\ell}=\tilde{\mu}_\ell\widetilde{\bS}_1^*\bbeta_{\bu,\ell}$. By Condition \ref{as:3}(ii), we know $\widetilde{\bS}_1$ and $\widetilde{\bS}_1^*$ are two invertible symmetric matrices. Hence, $\bbeta_{\bv,\ell}$ and $\bbeta_{\bu,\ell}$ are, respectively, also the eigenvectors of the eigenequations $\widetilde{\bS}_1^{-1}\widetilde{\bS}_2\bdelta=\tilde{\mu}_\ell\bdelta$ and $(\widetilde{\bS}_1^*)^{-1}\widetilde{\bS}_2^*\bdelta=\tilde{\mu}_\ell\bdelta$. For given $\bbeta_{\bv,\ell}$ and $\bbeta_{\bu,\ell}$, there exist two $d\times (d-1)$ matrices $\bR_{\bv,\ell}$ and $\bR_{\bu,\ell}$ such that $(\bbeta_{\bv,\ell},\bR_{\bv,\ell})$ and $(\bbeta_{\bu,\ell},\bR_{\bu,\ell})$ are two orthogonal matrices. For any $\ell\in[d]$, define
\begin{equation}\label{ev.b}
\theta_\ell=\sigma_{\min}(\bR_{\bv,\ell}^{\T}\widetilde{\bS}_1^{-1}\widetilde{\bS}_2\bR_{\bv,\ell}-\tilde{\mu}_\ell\bI_{d-1})~~\textrm{and}~~\theta_\ell^*=\sigma_{\min}\{\bR_{\bu,\ell}^{\T}(\widetilde{\bS}_1^*)^{-1}\widetilde{\bS}_2^*\bR_{\bu,\ell}-\tilde{\mu}_\ell\bI_{d-1}\}\,,
\end{equation}
 the smallest singular values of $\bR_{\bv,\ell}^{\T}\widetilde{\bS}_1^{-1}\widetilde{\bS}_2\bR_{\bv,\ell}-\tilde{\mu}_\ell\bI_{d-1}$ and $\bR_{\bu,\ell}^{\T}(\widetilde{\bS}_1^*)^{-1}\widetilde{\bS}_2^*\bR_{\bu,\ell}-\tilde{\mu}_\ell\bI_{d-1}$, respectively. Under Condition \ref{as:3}(i), we know $\min_{\ell\in[d]}\theta_\ell>0$ and $\min_{\ell\in[d]}\theta_\ell^*>0$. Such defined $\theta_\ell$ and $\theta_\ell^*$ can be viewed as the extension of the concept `eigen-gap' in symmetric matrices to non-symmetric matrices. If $\widetilde{\bS}_1^{-1}\widetilde{\bS}_2$ is a symmetric matrix, such defined $\theta_\ell$ is actually the eigen-gap $\min_{j:\,j\neq \ell}|\tilde{\mu}_j-\tilde{\mu}_\ell|$. Write $\widehat{\bA}=(\hat{\ba}_1,\ldots,\hat{\ba}_{\hat{d}})$ and $\widehat{\bB}=(\hat{\bb}_1,\ldots,\hat{\bb}_{\hat{d}})$. Define
\begin{align*}%\label{pi2n}
\Pi_{2,n}=(s_3s_4)^{1/2}\{ {n}^{-1}{\log(pq)} \}^{(1-\iota)/2}\,.
\end{align*}
Theorem \ref{a.exp} indicates that the columns of $\widehat{\bA}$ and $\widehat{\bB}$ defined in \eqref{eq:hatAhatB} are, respectively, consistent to those of $\bA$ and $\bB$ upto the reflection and permutation indeterminacy.

\begin{theorem}\label{a.exp}
	Let Conditions \ref{as:1}--\ref{sp2} hold and  the threshold levels $\delta_1=C_* \{n^{-1}\log(pq)\}^{1/2}$ and  $\delta_2=C_{**} \{n^{-1}\log(pq)\}^{1/2}$ for some sufficiently large constants $C_*>0$ and $C_{**}>0$.
	If $\hat{d}=d$, there exists a permutation of $(1,\ldots,d)$, denoted by $(j_1,\ldots,j_d)$, such that $
	| \kappa_{1,\ell}\hat\ba_{j_\ell}-\ba_\ell  |_2 = (1+\theta_\ell^{-1})\cdot O_{\rm p}(\Pi_{1,n}+\Pi_{2,n})
	$ and $
	| \kappa_{2,\ell}\hat{\bb}_{j_\ell}-\bb_\ell  |_2 =\{1+(\theta_\ell^*)^{-1}\}\cdot  O_{\rm p}(\Pi_{1,n}+\Pi_{2,n})
	$ for any $\ell\in[d]$ with some  $\kappa_{1,\ell},\kappa_{2,\ell}\in\{-1,1\}$, provided that $(\Pi_{1,n}+\Pi_{2,n}) \max\{1,d^{1/2}\theta_\ell^{-1},d^{1/2}(\theta_\ell^*)^{-1},d^{1/2}\theta_\ell^{-2},d^{1/2}(\theta_\ell^*)^{-2}\}=o(1) $ and $\log(pq)=o(n^{c})$ for some constant $c\in(0,1)$ depending only on $r_1$ and $r_2$ specified in Condition \ref{exp.tail}. Furthermore, it also holds that $1-|\hat{\ba}_{j_\ell}^{\H}\ba_\ell|^2=(1+\theta_\ell^{-1})^2\cdot O_{\rm p}(\Pi_{1,n}^2+\Pi_{2,n}^2)
	$ and $
	1-| \hat{\bb}_{j_\ell}^{\H}\bb_\ell  |^2 =\{1+(\theta_\ell^*)^{-1}\}^2\cdot  O_{\rm p}(\Pi_{1,n}^2+\Pi_{2,n}^2)
	$ for any $\ell\in[d]$. Here, the terms $O_{\rm p}(\Pi_{1,n}+\Pi_{2,n})$ and $O_{\rm p}(\Pi_{1,n}^2+\Pi_{2,n}^2)$ hold uniformly over $\ell\in[d]$. 	\end{theorem}

For $(j_1,\ldots,j_d)$ specified in Theorem \ref{a.exp}, Proposition \ref{conj} in Section \ref{sec31} shows that $\hat{\ba}_{j_\ell}$ and $\hat{\bb}_{j_\ell}$ may not be real vectors for some $\ell\in[d]$ although $\ba_\ell$ and $\bb_\ell$ are real vectors for all $\ell\in[d]$. When $\hat{d}=d$,  we can measure the difference between $\bA=(\ba_1,\ldots,\ba_d)$ and $\widehat{\bA}=(\hat{\ba}_1,\ldots,\hat{\ba}_d)$ by $\max_{\ell\in[d]}(1-|\hat{\ba}_{j_\ell}^{\H}\ba_\ell|^2)$ with $(j_1,\ldots,j_d)$ specified in Theorem \ref{a.exp}. In finite samples, $\hat{d}$ may not be exactly equal to $d$. In general scenario without assuming $\hat{d}=d$, we consider to measure the difference between $\bA=(\ba_1,\ldots,\ba_d)$ and $\widehat{\bA}=(\hat{\ba}_1,\ldots,\hat{\ba}_{\hat{d}})$ by
\begin{equation}\label{eq:dist}
\rho^2(\bA,\widehat{\bA})=\max_{\ell\in[d]}\min_{j\in[\hat{d}]}(1-|\hat{\ba}_{j}^{\H}\ba_\ell|^2)\,.
\end{equation}
Analogously, we can measure the difference between $\bB=(\bb_1,\ldots,\bb_d)$ and $\widehat{\bB}=(\hat{\bb}_1,\ldots,\hat{\bb}_{\hat{d}})$ by
\begin{equation}\label{eq:distB}
\rho^2(\bB,\widehat{\bB})=\max_{\ell\in[d]}\min_{j\in[\hat{d}]}(1-|\hat{\bb}_{j}^{\H}\bb_\ell|^2)\,.
\end{equation}
When $\hat{d}=d$, Theorem \ref{a.exp} yields that $\rho^2(\bA,\widehat{\bA})=\{1+(\min_{\ell\in[d]}\theta_\ell)^{-1}\}^2\cdot O_{\rm p}(\Pi_{1,n}^2+\Pi_{2,n}^2)$ and $\rho^2(\bB,\widehat{\bB})=\{1+(\min_{\ell\in[d]}\theta_\ell^*)^{-1}\}^2\cdot O_{\rm p}(\Pi_{1,n}^2+\Pi_{2,n}^2)$. Write $\varphi_n=\{1+(\min_{\ell\in[d]}\theta_\ell)^{-1}\}^2(\Pi_{1,n}^2+\Pi_{2,n}^2)$. For any $\epsilon>0$, there exists some constant $C_\epsilon>0$ such that $\mathbb{P}\{\rho^2(\bA,\widehat{\bA})>C_{\epsilon}\varphi_n\,|\,\hat{d}=d\}\leq \epsilon$. Together with Theorem \ref{hatd.exp}, we have
$
\mathbb{P}\{\rho^2(\bA,\widehat{\bA})>C_{\epsilon}\varphi_n\}
\leq\mathbb{P}\{\rho^2(\bA,\widehat{\bA})>C_{\epsilon}\varphi_n\,|\,\hat{d}=d\}\mathbb{P}(\hat{d}=d)+\mathbb{P}(\hat{d}\neq d)
\leq\epsilon+o(1)\rightarrow\epsilon$, which implies $\{1+(\min_{\ell\in[d]}\theta_\ell)^{-1}\}^2(\Pi_{1,n}^2+\Pi_{2,n}^2)$, the convergence rate of $\rho^2(\bA,\widehat{\bA})$ conditional on $\hat{d}=d$, is also the convergence rate of $\rho^2(\bA,\widehat{\bA})$. Identically, we also know $\{1+(\min_{\ell\in[d]}\theta_\ell^*)^{-1}\}^2(\Pi_{1,n}^2+\Pi_{2,n}^2)$ is the convergence rate of $\rho^2(\bB,\widehat{\bB})$.

\begin{remark}\label{remark2}
(i) If $p$ and $q$ are fixed constants, we can select the threshold levels  $\delta_1=\delta_2=0$ in \eqref{eq:mesthd} and \eqref{eq:estsigmazhd}. In this scenario, Conditions \ref{b.x} and \ref{sp2} hold automatically with $\iota=0$ and $(s_1,s_2,s_3,s_4)$ being some fixed constants, and Condition \ref{exp.tail} can be replaced by the weaker requirements that $\max_{i\in[p]}\max_{j\in[q]}\max_{t\in[n]}\mathbb{E}(|y_{i,j,t}|^{2\nu})=O(1)$, $\max_{t\in[n]}\mathbb{E}(|\xi_t|^{2\nu})=O(1)$, and $\sum_{k=1}^{\infty}\{\alpha(k)\}^{1-2/\nu}=O(1)$ for some constant $\nu>2$. Under these conditions, using the Davydov inequality, we have Theorem \ref{hatd.exp}, Proposition \ref{hatp.exp} and Theorem \ref{a.exp} hold with $\Pi_{1,n}^{*}=\Pi_{2,n}^{*}=n^{-1/2}$ and $\Pi_{1,n}^{*}\ll c_n \ll 1$, provided that $(\Pi_{1,n}^{*}+\Pi_{2,n}^{*})\max\{1, \theta_\ell^{-2},(\theta_\ell^*)^{-2}\}=o(1)$.

(ii) If $p$ and $q$ diverge at some polynomial rate of $n$,  we can replace Condition \ref{exp.tail} by the weaker requirements $\max_{i\in[p]}\max_{j\in[q]}\max_{t\in[n]}\mathbb{P}(|y_{i,j,t}|>x)=O\{x^{-2(l+\tau)}\}$ for any $x>0$, $\max_{t\in[n]}\mathbb{P}(|\xi_t|>x)=O\{x^{-2(l+\tau)}\}$ for any $x>0$,  and $\alpha(k)=O\{k^{-(l-1)(l+\tau)/\tau}\}$ as $k\rightarrow\infty$ with some constants $l>2$ and $\tau>0$. Under these conditions,  if the threshold levels $\delta_1=C_* (pq)^{1/l}n^{-1/2}$ and $\delta_2=C_{**} (pq)^{2/l}n^{-1/2}$ in \eqref{eq:mesthd} and \eqref{eq:estsigmazhd} for some sufficiently large constants $C_*>0$ and $C_{**}>0$, Theorem \ref{hatd.exp}, Proposition \ref{hatp.exp} and Theorem \ref{a.exp} hold with $\Pi_{1,n}^{*}=(s_1s_2)^{1/2}\{(pq)^{1/l}n^{-1/2}\}^{1-\iota}$, $\Pi_{2,n}^{*}=(s_3s_4)^{1/2}\{(pq)^{2/l}n^{-1/2}\}^{1-\iota}$ and $\Pi_{1,n}^{*}\ll c_n \ll 1$, provided that $(\Pi_{1,n}^{*}+\Pi_{2,n}^{*})\max\{1,d^{1/2}\theta_\ell^{-1},d^{1/2}(\theta_\ell^*)^{-1},d^{1/2}\theta_\ell^{-2},d^{1/2}(\theta_\ell^*)^{-2}\}=o(1) $.

\end{remark}

\section{Numerical studies}\label{sec:numerical}

\subsection{Simulation}

We illustrate the finite-sample performance of the
proposed methods by simulation based on model (\ref{b1}). % or, equivalently, model (\ref{b0}).
Let $\bA^* \equiv (a_{i,\ell}^*)_{p\times d} =(\ba_1^*, \ldots, \ba_d^*)$ and $\bB^* \equiv (b_{j,\ell}^*)_{q\times d} =(\bb_1^*, \ldots, \bb_d^*)$  with the elements drawn from the uniform distribution on $[-3,\, 3]$ independently satisfying the restriction rank$(\bA^*)=d={\rm rank}(\bB^*)$. Write $\tilde{\bx}_\ell=(\tilde{x}_{1,\ell},\ldots,\tilde{x}_{n,\ell})^{\T}$ and let $\tilde{\bx}_{1}, \ldots, \tilde{\bx}_{d}$ be independent AR(1) processes with independent $\mathcal{N}(0, 1)$ innovations, and the autoregressive coefficients drawn from the uniform distribution on $[-0.95,\, -0.6]\cup[0.6, \, 0.95]$. The elements of the error term $\eulbE$ in \eqref{b1} are drawn from $\mathcal{N}(0, 1)$ independently. Then, we generate the tensor $\eulbY = \sum_{\ell=1}^d \ba_\ell^* \circ \bb_\ell^* \circ \tilde{\bx}_\ell + \eulbE$. Let $\bA =(\ba_1, \ldots, \ba_d)$ and $\bB = (\bb_1, \ldots, \bb_d)$ with $\ba_{\ell}=\ba_{\ell}^*/|\ba_{\ell}^*|_2$ and $\bb_{\ell}=\bb_{\ell}^*/|\bb_{\ell}^*|_2$. Equivalently, we have
$%\label{eq:simu1}
\eulbY = \sum_{\ell=1}^d \ba_\ell \circ \bb_\ell \circ \bx_\ell + \eulbE$, 
where $\bx_{\ell} = |\ba_{\ell}^*|_2 |\bb_{\ell}^*|_2 \tilde{\bx}_\ell$. %In our simulation, we estimate $\bA$ and $\bB$ using the proposed estimation procedures.
%
%Recall $\bx_\ell=(x_{1,\ell},\ldots,x_{n,\ell})^{\T}$. In our simulation, the elements of
%the loading matrices $\bA$ and $\bB$ are drawn from the uniform distribution
%on $[-3,\, 3]$ independently with the
%restriction rank$(\bA)=d={\rm rank}(\bB)$, and $\bx_{1}, \ldots, \bx_{d}$
%are independent AR(1) processes
%with independent
%$\mathcal{N}(0, 1)$ innovations, and the autoregressive coefficients drawn from the uniform distribution
%on $[-0.95,\, -0.6]\cup[0.6, \, 0.95]$.
We set $d\in\{1, 3,6\}$,
$n\in\{300, 600,900\}$, and $p, q$ taking values between 4 and 256. %To specify $\xi_t$ in our procedure,
We consider the following two choices for $\xi_t$:
\begin{itemize}
\item (PCA) Let $\bY = \{{\rm vec}(\bY_1),\ldots,{\rm vec}(\bY_n)\}^{\T}$. Perform the principal component analysis for $\bY$ using the the R-function {\tt prcomp} in the R-package {\tt stats}, and select $\xi_t$ as the average of the first $m$ principal components corresponding to the eigenvalues which count for at least 99$\%$ of the total variations.

\item (Random weighting) Generate a $(pq)$-dimensional vector $\bh$ with its components randomly from the uniform distribution on $[0,1]$, and normalize $\bh$ as a unit vector, which is denoted by $\bh_0$. Then define $\xi_t = \bh_0^{\T}{\rm vec}(\bY_t)$.
\end{itemize}
For the refined method, $\hat{\eta}_t$ is specified in the
same manners with $\bY_t$ replaced by $\widehat{\bZ}_t$.
%Similarly, we also select $\hat{\eta}_t$ in the refined estimation procedure in the same way with replacing $\bY_t$ by $\widehat{\bZ}_t$ correspondingly.
We only present the results  for the cases  with $p\geq q$. More simulation results with $p<q$ can be found in the supplementary material. %{\color{red} The simulation is conducted on the centOS7(Linux) platform with two AMD 64-Core processors.}
%{\color{red} What is the point to consider $p<q$??? Maybe delete "with $p<q$?}

We first consider the finite-sample performance of the
estimation for $d$ by \eqref{eq:ratio1} of the direct estimation and by \eqref{eq:ratio} of the refined method. We set $\delta_1=0$ and $c_n=0$ in \eqref{eq:ratio1} and \eqref{eq:ratio}. %As indicated in Tables \ref{SM:Dhat_d1}--\ref{SM:Dhat_d6} in the supplementary material, the performance of the estimation procedure with $\xi_t$ determined by the PCA method is uniformly better than that with $\xi_t$ determined by the random weight method.
Table \ref{tb1} reports the relative frequency estimates of $\mathbb{P}(\hat{d} = d)$ based on 2000 repetitions with $\xi_t$ determined by PCA. %For $K=1$, the estimations of $d$ by the basic and refined procedure are equivalent, since $\wh\bK_1$ and $\wh\bM_1$ have the same non-zero eigenvalues.
When $d=1$, we observe
$\hat{d} \equiv d$ for both the direct and refined methods in all the simulation replications.
For $d>1$, the relative frequency estimates of $\mathbb{P}(\hat{d} = d)$ based on both the direct and refined methods increase as $n$, $p$ and $q$ grow in most of the cases. The refined method works uniformly better than the direct method except $(p,q,d,n)=(8,8,3,600)$, $(8,8,3,900)$ and $(16,16,3,900)$, and their performances in these three cases are  similar. As $d$ increases, the improvement from using the refined method also increases. Also, the refined method with larger $K$ has better performance in most of the cases. As shown in the proof of Theorem \ref{hatd.exp} in the supplementary material, the consistency of $\hat{d}$ depends on the convergence rate of $\|\widehat{\bM}_1-\bM_1\|_2$. Recall $\widehat{\bM}_1=\sum_{k=1}^KT_{\delta_1}\{\widehat{\bSigma}_{\bY,\xi}(k)\}T_{\delta_1}\{\widehat{\bSigma}_{\bY,\xi}(k)^{\T}\}$ and $\bM_1=\sum_{k=1}^K\bSigma_{\bY,\xi}(k)\bSigma_{\bY,\xi}(k)^{\T}$. The proof of Lemma 1 in the supplementary material indicates that the convergence rate of $|\widehat{\bSigma}_{\bY,\xi}(k)-\bSigma_{\bY,\xi}(k)|_\infty$ plays a key role in deriving the convergence rate of $\|\widehat{\bM}_1-\bM_1\|_2$. If $K$ is a fixed constant, $\max_{k\in[K]}|\widehat{\bSigma}_{\bY,\xi}(k)-\bSigma_{\bY,\xi}(k)|_\infty=O_{\rm p}[\{n^{-1}\log(pq)\}^{1/2}]$. If $K$ diverges with $n$, $K$ will appear in the convergence rate of $\max_{k\in[K]}|\widehat{\bSigma}_{\bY,\xi}(k)-\bSigma_{\bY,\xi}(k)|_\infty$. Then the convergence rate of $\|\widehat{\bM}_1-\bM_1\|_2$ with diverging $K$ will be slower than that with fixed $K$. Hence, we cannot select $K$ as large as possible since too large $K$ may lead to a bad estimate $\hat{d}$. We suggest to restrict $K\leq 10$ in practice. Table \ref{SM:originalDhat_random} in the supplementary material reports the results using randomly
weighted $\xi_t$; showing the similar patterns as
those in Table \ref{tb1}.  Note that using
PCA-based $\xi_t$ produces
uniformly more accurate estimates than using randomly weighted $\xi_t$. %determined by the PCA method uniformly outperforms those with $\xi_t$ determined by the random weight method.

\begin{figure}[htbp]
  \centering
  %\includegraphics[width=18cm]{Test1_plot_3modes.pdf}\\
  \includegraphics[width=18cm]{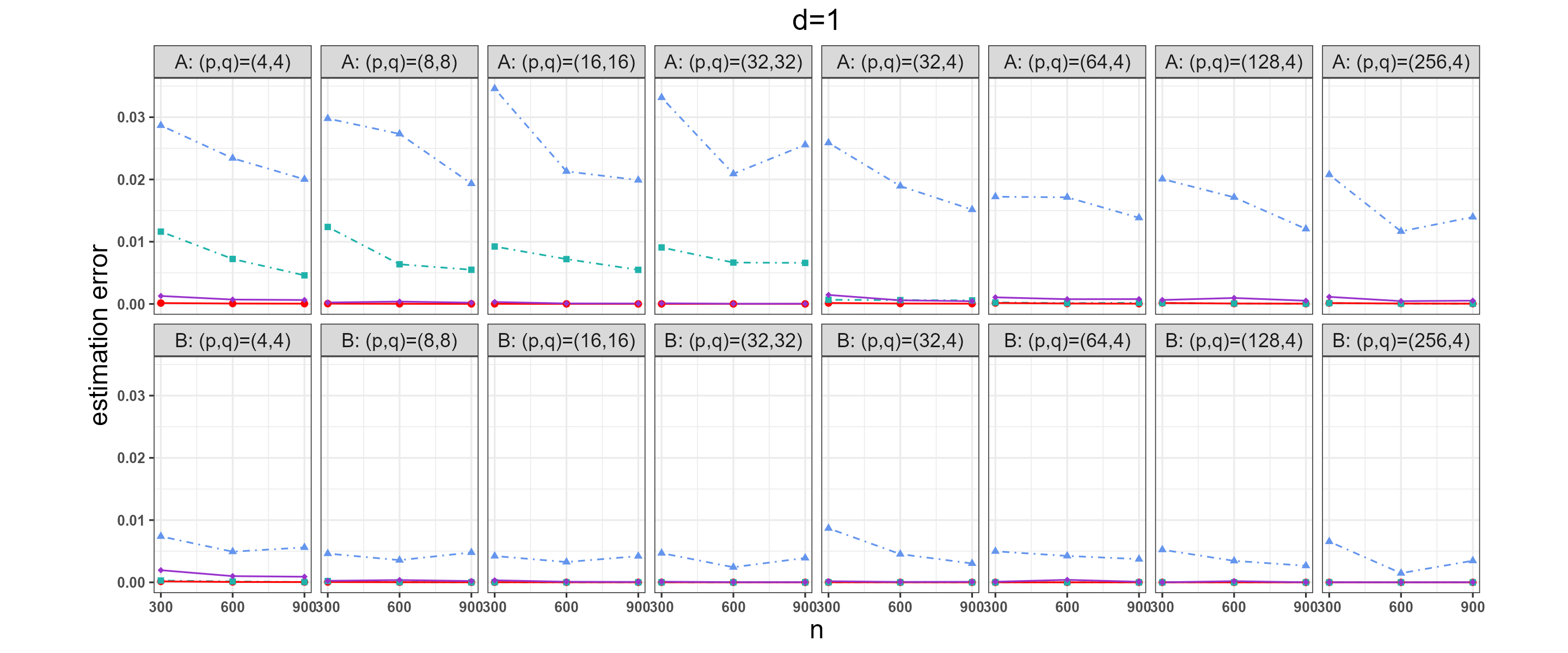}\\
  \includegraphics[width=18cm]{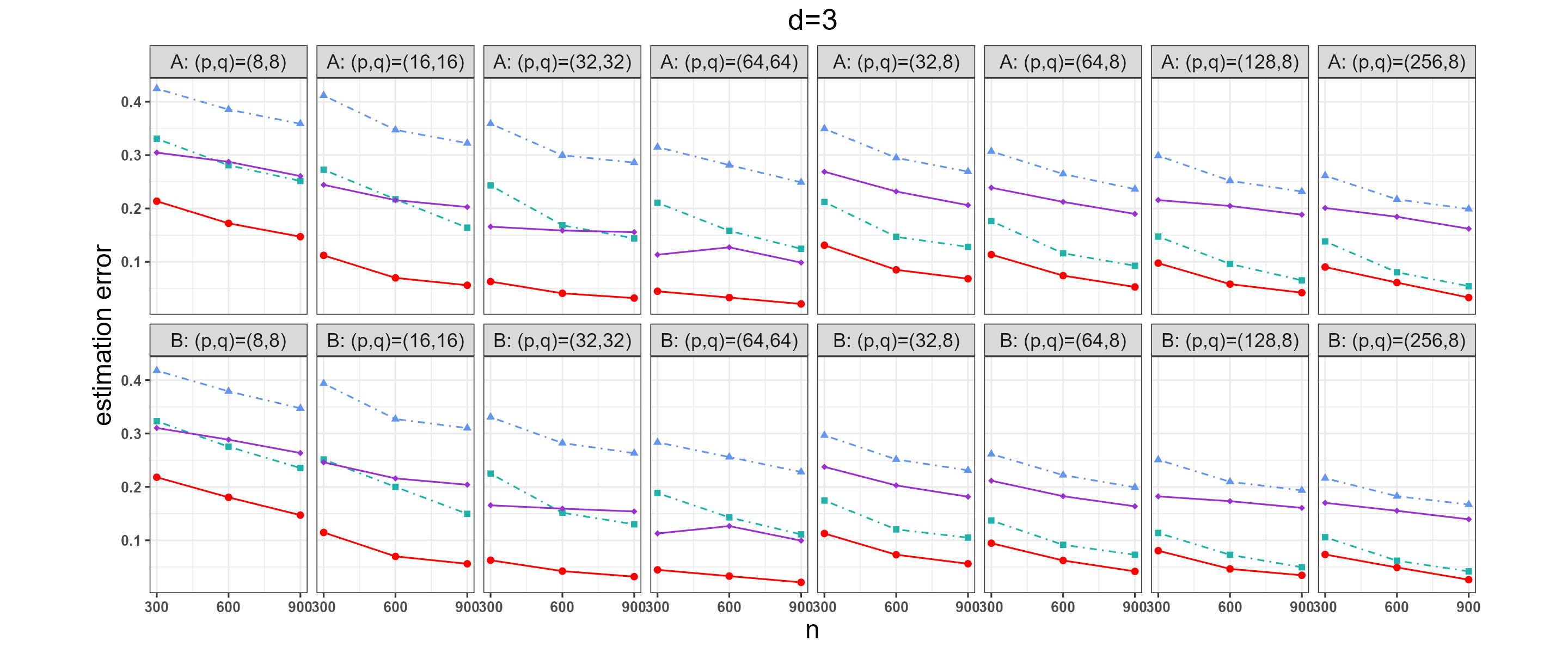}\\
  \includegraphics[width=18cm]{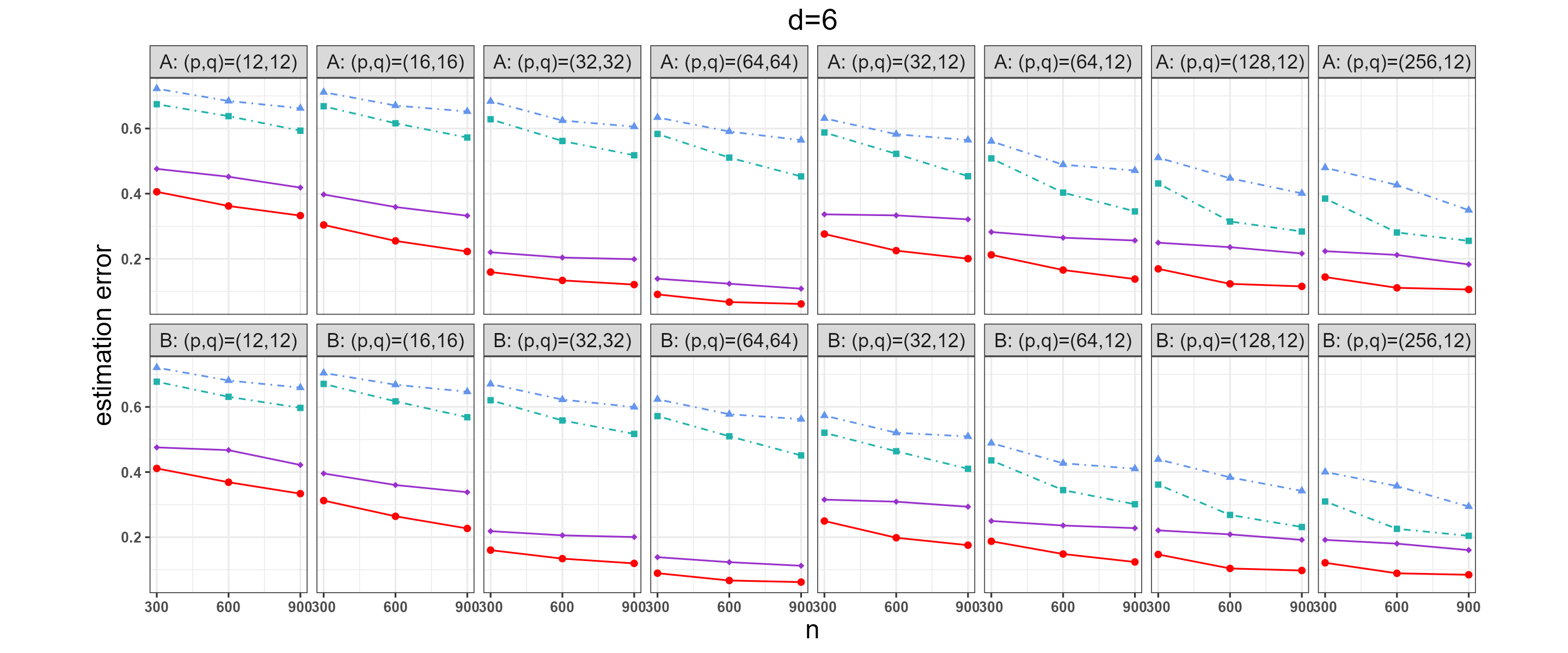}
  \caption{The lineplots for the averages of $\rho^2(\bA, \wh \bA)$ and $\rho^2(\bB, \wh \bB)$ based on 2000 repetitions. The legend is defined as follows: (i) the direct method with PCA-based $\xi_t$ ($ - \cdot - \blacksquare - \cdot -$); (ii) the direct method with randomly weighted
  $\xi_t$ ($ - \cdot - \blacktriangle - \cdot -$); (iii) the refined method ($K=3$) with PCA-based $\xi_t$ (---$\bullet$---); (iv) the refined method ($K=3$) with randomly weighted $\xi_t$ (---$\blacklozenge$---).  }\label{fig:rhoAB_PCArandom_basicK3}
\end{figure}

\begin{figure}[htbp]
  \centering
  %\includegraphics[width=18cm]{Test1_plot_3modes.pdf}\\
  \includegraphics[width=18cm]{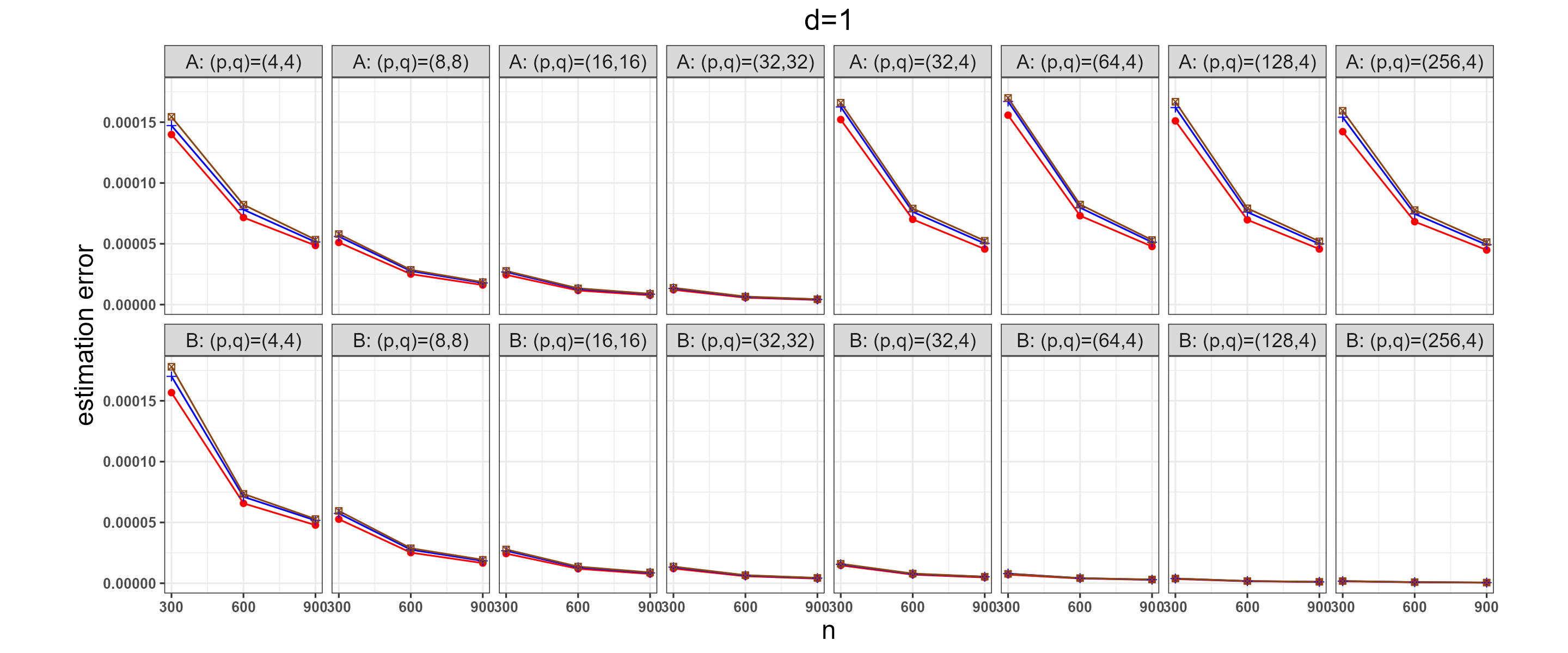}\\
  \includegraphics[width=18cm]{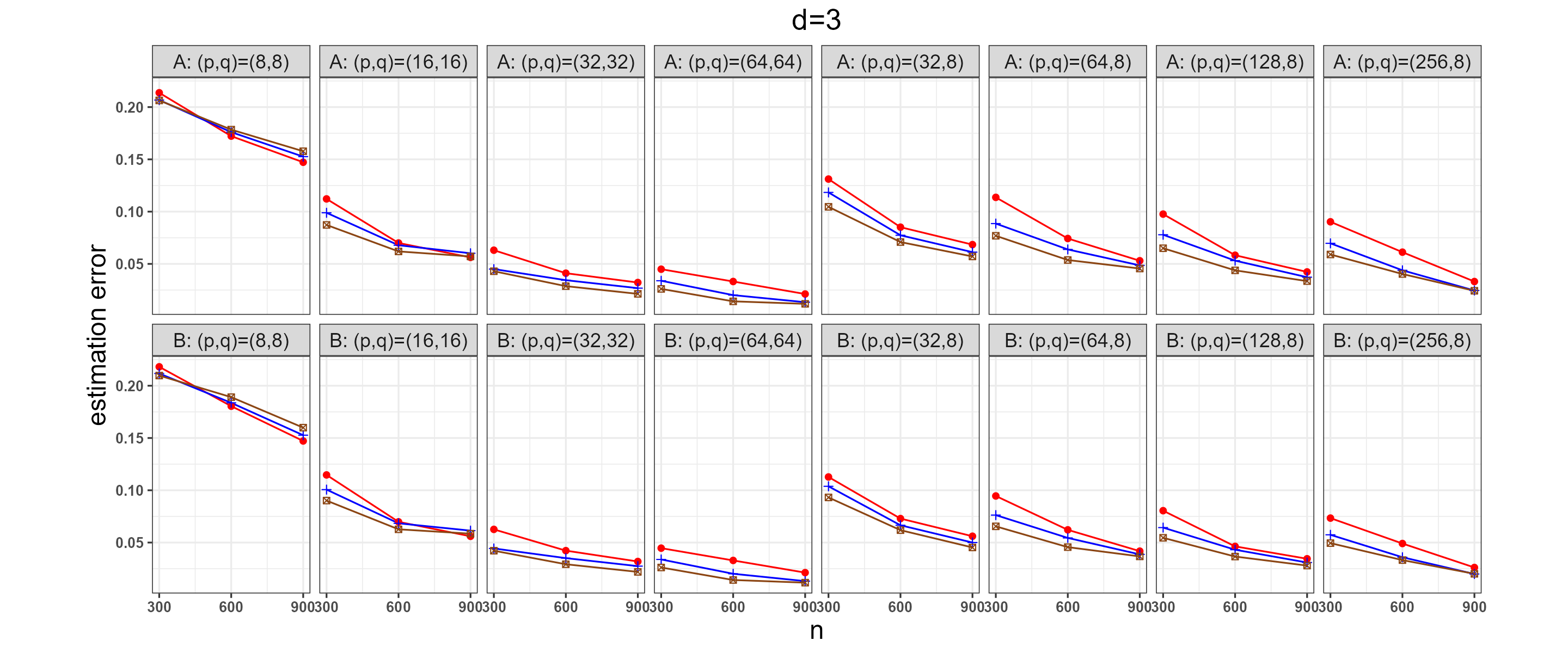}\\
  \includegraphics[width=18cm]{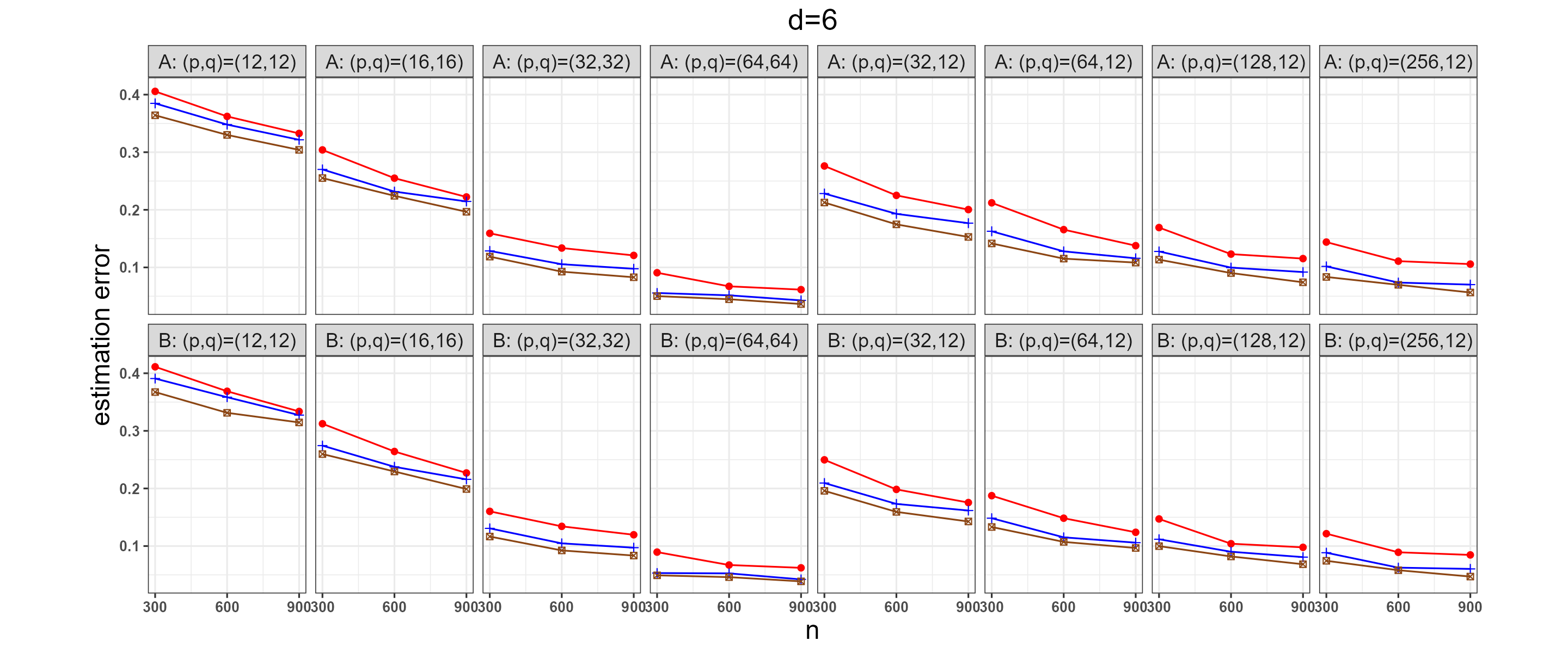}
  \caption{The lineplots for the averages of $\rho^2(\bA, \wh \bA)$ and $\rho^2(\bB, \wh \bB)$ based on 2000 repetitions, where $(\wh\bA,\wh\bB)$ is estimated by the refined method $(K=3, 5, 7)$ with PCA-based $\xi_t$. The legend is defined as follows: (i) $K=3$ (---$\bullet$---); (ii) $K=5$ (---$+$---); (iii) $K=7$ (---$\boxtimes$---).  }\label{fig:rhoAB_PCA_K357}
\end{figure}

Tables \ref{SM:tb1a}--\ref{SM:tb3} in the supplementary material present the averages and standard deviations of the estimation errors $\rho^2(\bA, \wh \bA)$ and $\rho^2(\bB, \wh \bB)$ defined in \eqref{eq:dist} and \eqref{eq:distB} based on 2000 repetitions. % for the direct and refined methods.
To highlight the key information,  Figure \ref{fig:rhoAB_PCArandom_basicK3} plots the results of the direct method and the refined method with $K=3$. It shows that (i) the refined method outperforms the direct method uniformly when $d>1$, (ii) two methods perform about
the same in some cases when $d=1$, and (iii) the PCA-based $\xi_t$ performs better than the randomly weighted $\xi_t$. Figure \ref{fig:rhoAB_PCA_K357} summarizes the performance of the refined method with $K\in\{3,5,7\}$ and $\xi_t$ determined by the PCA method. We can find that (i) the refined
method performs about the same for $K\in\{3,5,7\}$ when $d=1$, and (ii) the refined method with larger $K$ in general has slightly better performance when $d>1$,
%{\color{red} (except for $(p,q,d,n)=(8,8,3,600)$, $(8,8,3,900)$, $(16,16,3,900)$ and (256,8,3,900).)\footnote{The cases when $(p,q,d,n)=(8,8,3,600)$, $(8,8,3,900)$ and $(16,16,3,900)$ correspond to those when the relative frequency estimates of $\mathbb{P}(\hat{d} = d)$ do not increase as $K$ grows. When $(p,q,d,n)= (256,8,3,900)$, although $\rho(\bB,\wh\bB)$ for $K=7$ is slightly larger than that for $K=5$, $\rho(\bA,\wh\bA$, $\rho(\bB,\wh\bB)$ and the relative frequency estimates of $\mathbb{P}(\hat{d} = d)$ for $K=5$ and 7 are very similar. That is why I think that the better performance with respect to $K$ is due to the accuracy estimation of $d$.} }
mainly because larger $K$ is more likely to lead to more accurate estimate of $d$, see Table \ref{tb1}.

\begin{table}[!tbp]
\renewcommand\arraystretch{1.2}
\centering
\begin{threeparttable}
\footnotesize
\caption{\textit{ Relative frequency estimates of $\mathbb{P}(\hat{d} = d)$ based on $2000$ repetitions with PCA-based $\xi_t$, where the direct estimate and the refined estimate are given in \eqref{eq:ratio1} and \eqref{eq:ratio} respectively.}}\label{tb1}
\begin{tabular}{ccccccccccccc}
\hline\hline
& & & \multicolumn{3}{c}{Refined} & Direct & & & \multicolumn{3}{c}{Refined} & Direct \\
 & $(p,\,q)$ &$n$    & $K=3$    & $K=5$    & $K=7$  &   & $(p,\,q)$  &$n$    & $K=3$    & $K=5$    & $K=7$  & \\  \hline
\multirow{12}{*}{$d=1$} & $(4,\,4)$   & 300 & 100.00     & 100.00    & 100.00    & 100.00       & $(32,\,4)$   & 300 & 100.00     & 100.00    & 100.00    & 100.00       \\
                      &         & 600 & 100.00     & 100.00    & 100.00    & 100.00       &          & 600 & 100.00     & 100.00    & 100.00    & 100.00       \\
                      &         & 900 & 100.00     & 100.00    & 100.00    & 100.00       &          & 900 & 100.00     & 100.00    & 100.00    & 100.00       \\
                      & $(8,\,8)$   & 300 & 100.00     & 100.00    & 100.00    & 100.00       & $(64,\,4)$   & 300 & 100.00     & 100.00    & 100.00    & 100.00       \\
                      &         & 600 & 100.00     & 100.00    & 100.00    & 100.00       &          & 600 & 100.00     & 100.00    & 100.00    & 100.00       \\
                      &         & 900 & 100.00     & 100.00    & 100.00    & 100.00       &          & 900 & 100.00     & 100.00    & 100.00    & 100.00       \\
                      & $(16,\,16)$ & 300 & 100.00     & 100.00    & 100.00    & 100.00       & $(128,\,4)$  & 300 & 100.00     & 100.00    & 100.00    & 100.00       \\
                      &         & 600 & 100.00     & 100.00    & 100.00    & 100.00       &          & 600 & 100.00     & 100.00    & 100.00    & 100.00       \\
                      &         & 900 & 100.00     & 100.00    & 100.00    & 100.00       &          & 900 & 100.00     & 100.00    & 100.00    & 100.00       \\
                      & $(32,\,32)$ & 300 & 100.00     & 100.00    & 100.00    & 100.00       & $(256,\,4)$  & 300 & 100.00     & 100.00    & 100.00    & 100.00       \\
                      &         & 600 & 100.00     & 100.00    & 100.00    & 100.00       &          & 600 & 100.00     & 100.00    & 100.00    & 100.00       \\
                      &         & 900 & 100.00     & 100.00    & 100.00    & 100.00       &          & 900 & 100.00     & 100.00    & 100.00    & 100.00       \\ \hline
\multirow{12}{*}{$d=3$} & $(8,\,8)$   & 300 & 78.85      & 79.85     & 80.35     & 78.75        & $(32,\,8)$   & 300 & 88.65      & 90.20     & 91.55     & 85.65        \\
                      &         & 600 & 82.45      & 82.15     & 81.65     & 83.75        &          & 600 & 92.95      & 93.65     & 94.50     & 92.05        \\
                      &         & 900 & 85.55      & 85.05     & 84.50     & 86.55        &          & 900 & 94.45      & 95.25     & 96.00     & 93.20        \\
                      & $(16,\,16)$ & 300 & 89.45      & 90.95     & 92.10     & 88.00        & $(64,\,8)$   & 300 & 89.85      & 92.45     & 93.70     & 87.70        \\
                      &         & 600 & 93.85      & 94.35     & 95.05     & 92.35        &          & 600 & 93.55      & 94.60     & 95.75     & 92.75        \\
                      &         & 900 & 94.95      & 94.75     & 95.10     & 94.80        &          & 900 & 95.65      & 96.05     & 96.50     & 94.40        \\
                      & $(32,\,32)$ & 300 & 94.30      & 96.25     & 96.45     & 91.25        & $(128,\,8)$  & 300 & 91.40      & 93.55     & 94.90     & 88.85        \\
                      &         & 600 & 96.20      & 96.95     & 97.60     & 94.90        &          & 600 & 95.05      & 95.65     & 96.55     & 93.60        \\
                      &         & 900 & 97.20      & 97.80     & 98.35     & 96.20        &          & 900 & 96.45      & 97.05     & 97.35     & 95.95        \\
                      & $(64,\,64)$ & 300 & 95.80      & 96.95     & 97.80     & 92.10        & $(256,\,8)$  & 300 & 91.90      & 94.00     & 95.05     & 88.85        \\
                      &         & 600 & 96.95      & 98.25     & 98.90     & 95.15        &          & 600 & 94.65      & 96.30     & 96.65     & 93.50        \\
                      &         & 900 & 98.15      & 98.90     & 99.10     & 97.40        &          & 900 & 97.25      & 98.00     & 98.10     & 96.40        \\ \hline
\multirow{12}{*}{$d=6$} & $(12,\,12)$ & 300 & 73.35      & 78.15     & 81.80     & 60.70        & $(32,\,12)$  & 300 & 85.25      & 90.85     & 94.55     & 66.45        \\
                      &         & 600 & 77.85      & 81.50     & 84.85     & 68.00        &          & 600 & 89.55      & 93.75     & 95.00     & 76.70        \\
                      &         & 900 & 80.15      & 82.90     & 85.10     & 73.05        &          & 900 & 90.65      & 93.50     & 95.75     & 81.20        \\
                      & $(16,\,16)$ & 300 & 81.50      & 87.00     & 89.05     & 66.30        & $(64,\,12)$  & 300 & 88.35      & 93.55     & 95.50     & 69.25        \\
                      &         & 600 & 85.35      & 89.60     & 91.40     & 74.55        &          & 600 & 92.00      & 95.70     & 97.00     & 80.25        \\
                      &         & 900 & 88.45      & 90.90     & 93.20     & 79.40        &          & 900 & 93.75      & 96.50     & 97.70     & 85.95        \\
                      & $(32,\,32)$ & 300 & 90.65      & 94.90     & 96.40     & 76.65        & $(128,\,12)$ & 300 & 90.80      & 95.10     & 96.80     & 72.05        \\
                      &         & 600 & 92.50      & 96.40     & 97.80     & 81.50        &          & 600 & 94.05      & 96.45     & 97.60     & 83.30        \\
                      &         & 900 & 93.40      & 96.00     & 97.55     & 85.40        &          & 900 & 94.60      & 96.85     & 98.40     & 85.65        \\
                      & $(64,\,64)$ & 300 & 94.30      & 98.35     & 99.20     & 79.30        & $(256,\,12)$ & 300 & 90.85      & 95.40     & 97.65     & 71.95        \\
                      &         & 600 & 96.15      & 98.20     & 99.10     & 85.70        &          & 600 & 93.90      & 97.40     & 98.25     & 81.95        \\
                      &         & 900 & 96.30      & 98.35     & 99.10     & 89.15        &          & 900 & 93.85      & 97.35     & 98.75     & 84.50        \\
\hline\hline
\end{tabular}
%\begin{tablenotes}
%  \item[*] ``New'' represents the proposed new method, and ``Original'' represents the proposed original method.
%\end{tablenotes}
\end{threeparttable}
\end{table}

%\begin{figure}[htbp]
%  \centering
%  % Requires \usepackage{graphicx}
%  %\includegraphics[width=18cm]{Test1_plot_3modes.pdf}\\
%  \includegraphics[width=18cm]{Rho_PCArandom_basicK5_d1.png}\\
%  \includegraphics[width=18cm]{Rho_PCArandom_basicK5_d3.png}\\
%  \includegraphics[width=18cm]{Rho_PCArandom_basicK5_d6.png}
%  \caption{The lineplots for the averages of $\rho(\bA, \wh \bA)$ and $\rho(\bB, \wh \bB)$ based on 2000 repetitions. The legend is defined as follows: (i) the basic method + PCA ($ - \cdot - \blacksquare - \cdot -$); (ii) the basic method + random weight ($ - \cdot - \blacktriangle - \cdot -$); (iii) the refined method ($K=5$) + PCA (---$\bullet$---); (iv) the refined method ($K=5$) + random weight (---$\blacklozenge$---).  }\label{fig:rhoAB_PCArandom_basicK5}
%\end{figure}
%
%\begin{figure}[htbp]
%  \centering
%  % Requires \usepackage{graphicx}
%  %\includegraphics[width=18cm]{Test1_plot_3modes.pdf}\\
%  \includegraphics[width=18cm]{Rho_PCArandom_basicK7_d1.png}\\
%  \includegraphics[width=18cm]{Rho_PCArandom_basicK7_d3.png}\\
%  \includegraphics[width=18cm]{Rho_PCArandom_basicK7_d6.png}
%  \caption{The lineplots for the averages of $\rho(\bA, \wh \bA)$ and $\rho(\bB, \wh \bB)$ based on 2000 repetitions. The legend is defined as follows: (i) the basic method + PCA ($ - \cdot - \blacksquare - \cdot -$); (ii) the basic method + random weight ($ - \cdot - \blacktriangle - \cdot -$); (iii) the refined method ($K=7$) + PCA (---$\bullet$---); (iv) the refined method ($K=7$) + random weight (---$\blacklozenge$---).  }\label{fig:rhoAB_PCArandom_basicK7}
%\end{figure}

\subsection{A real data analysis}

In this section, we analyze the monthly average value weighted returns of the 100 portfolios
from January 1990 to December 2017.
The portfolios include all NYSE, AMEX, and NASDAQ stocks, which are constructed by the
intersections of 10 levels of size (market equity) and 10 levels of the
book equity to market equity ratio (BE).
The data were downloaded
from {\tt
http://mba.tuck.dartmouth.edu/pages/faculty/ken.french/data\_library.html}.
Although this website provides monthly return data from July 1926 to June
2021, there are many missing values in the early years. We restrict the
time period from January 1990 to December 2017
to avoid the large numbers
of missing data and large fluctuations. The data can be represented
as a $10 \times 10$ matrix $\bY_{t} = (y_{i,j,t})$ for $t=1,\ldots,336$
(i.e. $p=q=10$, $n=336$),
% with $p=q=10$, $T=492$ and the element
where $y_{i,j,t}$ is the return of the
portfolio at the $i$-th level of size and $j$-th level of the BE-ratio at time $t$. We impute the missing values by the
weighted averages of the three previous months, i.e. set
$y_{i,j,t}=0.5y_{i,j,t-1}+0.3y_{i,j,t-2}+0.2y_{i,j,t-3}$ for missing $y_{i,j,t}$.
% with the initial values $y_{ij(-1)}=y_{ij(-2)}=0$.
% {\color{red} In Wang et al. (2019)'s paper, they have mentioned that they subtracted the corresponding monthly excess market return from each of the series, which results in 100 market-adjusted return series. But I did not find the market return data from the same website as they indicated. Since the return series are stationary themselves and do not have any seasonal patterns in the auto-correlation functions, I used the original series in this analysis to check the performance of the proposed method. }

\begin{figure}[htbp]
  \centering
  %\includegraphics[width=18cm]{Test1_plot_3modes.pdf}\\
  \includegraphics[width=16.2cm]{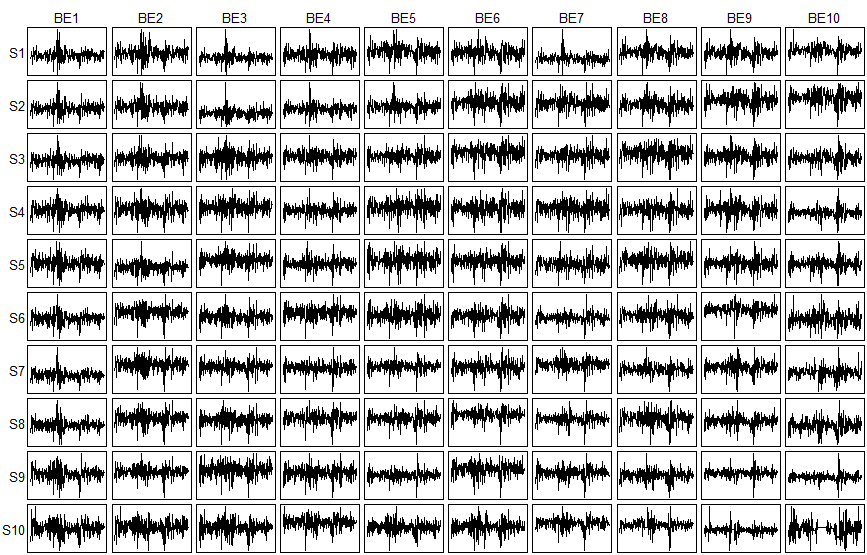}\\
  \caption{The plots of the return series of the portfolios formed on different levels of size (by rows) and book equity to market equity ratio (by columns). The horizontal axis represents time and the vertical axis represents the monthly returns. The ranges of the vertical values are not the same.  }\label{fig2}
\end{figure}

We standardize each of the 100 component time series $\{y_{i,j,t}\}_{t=1}^{n}$ so that they have
mean zero and unit variance. To economize the notation, we still use $y_{i,j,t}$ to denote the standardized data.
Figure \ref{fig2} shows the plots of the standardized
return series $\{y_{i,j,t}\}_{t=1}^{n}$, for $i,j=1,\ldots,10$. The rows in Figure \ref{fig2} correspond to the ten levels of size and the
columns correspond to the ten levels of the BE-ratio. Notice that the ranges of the vertical values are
not the same, and the figures are not directly comparable. All the 100 return
series appear to be stationary. The ACF (autocorrelation functions) plots of these 100 time series indicate that most
series have significant ACF at the first lag, and all series do not
show any seasonal patterns. The cross correlations between different time
series are mostly significant at time lags 0 and 1. %Due to the page limit, we do not report the
%ACF plot here.

We apply our model \eqref{b3} to fit the standardized matrix time series $\{\bY_{t}\}_{t=1}^{336}$ using the refined estimation method with PCA-based $\xi_t$; leading to $\hat{d}\equiv1$ with $K=3, 5$ or  7. See
\eqref{eq:ratio}. In the sequel, we only present the results with $K=5$. The results based on $K\in\{3,7\}$ are similar and thus omitted here. Based on \eqref{eq:hatAhatB}, we obtain $\wh{\bA}=(0.44, 0.34, 0.32, 0.32, 0.29, 0.25, 0.30, 0.30, 0.33, 0.23)^{\T}$ and $\wh{\bB}=(0.20, 0.26, 0.27, 0.29, 0.35, 0.33, 0.34,0.31, 0.37, 0.39)^{\T}$. Following the arguments above Proposition \ref{conj} in Section \ref{sec31}, we can recover the latent time series $\{\hat{x}_{t,1}\}_{t=1}^{336}$. Figure \ref{fig5} displays the plots of time series $\{\hat{x}_{t,1}\}_{t=1}^{336}$ and its ACF, which shows that
the autocorrelations of $\{\hat{x}_{t,1}\}_{t=1}^{336}$ is significant at the first lag that is consistent to the ACF patterns of $\bY_t$.
%We can fit $\{\hat{x}_{t,1}\}_{t=1}^{336}$ by an ARMA model and then model $\bY_t$ by $\widehat{\bA}\hat{x}_{t,1}\widehat{\bB}^{\T}$.
The Akaike information criterion (AIC) suggests to fit $\{\hat{x}_{t,1}\}_{t=1}^{336}$ by an AR(1) model. Hence, to model this $10\times 10$ matrix time series $\bY_t$, our method essentially only needs to estimate one parameter in an AR(1) model. We also consider to fit the matrix time series $\bY_t$ by following methods:
\begin{figure}[htbp]
  \centering
  %\includegraphics[width=18cm]{Test1_plot_3modes.pdf}\\
  \includegraphics[width=8cm]{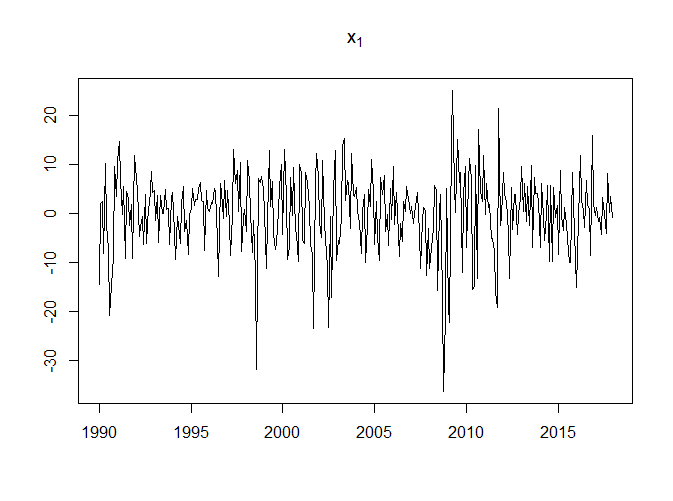}
  \includegraphics[width=8cm]{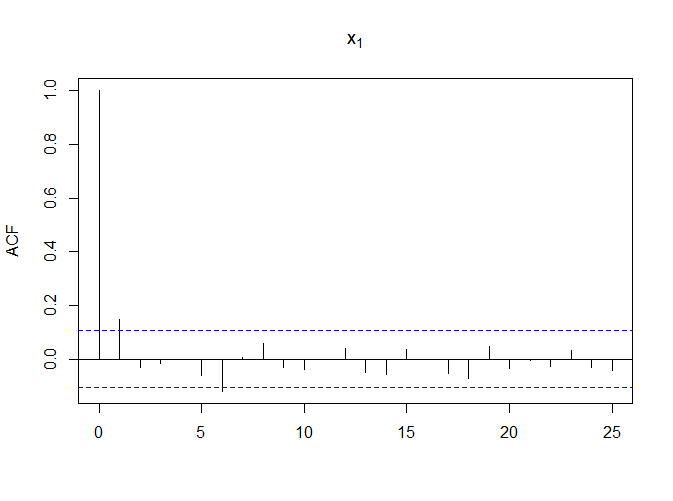}
  \caption{The plots of the latent time series $\{\hat{x}_{t,1}\}_{t=1}^{336}$ and its ACF.  }\label{fig5}
\end{figure}
\begin{itemize}
\item (UniARMA) For each of 100 component time series $\{y_{i,j,t}\}_{t=1}^{336}$, we fit an ARMA model specified by the AIC;
%This method requires to fit 100 ARMA models. Based on the models determined by the AIC, we find there are
leading to the estimation for 135 coefficient parameters in the total 100 models.

\item (SVAR) Fit a sparse VAR($\ell$) model to $\{{\rm vec}(\bY_t)\}_{t=1}^{336}$ using the R-function {\tt sparseVAR} in the R-package {\tt bigtime} with the standard lasso penalization and the optimal sparsity parameter selected by the time series cross validation procedure. The program selects $\ell=27$  automatically based on the time series length, and there are 270000 parameters to be estimated.

\item (MAR) Fit $\{\bY_t\}_{t=1}^{336}$ by the matrix-AR(1) of \cite{ChenXiaoYang_2021}, which involves 200 parameters.

\item (TS-PCA) Apply the principle component analysis for time series suggested in \cite{ChangGuoYao_2018} to the 100-dimensional time series $\{{\rm vec}(\bY_t)\}_{t=1}^{336}$ using the R-package {\tt HDTSA}, leading
to 98 univariate time series and one two-dimensional time series. For the obtained
univariate time series, we fit it by an ARMA model with the order determined by the AIC.
For the obtained two-dimensional time series, we fit it by an VAR model with the order
determined by the AIC. There are in total 93 parameters in the models. 

\item (FAC) Apply the factor model of \cite{WangLiuChen_2019} to matrix time series  $\{\bY_t\}_{t=1}^{336}$ with the pre-determined parameter $h_0=1$ as suggested in the real data analysis part of their paper. Based on their method, we find there is only one factor. We fit the latent factor series by an AR(1) model specified by the AIC which only needs to estimate one parameter.

\end{itemize}
While UniARMA, SVAR and MAR model $\bY_t$ or $\mathrm{vec}(\bY_t)$ directly,  our proposed method,  TS-PCA and FAC seek dimension reduction first and then model  the resulting low-dimensional time series. %{\color{red} The real data analysis is conducted on the Windows 10 platform with Intel(R) Core(TM) i7-8550U CPU.} 
Both RMSE and MAE, defined as below, of the fitted models are listed in Table \ref{tbC}:
\[ \mathrm{RMSE} = \bigg\{\frac{1}{33600}\sum_{t=1}^{336}\sum_{i=1}^{10}\sum_{j=1}^{10}(\hat{y}_{i,j,t}-y_{i,j,t})^2\bigg\}^{1/2}\,,~~~\mathrm{MAE} = \frac{1}{33600}\sum_{t=1}^{336}\sum_{i=1}^{10}\sum_{j=1}^{10}|\hat{y}_{i,j,t}-y_{i,j,t}| \,.\]
%where $\hat{y}_{i,j,t}$ is the fitting value of the model for $y_{i,j,t}$.
%The fitting errors are reported in Table \ref{tbC}.
Among the three dimension-reduction methods, our proposed method has the smallest RMSE and MAE, while MAR achieves the overall minimum RMSE and MAE.

%In comparison to the methods focused on modelling $\bY_t$ or $\mathrm{vec}(\bY_t)$ directly, our proposed method is worse than MAR, slightly worse than UniARMA and slightly better than SVAR. These three methods need to estimate much more unknown parameters than ours and require more computational time. As we will show later in the comparison of prediction, the good performance of MAR may be due to overfitting.

\begin{table}[!tbp]
\renewcommand\arraystretch{1.2}
\centering
\begin{threeparttable}
\caption{Fitting errors for the monthly data from year 1990 to 2017. The computational time is conducted on the Windows platform with Intel(R) Core(TM) i7-8550U CPU at 1.99 GHz.}\label{tbC}
\footnotesize
\begin{tabular}{c|cccccc}
\hline
&Proposed&TS-PCA&FAC&UniARMA&SVAR&MAR \\ \hline
RMSE&0.9913&0.9935&0.9923&0.9895&0.9985&{\bf 0.9613} \\
MAE&0.7432&0.7456&0.7436&0.7417&0.7444&{\bf 0.7235}  \\
$\#$ Parameters&1&93&1&135&270000&200 \\
time (seconds)&{\bf 0.3172}&6.4618&0.6596&6.7335&1689.1860&1.8470\\ \hline
\end{tabular}
\end{threeparttable}
\end{table}

We also evaluate the post-sample forecasting performance of these methods by performing the one-step and two-step ahead rolling forecasts for the 24 monthly readings in the last two years (i.e. 2016 and 2017). For each $s=1,\ldots,24$, we use our proposed method and the other five methods to fit $\{\bY_t\}_{t=s}^{311+s}$ and then obtain the one-step forecast of $\bY_{312+s}$ denoted by $\widehat{\bY}_{312+s}=\{\hat{y}_{i,j,312+s}^{(s)}\}_{10\times10}$. For the two-step ahead forecast, we fit $\{\bY_t\}_{t=s}^{310+s}$ by the six methods, and the two-step ahead forecast $\widehat{\bY}_{312+s}$ can be obtained by plug-in the one-step forecast into the models. 
For our proposed method, TS-PCA and FAC, if the dimension of the obtained latent time series is larger than $1$ we fit it by a VAR model with the order determined by the AIC, otherwise we fit it by an ARMA model with the order determined by the AIC. The post-sample forecasting performance is evaluated by the rRMSE and rMAE defined as
\begin{align*}
 {\rm rRMSE} =&~\bigg[\frac{1}{2400}\sum_{s=1}^{24}\sum_{i=1}^{10}\sum_{j=1}^{10}\{\hat{y}_{i,j,312+s}^{(s)}-y_{i,j,312+s}\}^2\bigg]^{1/2}\,,\\
 {\rm rMAE} =&~\frac{1}{2400}\sum_{s=1}^{24}\sum_{i=1}^{10}\sum_{j=1}^{10}\big|\hat{y}_{i,j,312+s}^{(s)}-y_{i,j,312+s}\big|\,.
 \end{align*}
Table \ref{tbD} summarizes the post-sample forecasting rRMSE and rMAE. The newly proposed method, in spite of its simplicity, exhibits the promising
post-sample forecasting performance, as its rRMSE and rMAE are the
smallest in one-step ahead forecasting among all the methods concerned, and
are the 2nd smallest in the two-step ahead forecast for which only TS-PCA has smaller rRMSE and rMAE.

\begin{table}[!tbp]
\renewcommand\arraystretch{1.2}
\centering
\begin{threeparttable}
\caption{One-step and two-step ahead forecasting errors for the monthly readings in the last two years 2016 and 2017. }\label{tbD}
\footnotesize
\begin{tabular}{c|cccccc}
\hline
&Proposed&TS-PCA&FAC&UniARMA&SVAR&MAR \\ \hline
&\multicolumn{6}{c}{one-step ahead forecast}\\
rRMSE&{\bf 0.7678}&0.7802&0.7701&0.7724&0.7690&0.8067 \\
rMAE&{\bf 0.5609}&0.5696&0.5649&0.5652&0.5614&0.5948  \\ \hline
&\multicolumn{6}{c}{two-step ahead forecast}\\
rRMSE&0.7668&{\bf 0.7526}&0.7683&0.7707&0.7693&0.7728 \\
rMAE&0.5590&{\bf 0.5512}&0.5610&0.5638&0.5616&0.5627  \\ \hline
\end{tabular}

\end{threeparttable}
\end{table}

\section*{Acknowledgements}

We thank the editor, the associate editor and two referees for their constructive comments. Chang, He and Yang were supported in part by the National Natural Science Foundation of China (grant nos. 71991472, 72125008, 11871401 and 11701466). Yao was supported in part by the U.K. Engineering and Physical Sciences Research Council. Chang was also supported by the Center of Statistical Research at Southwestern University of Finance and Economics.

\newpage
%
%\setcounter{page}{1}
%\pagestyle{fancy}
%\fancyhf{}
%\rhead{\bfseries\thepage}
%\lhead{\bfseries SUPPLEMENTARY MATERIAL}
%
%
\setcounter{page}{1}
\begin{center}
{\bf\Large Supplementary Material for ``Modelling Matrix Time Series via a Tensor CP-Decomposition'' by Jinyuan Chang, Jing He, Lin Yang and Qiwei Yao}
\end{center}
%
%\appendix
%
%
\renewcommand{\theequation}{S.\arabic{equation}}
\setcounter{table}{0}
\setcounter{equation}{0}
\renewcommand {\thetable} {S\arabic{table}}

Throughout the supplementary material, we use $C \in (0,\infty) $ to denote a generic finite constant that does not depend on $(n,p,q,d)$, and  may be different in different uses.
For two sequences of positive numbers $\{a_n\}$ and $\{b_n\}$, we write $a_n\lesssim b_n$ or $b_n\gtrsim a_n$ if $\limsup_{n\rightarrow\infty}a_n/b_n\leq c_0$ for some constant $c_0>0$.

%For an $m_1\times m_2$ matrix $\bA=(a_{i,j})_{m_1\times m_2}$, let $\|\bA\|_2$ and $\|\bA\|_{\rm F}$ be, respectively, the spectral norm of $\bA$ and the Frobenius norm of $\bA$.  Specifically, if $m_2=1$, we use $|\bA|_2=(\sum_{i=1}^{m_1}a_{i,1}^2)^{1/2}$ to denote the $\ell^2$-norm of the $m_1$-dimensional vector $\bA$.

\appendix

\section{Proof of Proposition \ref{conj}}
Recall $\widehat{\bK}_{2,q}\hat{\bb}^\ell=\lambda_*\widetilde{\bK}_{1,q}\hat{\bb}^\ell$. Then $\widehat{\bK}_{2,q}\overline{\hat{\bb}^{\ell}}=\overline{\lambda_*}\widetilde{\bK}_{1,q}\overline{\hat{\bb}^{\ell}}$, where $\overline{\hat{\bb}^{\ell}}$ is the complex conjugate of $\hat{\bb}^\ell$. Hence, $\overline{\lambda_*}$ and $\overline{\hat{\bb}^{\ell}}$ are, respectively, the eigenvalue and the associated eigenvector of the generalized eigenequation $\widehat{\bK}_{2,q}\bb=\lambda\widetilde{\bK}_{1,q}\bb$. Since the eigenvalues of $\widehat{\bK}_{2,q}\bb=\lambda\widetilde{\bK}_{1,q}\bb$ are distinct, there exists $\tilde{\ell}\in[\hat{d}]$ and a constant $\kappa\in\{-1,1\}$ such that $\overline{\hat{\bb}^\ell}=\kappa\hat{\bb}^{\tilde{\ell}}$. By \eqref{eq:esta}, we have
\begin{equation}\label{eq:hatahata}
\hat{\ba}_{\tilde{\ell}}=\frac{\widehat{\bSigma}_1\hat{\bb}^{\tilde{\ell}}}{|\widehat{\bSigma}_1\hat{\bb}^{\tilde{\ell}}|_2}=\frac{\kappa\widehat{\bSigma}_1\overline{\hat{\bb}^{\ell}}}{|\widehat{\bSigma}_1\overline{\hat{\bb}^{\ell}}|_2}=\kappa\overline{\hat{\ba}_\ell}\,,
\end{equation}
where $\overline{\hat{\ba}_\ell}$ is the complex conjugate of $\hat{\ba}_\ell$.

Recall $\widehat{\bA}^+=(\hat{\ba}^1,\ldots,\hat{\ba}^{\hat{d}})^{\t}$ is the Moore-Penrose inverse of $\widehat{\bA}=(\hat{\ba}_1,\ldots,\hat{\ba}_{\hat{d}})$. To show $\hat{\bb}_{\tilde{\ell}}=\kappa\overline{\hat{\bb}_\ell}$,  by \eqref{eq:estb}, we only need to show $\hat{\ba}^{\tilde{\ell}}=\kappa\overline{\hat{\ba}^\ell}$. Assume the generalized eigenequation $\widehat{\bK}_{2,q}\bb=\lambda\widetilde{\bK}_{1,q}\bb$ has $s$ real eigenvalues. Then there are $\hat{d}-s$ complex eigenvalues. Since the complex eigenvalues always occur in complex conjugate pairs, $\hat{d}-s$ is an even integer. Write $\hat{d}-s=2m$. Let $\lambda_1,\overline{\lambda_1},\ldots,\lambda_m,\overline{\lambda_m}$ be the $\hat{d}-s$ complex eigenvalues of the generalized eigenequation $\widehat{\bK}_{2,q}\bb=\lambda\widetilde{\bK}_{1,q}\bb$. For each $j\in[m]$, there exist $\ell_j, \tilde{\ell}_j\in[\hat{d}]$ such that the eigenvectors associated with $\lambda_j$ and $\overline{\lambda_j}$ are, respectively, $\hat{\bb}^{\ell_j}$ and $\hat{\bb}^{\tilde{\ell}_j}$. Then there exists $(\kappa_1,\ldots,\kappa_m)\in\{-1,1\}^m$ such that $\hat{\bb}^{\tilde{\ell}_j}=\kappa_j\overline{\hat{\bb}^{\ell_j}}$ for each $j\in[m]$. Same as \eqref{eq:hatahata}, we have $\hat{\ba}_{\tilde{\ell}_j}=\kappa_j\overline{\hat{\ba}_{\ell_j}}$ for each $j\in[m]$. Write $\boldsymbol{\varpi}_j={\rm Re}(\hat{\ba}_{\ell_j})$ and $\boldsymbol{\varsigma}_j={\rm Im}(\hat{\ba}_{\ell_j})$ for each $j\in[m]$. Then $\hat{\ba}_{\tilde{\ell}_j}=\kappa_j\boldsymbol{\varpi}_j-{\rm i}\kappa_j\boldsymbol{\varsigma}_j$ for each $j\in[m]$, where ${\rm i}=\sqrt{-1}$. Let $\{j_1,\ldots,j_s\}=[\hat{d}]\setminus\{\ell_1,\tilde{\ell}_1,\ldots,\ell_m,\tilde{\ell}_m\}$. For each $r\in[s]$, $\hat{\bb}^{j_r}$ are the eigenvector of the generalized eigenequation $\widehat{\bK}_{2,q}\bb=\lambda\widetilde{\bK}_{1,q}\bb$ associated with the real eigenvalue $\tilde{\lambda}_{r}$. Write $\hat{\bb}^{j_r}={\rm Re}(\hat{\bb}^{j_r})+{\rm i}\,{\rm Im}(\hat{\bb}^{j_r})$. Since $\widetilde{\bK}_{1,q}$ and $\widehat{\bK}_{2,q}$ are two real matrices, then ${\rm Re}(\hat{\bb}^{j_r})$ and ${\rm Im}(\hat{\bb}^{j_r})$ are both the eigenvectors of the generalized eigenequation $\widehat{\bK}_{2,q}\bb=\lambda\widetilde{\bK}_{1,q}\bb$ associated with the real eigenvalue $\tilde{\lambda}_{r}$. Recall the eigenvalues of the generalized eigenequation $\widehat{\bK}_{2,q}\bb=\lambda\widetilde{\bK}_{1,q}\bb$ are distinct. There exists a constant $c_r\in\mathbb{R}$ such that ${\rm Im}(\hat{\bb}^{j_r})=c_r{\rm Re}(\hat{\bb}^{j_r})$. Thus, any eigenvector of the generalized eigenequation $\widehat{\bK}_{2,q}\bb=\lambda\widetilde{\bK}_{1,q}\bb$ associated with the real eigenvalue $\tilde{\lambda}_{r}$ can be formulated as $\tilde{c}_r{\rm Re}(\hat{\bb}^{j_r})$ for some $\tilde{c}_r\in\mathbb{C}$. Without loss of generality, we can assume $\hat{\bb}^{j_1},\ldots,\hat{\bb}^{j_s}$ are all real vectors. Notice that there exists a $\hat{d}\times \hat{d}$ real matrix $\bF$ such that
\begin{align}
\widehat{\bA}\bF=&\,(\hat{\ba}_{\ell_1},\hat{\ba}_{\tilde{\ell}_1},\ldots,\hat{\ba}_{\ell_m},\hat{\ba}_{\tilde{\ell}_m}, \hat{\ba}_{j_1},\ldots,\hat{\ba}_{j_s})\notag\\
=&\,\underbrace{(\boldsymbol{\varpi}_{1},\boldsymbol{\varsigma}_1,\ldots,\boldsymbol{\varpi}_{m},\boldsymbol{\varsigma}_m, \hat{\ba}_{j_1},\ldots,\hat{\ba}_{j_s})}_{\bS}{\rm diag}(\bK_1,\ldots,\bK_m,\bI_s)\,,\label{eq:hatAF}
\end{align}
where
\[
\bK_j=\left(
\begin{array}{cc}
1 & 1 \\
\rm{i} & -\rm{i} \\
\end{array}	
\right)\left(
\begin{array}{cc}
1 & 0 \\
0 & \kappa_j \\
\end{array}	
\right)\,,~~~j\in[m]\,.
\]
The matrix $\bF$ essentially exchanges the columns of $\widehat{\bA}$ which is invertible. Without loss of generality, we assume $(\ell,\tilde{\ell})=(\ell_1,\tilde{\ell}_1)$. Then $\kappa_1=\kappa$. Since $\widehat{\bA}$ has full column rank, $\widehat{\bA}^+=(\widehat{\bA}^{\H}\widehat{\bA})^{-1}\widehat{\bA}^{\H}$, where $^{\H}$ denotes the conjugate transpose. Notice that $\bS$ is a real matrix. By \eqref{eq:hatAF}, it holds that
\[
\widehat{\bA}^+=\bF{\rm diag}(\bK_1^{-1},\ldots,\bK_m^{-1},\bI_s)(\bS^{\t}\bS)^{-1}\bS^{\t}\,.
\]
Since the first two rows of $\bF^{\t}\widehat{\bA}^+$ are $(\hat{\ba}^{\ell})^{\t}$ and $(\hat{\ba}^{\tilde{\ell}})^{\t}$, due to the fact $\bF^{\t}\bF=\bI_{\hat{d}}$, we have
$
(\hat{\ba}^{\ell},\hat{\ba}^{\tilde{\ell}})^{\t}=(\bK_1^{-1},\bzero)(\bS^{\t}\bS)^{-1}\bS^{\t}$, which implies
\[
\hat{\ba}^\ell=\frac{1}{2}\bS(\bS^{\t}\bS)^{-1}(1,-{\rm i},\bzero)^{\t}~~\textrm{and}~~\hat{\ba}^{\tilde{\ell}}=\frac{1}{2}\bS(\bS^{\t}\bS)^{-1}(\kappa,{\rm i}\kappa,\bzero)^{\t}\,.
\]
Hence, $\hat{\ba}^{\tilde{\ell}}=\kappa \overline{\hat{\ba}^{\ell}}$.

Recall $(\hat{x}_{t,1},\ldots,\hat{x}_{t,\hat{d}})^{\t}= \wh\bH^{+}\mathrm{vec}(\bY_t)$ with $
\widehat{\bH}=(\hat{\bb}_1\otimes\hat{\ba}_1,\ldots,\hat{\bb}_{\hat{d}}\otimes\hat{\ba}_{\hat{d}})
$. Since $\hat{\bb}_{\tilde{\ell}}\otimes\hat{\ba}_{\tilde{\ell}}$ is the complex conjugate of $\hat{\bb}_{{\ell}}\otimes\hat{\ba}_{{\ell}}$, following the same arguments stated above, we know the $\tilde{\ell}$-th row of $ \wh\bH^{+}$ is the complex conjugate of the ${\ell}$-th row of $ \wh\bH^{+}$, which yields that $\hat{x}_{t,\tilde{\ell}}=\overline{\hat{x}_{t,{\ell}}}$. $\hfill\Box$

\section{Proofs of Theorem \ref{hatd.exp} and Proposition \ref{hatp.exp}}

Recall $\Pi_{1,n}=(s_1s_2)^{1/2}\{ {n}^{-1}{\log(pq)} \}^{(1-\iota)/2}$. To construct Theorem \ref{hatd.exp} and Proposition \ref{hatp.exp}, we need the following lemma whose proof is given in Section \ref{sec:pfpn2}.

\begin{lemma}\label{m1.exp}
	Under Conditions %\ref{as:1},
	\ref{b.x} and \ref{exp.tail}, if the threshold level $\delta_1=C_* \{n^{-1}\log(pq)\}^{1/2}$ for some sufficiently large constant $C_*>0$, we have $
	\| \widehat{\bM}_{1}-\bM_1  \|_2 = O_{\rm p}(\Pi_{1,n})=
	\| \widehat{\bM}_{2}-\bM_2  \|_2 $, provided that $\Pi_{1,n}=o(1)$ and $\log(pq)=o(n^{c})$ for some constant $c\in(0,1)$ depending only on $r_1$ and $r_2$ specified in Condition \ref{exp.tail}.
\end{lemma}

\subsection{Proof of Theorem \ref{hatd.exp}}

Denote by $\hat{\lambda}_{1}\geq\cdots\geq\hat{\lambda}_{p}$ and ${\lambda}_{1}\geq\cdots\geq\lambda_{p}$, respectively, the eigenvalues of $\widehat{\bM}_1$ and $\bM_1$. By Lemma \ref{m1.exp}, $\max_{i\in[p]}|\hat{\lambda}_{i}-\lambda_{i}|\leq \|\widehat{\bM}_1-\bM_1  \|_2=o_{\rm p}(c_n)$ for some $c_n=o(1)$ satisfying $\Pi_{1,n}=o(c_n)$. By Condition \ref{as:2}, we know $\lambda_{d}\geq C$, and $\lambda_{d+1}=\cdots=\lambda_{p}=0$. Hence, for any $j<d$,
$
(\hat\lambda_{j+1}+c_n)/(\hat\lambda_{j}+c_n)\stackrel{\mathbb{P}}{\rightarrow}  \lambda_{j+1}/\lambda_{j}>0$.
For any $j>d$, note that $|\hat{\lambda}_j|=o_{\rm p}(c_n)$, then
$
(\hat\lambda_{j+1}+c_n)/(\hat\lambda_{j}+c_n)\stackrel{\mathbb{P}}{\rightarrow}  1$.
On the other hand,
$
(\hat\lambda_{d+1}+c_n)/(\hat\lambda_{d}+c_n)\stackrel{\mathbb{P}}{\rightarrow}  0$,
which implies
$(\hat{\lambda}_{d+1}+c_n)/(\hat{\lambda}_{d}+c_n)=\min_{j\in[R]}(\hat{\lambda}_{j+1}+c_n)/(\hat{\lambda}_{j}+c_n)$ with probability approaching one. This indicates that $\mathbb{P}(\hat{d}=d)\rightarrow1$ as $n\rightarrow\infty$. $\hfill\Box$%We complete the proof of Theorem \ref{hatd.exp}.

\subsection{Proof of Proposition \ref{hatp.exp}}
Denote by $\lambda_{1,1}\geq\cdots\geq\lambda_{1,p}$ and ${\lambda}_{2,1}\geq\cdots\geq\lambda_{2,q}$, respectively, the eigenvalues of $\bM_1$ and $\bM_2$. It follows from Condition \ref{as:2} that $\lambda_{1,1}\geq\cdots\geq\lambda_{1,d}>0=\lambda_{1,d+1}=\cdots=\lambda_{1,p}$ and $\lambda_{2,1}\geq\cdots\geq\lambda_{2,d}>0=\lambda_{2,d+1}=\cdots=\lambda_{2,q}$. Notice that $\lambda_{1,d}$ and $\lambda_{2,d}$ are uniformly bounded away from zero. Lemma 1 of \cite{ChangGuoYao_2018} implies that
$\|\widehat{\bP}\bE_1-\bP\|_2\leq C\|\widehat{\bM}_1-\bM_1\|_2$ and $\|\widehat{\bQ}\bE_2-\bQ\|_2\leq C\|\widehat{\bM}_2-\bM_2\|_2$,
where $\bE_1$ and $\bE_2$ are two orthogonal matrices. Together with Lemma \ref{m1.exp},  we complete the proof of Proposition \ref{hatp.exp}. $\hfill\Box$

\section{Proof of Theorem \ref{a.exp}}

Recall $\Pi_{1,n}=(s_1s_2)^{1/2}\{ {n}^{-1}{\log(pq)} \}^{(1-\iota)/2}$ and $\Pi_{2,n}=(s_3s_4)^{1/2}\{ {n}^{-1}{\log(pq)} \}^{(1-\iota)/2}$. Write $\Pi_n=\Pi_{1,n}+\Pi_{2,n}$. For $\widehat{\bSigma}_{\bZ,\eta}(k)$ defined as \eqref{eq:estsigmazhd}, we write
\begin{align*}
&\widehat{\bS}_1=\widehat{\bSigma}_{\bZ,\eta}(1)^{\T}\widehat{\bSigma}_{\bZ,\eta}(1)\,,~~~\widehat{\bS}_2=\widehat{\bSigma}_{\bZ,\eta}(1)^{\T}\widehat{\bSigma}_{\bZ,\eta}(2)\,,\\
&\widehat{\bS}_1^*=\widehat{\bSigma}_{\bZ,\eta}(1)\widehat{\bSigma}_{\bZ,\eta}(1)^{\T}\,,~~~\widehat{\bS}_2^*=\widehat{\bSigma}_{\bZ,\eta}(1)\widehat{\bSigma}_{\bZ,\eta}(2)^{\T}\,.
\end{align*}
Recall $
\widetilde{\bS}_1={\bSigma}_{\bZ,\tilde{\eta}}(1)^{\T}{\bSigma}_{\bZ,\tilde{\eta}}(1)$, $\widetilde{\bS}_2={\bSigma}_{\bZ,\tilde{\eta}}(1)^{\T}{\bSigma}_{\bZ,\tilde{\eta}}(2)$,
$\widetilde{\bS}_1^*={\bSigma}_{\bZ,\tilde{\eta}}(1){\bSigma}_{\bZ,\tilde{\eta}}(1)^{\T}$ and $\widetilde{\bS}_2^*={\bSigma}_{\bZ,\tilde{\eta}}(1){\bSigma}_{\bZ,\tilde{\eta}}(2)^{\T}$. To construct Theorem \ref{a.exp}, we need the following lemmas. The proofs of Lemmas  \ref{hatz.exp} and \ref{s1.exp} are given in Sections  \ref{sec:pflemma1} and \ref{sec:pfpn3}, respectively. Lemma \ref{ev.sens}  is Corollary 7.2.6 of \cite{GolubVan_2013}.

\begin{lemma}\label{hatz.exp}
	Under Conditions \ref{as:1}--\ref{exp.tail},  and \ref{sp2}, if the threshold levels $\delta_1=C_* \{n^{-1}\log(pq)\}^{1/2}$ and $\delta_2=C_{**}\{n^{-1}\log(pq)\}^{1/2}$ for some sufficiently large constants $C_*>0$ and $C_{**}>0$, we have $\max_{k\in\{1,2\}}\|\bE_1^{\T}\widehat{\bSigma}_{\bZ,\eta}(k)\bE_2-{\bSigma}_{\bZ,\tilde{\eta}}(k)  \|_2=O_{\rm p}(\Pi_{n})$ for $(\bE_1,\bE_2)$  specified in Proposition \ref{hatp.exp}, provided that $\Pi_{n}=o(1)$ and $\log(pq)=o(n^{c})$ for some constant $c\in(0,1)$ depending only on $r_1$ and $r_2$ specified in Condition \ref{exp.tail}.
\end{lemma}

\begin{lemma}\label{s1.exp}
	Under conditions of Lemma \ref{hatz.exp},   we have  $
	\| \bE_2^{\T}\widehat{\bS}_1\bE_2-\widetilde\bS_1  \|_2 = O_{\rm p}(\Pi_{n})=
	\| \bE_2^{\T}\widehat{\bS}_2\bE_2-\widetilde\bS_2  \|_2 $ and $\| \bE_1^{\T}\widehat{\bS}_1^*\bE_1-\widetilde\bS_1^*  \|_2= O_{\rm p}(\Pi_{n})=
	\| \bE_1^{\T}\widehat{\bS}_2^*\bE_1-\widetilde\bS_2^*  \|_2$ for $(\bE_1,\bE_2)$  specified in Proposition \ref{hatp.exp}.
\end{lemma}

\begin{lemma}\label{ev.sens}
	Suppose $\bW,\bDelta\in \mathbb{C}^{d\times d}$ and that $\bR=(\br_1,\bR_2)\in \mathbb{C}^{d\times d}$ is unitary with $\br_1\in \mathbb{C}^d$. Assume
	\begin{align*}
	\bR^{\H}\bW\bR=\left(
	\begin{array}{cc}
	\lambda & \bv^{\H} \\
	\bzero & \bD_{22} \\
	\end{array}	
	\right) ~~\textrm{and}~~\bR^{\H}\bDelta\bR=\left(
	\begin{array}{cc}
	\epsilon & \bgamma^{\H} \\
	\bdelta & \bDelta_{22} \\
	\end{array}	
	\right),
	\end{align*}
	where $^{\H}$ denotes the conjugate transpose. Let $\theta$ be the smallest singular value of $\bD_{22}-\lambda\bI_{d-1}$, and denote by $\|\cdot\|_{\rm F}$ the Frobenius norm of $\cdot$. If $\theta>0$ and $5\|\bDelta\|_{\rm F}(1+5\theta^{-1}|\bv|_2)\leq \theta$, then there exists $\bu\in\mathbb{C}^{d-1}$ with $|\bu|_2\leq 4\theta^{-1}|\bdelta|_2$ such that $\tilde{\br}_1=(\br_1+\bR_2\bu)/\sqrt{1+\bu^{\H}\bu}$ is a unit $\ell^2$-norm eigenvector for $\bW+\bDelta$. %Moreover,
%	$
%	1-|\br_1^{\H}\tilde{\br}_1|^2\leq 16\theta^{-2}|\bdelta|_2^2$.
\end{lemma}

Now we begin to prove Theorem \ref{a.exp}. Let $\{\hat{\bbeta}_{\bv,1},\ldots,\hat{\bbeta}_{\bv,d}\}$ and $\{\bbeta_{\bv,1},\ldots,\bbeta_{\bv,d}\}$ be, respectively,  the  eigenvectors of $\widehat{\bS}_1^{-1}\widehat{\bS}_2$ and $\widetilde{\bS}_1^{-1}\widetilde{\bS}_2$ with unit $\ell^2$-norm, i.e., $\widehat{\bS}_1^{-1}\widehat{\bS}_2\hat{\bbeta}_{\bv,\ell}=\hat{\mu}_\ell \hat{\bbeta}_{\bv,\ell}$ and $\widetilde{\bS}_1^{-1}\widetilde{\bS}_2\bbeta_{\bv,\ell}=\tilde{\mu}_\ell\bbeta_{\bv,\ell}$ for any $\ell\in[d]$. Since $\hat{d}=d$, as we mentioned below \eqref{eq:estsigmay}, we have $\hat{\bbeta}_{\bv,\ell}=\hat{\bv}^\ell$ for any $\ell\in[d]$. Write  $\hat\bbeta^*_{\bv,\ell}=\bE_2^{\T}\hat\bbeta_{\bv,\ell}$ with $\bE_2$ specified in Proposition \ref{hatp.exp}. Then $\hat{\bbeta}^*_{\bv,\ell}$ is the  eigenvector of $\bE_2^{\T}\widehat{\bS}_1^{-1}\widehat{\bS}_2\bE_2$ associated with eigenvalue $\hat\mu_\ell$. Under Condition \ref{as:3}(i), applying Lemma \ref{ev.sens} with $\bW=\widetilde\bS_1^{-1}\widetilde\bS_2$, $\bW+\bDelta=\bE_2^{\T}\widehat{\bS}_1^{-1}\widehat{\bS}_2\bE_2$ and $\br_1=\bbeta_{\bv,\ell}$, it holds that
\[
|\kappa^{\ell} \hat{\bbeta}^*_{\bv,j_\ell}-\bbeta_{\bv,\ell} |_2\lesssim \theta_\ell^{-1}\|\bE_2^{\T}\widehat{\bS}_1^{-1}\widehat{\bS}_2\bE_2-\widetilde{\bS}_1^{-1}\widetilde{\bS}_2\|_2
\]
for any $\ell\in[d]$ provided that $5\|\bE_2^{\T}\widehat{\bS}_1^{-1}\widehat{\bS}_2\bE_2-\widetilde{\bS}_1^{-1}\widetilde{\bS}_2\|_{\rm F}(1+5\theta_\ell^{-1}\|\widetilde\bS_1^{-1}\widetilde\bS_2\|_2)\leq\theta_\ell$, where $\kappa^\ell\in\{-1,1\}$, $(j_1,\ldots,j_d)$ is a permutation of $(1,\ldots,d)$, and $\theta_\ell$ is given in \eqref{ev.b}. Without loss of generality, we assume $(j_1,\ldots,j_d)=(1,\ldots,d)$. For any $\ell\in[d]$, let
\begin{align*}
\Phi_{\ell}=\bigg|\frac{\kappa^{\ell}\bE_1^{\T}\widehat{\bSigma}_{\bZ,\eta}(1)\hat{\bbeta}_{\bv,\ell}}{|\widehat{\bSigma}_{\bZ,\eta}(1)\hat{\bbeta}_{\bv,\ell}|_2}-\frac{\bSigma_{\bZ,\tilde{\eta}}(1)\bbeta_{\bv,\ell}}{|\bSigma_{\bZ,\tilde{\eta}}(1)\bbeta_{\bv,\ell}|_2}\bigg|_2
\end{align*}
with $\bE_1$ specified in Proposition \ref{hatp.exp}. In the sequel, we will specify the convergence rate of $\Phi_{\ell}$.

%Let  $\rho=\max\{1,d^{1/2}\theta_\ell^{-2},d^{1/2}(\theta_\ell^*)^{-2},\theta_\ell^{-1},(\theta_\ell^*)^{-1}\}$. Define an event $\mathcal{E}:=\{ \rho(\Pi_{1,n}+\Pi_{2,n})=o(1)\, {\textrm{and}}\, \log(pq)=o(n^c) \}$ for some $c>0$ depending only on $(r_1,r_2)$.

Recall $\bSigma_{\bZ,\tilde{\eta}}(k)=\bP^{\T}\widetilde{\bTheta}^{\T}\bSigma_{\mathring{\bY}}(k)\bQ$, where  $\widetilde\bTheta=\bI_p\otimes \{(\bQ\otimes\bP)(\bE_2\otimes\bE_1)^{\T}\bw\}$. Note that $\|\widetilde{\bTheta}\|_2=|\bw|_2=O(1)$. By Condition \ref{sp2}(i),  $\|\bSigma_{\bZ,\tilde{\eta}}(k)\|_2\leq \|\bP\|_2\|\widetilde\bTheta\|_2\|\bQ\|_2\| \bSigma_{\mathring{\bY}}(k)\|_2\leq C$ for any $k=1,2$, which implies  $\|\widetilde\bS_2\|_2\leq \|\bSigma_{\bZ,\tilde{\eta}}(1)\|_2\|\bSigma_{\bZ,\tilde{\eta}}(2)\|_2\leq C$. Recall $\widetilde\bS_1$ and $\bE_2^{\T}\widehat{\bS}_1\bE_2$ are two symmetric matrices. Denote by $\lambda_1\geq\cdots\geq\lambda_d$ and $\hat{\lambda}_1\geq\cdots\geq\hat{\lambda}_d$, respectively, the eigenvalues of $\widetilde\bS_1$ and $\bE_2^{\T}\widehat{\bS}_1\bE_2$. By Condition \ref{as:3}(ii), we have $\lambda_d>0$ is uniformly bounded away from zero. By Lemma \ref{s1.exp}, $|\hat{\lambda}_d-\lambda_d|\leq \|\bE_2^{\T}\widehat\bS_1\bE_2-\widetilde\bS_1\|_2=o_{\rm p}(1)$, which implies
\[
\|\bE_2^{\T}\widehat\bS_1^{-1}\bE_2-\widetilde\bS_1^{-1} \|_2\leq \| \bE_2^{\T}\widehat\bS_1^{-1}\bE_2 \|_2\| \bE_2^{\T}\widehat\bS_1\bE_2-\widetilde\bS_1 \|_2\| \widetilde\bS_1^{-1}  \|_2=O_{\rm p}(\Pi_n)\,.
\]
By Triangle inequality and Lemma \ref{s1.exp},
\begin{align*}
\|\bE_2^{\T}\widehat\bS_1^{-1}\widehat\bS_2\bE_2-\widetilde\bS_1^{-1}\widetilde\bS_2  \|_2 \leq&~\|\bE_2^{\T}\widehat\bS_1^{-1}\bE_2-\widetilde\bS_1^{-1} \|_2\|\bE_2^{\T}\widehat\bS_2\bE_2-\widetilde\bS_2 \|_2+\|\widetilde\bS_1^{-1}\|_2\|\bE_2^{\T}\widehat\bS_2\bE_2-\widetilde\bS_2 \|_2\\
&+\|\bE_2^{\T}\widehat\bS_1^{-1}\bE_2-\widetilde\bS_1^{-1} \|_2\|\widetilde\bS_2\|_2\\
=&~O_{\rm p}(\Pi_n)\,,
\end{align*}
which implies $| \kappa^{\ell}\hat{\bbeta}^*_{\bv,\ell}-\bbeta_{\bv,\ell} |_2=\theta_\ell^{-1}\cdot O_{\rm p}(\Pi_n)$ for any $\ell\in[d]$. Note that $\kappa^{\ell}\bE_1^{\T}\widehat{\bSigma}_{\bZ,\eta}(1)\hat{\bbeta}_{\bv,\ell}=\kappa^{\ell}\bE_1^{\T}\widehat{\bSigma}_{\bZ,\eta}(1)\bE_2\hat{\bbeta}^*_{\bv,\ell}$. By Lemma \ref{hatz.exp} and Triangle inequality, for any $\ell\in[d]$, it holds that
\begin{align*}
|\kappa^{\ell}\bE_1^{\T}\widehat{\bSigma}_{\bZ,\eta}(1)\hat{\bbeta}_{\bv,\ell}-\bSigma_{\bZ,\tilde{\eta}}(1)\bbeta_{\bv,\ell}|_2\leq&~\| \bE_1^{\T}\widehat\bSigma_{\bZ,\eta}(1)\bE_2-\bSigma_{\bZ,\tilde{\eta}}(1) \|_2|\kappa^{\ell}\hat\bbeta^*_{\bv,\ell}|_2\\
&+\| \bSigma_{\bZ,\tilde{\eta}}(1) \|_2|\kappa^{\ell}\hat\bbeta^*_{\bv,\ell}-\bbeta_{\bv,\ell}|_2\\
=&~(1+\theta_\ell^{-1})\cdot O_{\rm p}(\Pi_{n})\,.
\end{align*}
Then   $|\kappa^{\ell}\bE_1^{\T}\widehat{\bSigma}_{\bZ,\eta}(1)\hat{\bbeta}_{\bv,\ell}|_2\geq |\bSigma_{\bZ,\tilde{\eta}}(1)\bbeta_{\bv,\ell}|_2-o_{\rm p}(1)$.
By Condition \ref{as:3}(ii), $|\bSigma_{\bZ,\tilde{\eta}}(1)\bbeta_{\bv,\ell}|_2^2=\bbeta_{\bv,\ell}^{\T}\widetilde{\bS}_1\bbeta_{\bv,\ell}\geq \lambda_d>C$, which implies $|\kappa^{\ell}\bE_1^{\T}\widehat{\bSigma}_{\bZ,\eta}(1)\hat{\bbeta}_{\bv,\ell}|_2\geq C$ with probability approaching one. Due to $|\widehat{\bSigma}_{\bZ,\eta}(1)\hat{\bbeta}_{\bv,\ell}|_2=|\kappa^{\ell}\bE_1^{\T}\widehat{\bSigma}_{\bZ,\eta}(1)\hat{\bbeta}_{\bv,\ell}|_2$, by Triangle inequality,
\begin{align*}
&~||\widehat\bSigma_{\bZ,\eta}(1)\hat\bbeta_{\bv,\ell}|_2^{-1}-|\bSigma_{\bZ,\tilde{\eta}}(1)\bbeta_{\bv,\ell}|_2^{-1}|\\
&~~~~~~~~~~~\leq|\kappa^{\ell}\bE_1^{\T}\widehat{\bSigma}_{\bZ,\eta}(1)\hat{\bbeta}_{\bv,\ell}|_2^{-1}|\bSigma_{\bZ,\tilde{\eta}}(1)\bbeta_{\bv,\ell}|_2^{-1} |\kappa^{\ell}\bE_1^{\T}\widehat{\bSigma}_{\bZ,\eta}(1)\hat{\bbeta}_{\bv,\ell}-\bSigma_{\bZ,\tilde{\eta}}(1)\bbeta_{\bv,\ell}|_2\\
&~~~~~~~~~~~=(1+\theta_\ell^{-1})\cdot O_{\rm p}(\Pi_{n})
\end{align*}
for any $\ell\in[d]$.
Hence,
\begin{align*}
\Phi_{\ell}
\leq&~|\kappa^{\ell}\bE_1^{\T}\widehat{\bSigma}_{\bZ,\eta}(1)\hat{\bbeta}_{\bv,\ell}|_2^{-1} |\kappa^{\ell}\bE_1^{\T}\widehat{\bSigma}_{\bZ,\eta}(1)\hat{\bbeta}_{\bv,\ell}-\bSigma_{\bZ,\tilde{\eta}}(1)\bbeta_{\bv,\ell}|_2\\
&+||\kappa^{\ell}\bE_1^{\T}\widehat{\bSigma}_{\bZ,\eta}(1)\hat{\bbeta}_{\bv,\ell}|_2^{-1}-|\bSigma_{\bZ,\tilde{\eta}}(1)\bbeta_{\bv,\ell}|_2^{-1}||\bSigma_{\bZ,\tilde{\eta}}(1)\bbeta_{\bv,\ell}|_2\\
=&~(1+\theta_\ell^{-1})\cdot O_{\rm p}(\Pi_{n})\,.
\end{align*}

Write $\widehat{\bA}=(\hat{\ba}_1,\ldots,\hat{\ba}_d)$ and $\widetilde{\bP}=\widehat{\bP}\bE_1$. For any $\ell\in[d]$, we have
$$
\kappa^{\ell}\hat\ba_{\ell}=\kappa^{\ell}\widehat\bP\hat\bu_{\ell}=|\widehat{\bSigma}_{\bZ,\eta}(1)\hat{\bbeta}_{\bv,\ell}|_2^{-1}\kappa^{\ell}\widetilde\bP\bE_1^{\T}\widehat{\bSigma}_{\bZ,\eta}(1)\hat{\bbeta}_{\bv,\ell}\,.$$
Recall $\bA=(\ba_1,\ldots,\ba_d)$ with $\ba_{\ell}=\bP\bu_{\ell}=|\bSigma_{\bZ,\tilde{\eta}}(1)\bbeta_{\bv,\ell}|_2^{-1}\bP\bSigma_{\bZ,\tilde{\eta}}(1)\bbeta_{\bv,\ell}$. By Proposition \ref{hatp.exp} and  Triangle inequality,  we have
$
|\kappa^{\ell}\hat\ba_{\ell}-\ba_{\ell}|_2\leq\|\widetilde{\bP}-\bP\|_2+\Phi_{\ell}=(1+\theta_\ell^{-1})\cdot O_{\rm p}(\Pi_{n})$
for any $\ell\in[d]$. Notice that
$
|\kappa^{\ell}\hat\ba_{\ell}-\ba_{\ell}|_2^2=2-(\kappa^{\ell}\hat\ba_{\ell})^{\H}\ba_{\ell}-\ba_{\ell}^{\H}\kappa^{\ell}\hat\ba_{\ell}\geq 2-2|\hat{\ba}_{\ell}^{\H}\ba_\ell|$. Hence,
$
1-|\hat{\ba}_{\ell}^{\H}\ba_\ell|=(1+\theta_\ell^{-1})^2\cdot O_{\rm p}(\Pi_{n}^2)$, which implies
\[
1-|\hat{\ba}_{\ell}^{\H}\ba_\ell|^2\leq 2(1-|\hat{\ba}_{\ell}^{\H}\ba_\ell|)=(1+\theta_\ell^{-1})^2\cdot O_{\rm p}(\Pi_{n}^2)\,.
\] Write $\widehat{\bB}=(\hat\bb_1,\ldots,\hat\bb_d)$.  Analogously, we can also prove  $|\kappa^{\ell}_{*}\hat{\bb}_{\ell}-\bb_\ell|_2=\{1+(\theta_\ell^*)^{-1}\}\cdot O_{\rm p}(\Pi_{n})$ for any $\ell\in[d]$, where $\kappa_*^{\ell}\in\{-1,1\}$. We complete the proof of Theorem \ref{a.exp}. $\hfill\Box$
\section{Proofs of auxiliary lemmas}

\subsection{Proof of Lemma \ref{m1.exp}}\label{sec:pfpn2}

We first show that, with $\delta_1=C_* \{n^{-1}\log(pq)\}^{1/2}$ for some sufficiently large constant $C_*>0$,
\begin{align}\label{eq:toshow}
\|T_{\delta_1}\{\widehat{\bSigma}_{\bY,\xi}(k)\}-{\bSigma}_{\bY,\xi}(k)  \|_2=O_{\rm p}(\Pi_{1,n})
\end{align}
 for any $k\in[K]$, provided that $\log(pq)=o(n^{c})$ for some constant $c\in(0,1)$.

To simplify the notation, we write $\widehat{\bSigma}_{\bY,\xi}(k)=(\hat{\sigma}_{i,j}^{(k)})_{p\times q}$ and ${\bSigma}_{\bY,\xi}(k)=(\sigma_{i,j}^{(k)})_{p\times q}$.
Write $n_k=n-k$ and $\bar{y}_{i,j}=n^{-1}\sum_{t=1}^ny_{i,j,t}$. Then we have
\begin{align*}
\hat\sigma_{i,j}^{(k)}-\sigma_{i,j}^{(k)}=&~ \frac{1}{n_k}\sum_{t=k+1}^n\{y_{i,j,t}\xi_{t-k}-\mathbb{E}(y_{i,j,t}\xi_{t-k})\}-\bigg\{\frac{\bar{y}_{i,j}}{n_k}\sum_{t=k+1}^n\xi_{t-k}- \frac{\mathbb{E}(\bar{y}_{i,j})}{n_k}\sum_{t=k+1}^n\mathbb{E}(\xi_{t-k})\bigg\}\\
&~- \bigg\{ \frac{\bar{\xi}}{n_k}\sum_{t=k+1}^ny_{i,j,t}-\frac{\mathbb{E}(\bar{\xi})}{n_k}\sum_{t=k+1}^n\mathbb{E}(y_{i,j,t}) \bigg\}+\{ \bar{y}_{i,j}\bar{\xi}-\mathbb{E}(\bar{y}_{i,j})\mathbb{E}(\bar{\xi}) \}\\
=&: {\rm I}_{i,j}(1)+{\rm I}_{i,j}(2)+{\rm I}_{i,j}(3)+{\rm I}_{i,j}(4)\,.
\end{align*}
Under Condition \ref{exp.tail}(i), applying Lemma 2 of \cite{ChangTangWu_2013}, we have $\mathbb{P}\{| y_{i,j,t}\xi_{t-k}-\mathbb{E}(y_{i,j,t}\xi_{t-k})|>x  \} \lesssim \exp(-Cx^{r_1/2})$ for any $x>0$. Write $|\cdot|_+={\rm max}(\cdot,0)$. Notice that $\{ y_{i,j,t}\xi_{t-k}-\mathbb{E}(y_{i,j,t}\xi_{t-k}) \}_{t=k+1}^n$ is an $\alpha$-mixing sequence with $\alpha$-mixing coefficients $\{\alpha(|m-k|_+)\}_{m\geq 1}$, where $\alpha(\cdot)$ is given in \eqref{alpha.def} in Section \ref{sec4}. Together with Condition \ref{exp.tail}(ii), Lemma L5 in the supplementary material of \cite{ChangChenWu2021} implies
\[
\mathbb{P}\{|{\rm I}_{i,j}(1)|\geq x\}\lesssim \exp(-Cnx^2)+\exp(-Cn^{\tilde{r}}x^{\tilde{r}})
\]
for any $x\in(0,1)$, where $\tilde{r}^{-1}=1+2r_1^{-1}+r_2^{-1}$. Since $\mathbb{E}(y_{i,j,t})=O(1)$ and $\mathbb{E}(\xi_t)=O(1)$ for any $t\in[n]$, applying Lemma L5 in the supplementary material of \cite{ChangChenWu2021} again,
\[
\mathbb{P}\{|{\rm I}_{i,j}(2)+{\rm I}_{i,j}(3)+{\rm I}_{i,j}(4)|\geq x\}\lesssim \exp(-Cnx^2)+\exp(-Cn^{\check{r}}x^{\check{r}})
\]
for any $x\in(0,1)$, where $\check{r}^{-1}=2+|r_1^{-1}-1|_{+}+r_2^{-1}$. Then  $Z:=\max_{i\in[p]}\max_{j\in[q]}|\hat\sigma_{i,j}^{(k)}-\sigma_{i,j}^{(k)}|=O_{\rm p}[\{ n^{-1}\log(pq) \}^{1/2}]$, provided that $\log(pq)=o(n^{c})$ for some constant $c\in(0,1)$ depending only on $r_1$ and $r_2$.

By Triangle inequality, we have
$$
\|T_{\delta_1}\{\widehat{\bSigma}_{\bY,\xi}(k)\}-{\bSigma}_{\bY,\xi}(k)  \|_2\leq \|T_{\delta_1}\{{\bSigma}_{\bY,\xi}(k)\}-{\bSigma}_{\bY,\xi}(k)  \|_2+\|T_{\delta_1}\{\widehat{\bSigma}_{\bY,\xi}(k)\}-T_{\delta_1}\{{\bSigma}_{\bY,\xi}(k)\} \|_2\,,$$
where $T_{\delta_1}\{ {\bSigma}_{\bY,\xi}(k) \}= (\sigma_{i,j}^{(k)} I\{ |\sigma_{i,j}^{(k)}|\geq \delta_1 \})_{p\times q}$. On the one hand,
\[
\|T_{\delta_1}\{{\bSigma}_{\bY,\xi}(k)\}-{\bSigma}_{\bY,\xi}(k)  \|_2^2
\leq \bigg[\max_{i\in[p]}\sum_{j=1}^q|\sigma_{i,j}^{(k)}|I\{|\sigma_{i,j}^{(k)}|<\delta_1 \}\bigg]\bigg[\max_{j\in[q]}\sum_{i=1}^p|\sigma_{i,j}^{(k)}|I\{|\sigma_{i,j}^{(k)}|<\delta_1 \}\bigg]\,.
\]
By Condition \ref{b.x}(ii), we have $\sum_{j=1}^q|\sigma_{i,j}^{(k)}|I\{|\sigma_{i,j}^{(k)}|<\delta_1 \}\leq \delta_1^{1-\iota}s_1$ and $\sum_{i=1}^p|\sigma_{i,j}^{(k)}|I\{|\sigma_{i,j}^{(k)}|<\delta_1 \}\leq \delta_1^{1-\iota}s_2$, which implies %that
$
\|T_{\delta_1}\{{\bSigma}_{\bY,\xi}(k)\}-{\bSigma}_{\bY,\xi}(k)  \|_2 \leq \delta_1^{1-\iota}(s_1s_2)^{1/2}
$.
On the other hand, we have
\begin{align}\label{div.y.i5}
 \|T_{\delta_1}\{\widehat{\bSigma}_{\bY,\xi}(k)\}-T_{\delta_1}\{{\bSigma}_{\bY,\xi}(k)\} \|_2^2
&\,\leq \bigg[\max_{i\in[p]}\sum_{j=1}^q\big|\hat\sigma_{i,j}^{(k)}I\{|\hat\sigma_{i,j}^{(k)}|\geq \delta_1 \}-\sigma_{i,j}^{(k)}I\{|\sigma_{i,j}^{(k)}|\geq \delta_1 \}\big|\bigg]\notag\\
&~~~\times \bigg[\max_{j\in[q]}\sum_{i=1}^p\big|\hat\sigma_{i,j}^{(k)}I\{|\hat\sigma_{i,j}^{(k)}|\geq \delta_1 \}-\sigma_{i,j}^{(k)}I\{|\sigma_{i,j}^{(k)}|\geq \delta_1 \}\big|\bigg]\notag\\
&\,=:  {\rm I}(1)\times {\rm I}(2)\,.
\end{align}
By Triangle inequality,
\begin{align*}
	{\rm I}(1) \leq&~ \max_{i\in[p]}\sum_{j=1}^q|\hat\sigma_{i,j}^{(k)}-\sigma_{i,j}^{(k)}|I\{ |\hat\sigma_{i,j}^{(k)}|\geq \delta_1,|\sigma_{i,j}^{(k)}|\geq \delta_1 \}
	+ \max_{i\in[p]}\sum_{j=1}^q|\hat\sigma_{i,j}^{(k)}|I\{ |\hat\sigma_{i,j}^{(k)}|\geq \delta_1,|\sigma_{i,j}^{(k)}|< \delta_1 \}\\
&+ \max_{i\in[p]}\sum_{j=1}^q|\sigma_{i,j}^{(k)}|I\{ |\hat\sigma_{i,j}^{(k)}|< \delta_1,|\sigma_{i,j}^{(k)}|\geq \delta_1 \}\\
	=&:{\rm I}(1,1)+{\rm I}(1,2)+{\rm I}(1,3)\,.
\end{align*}
Recall $Z=\max_{i\in[p]}\max_{j\in[q]}|\hat\sigma_{i,j}^{(k)}-\sigma_{i,j}^{(k)}|$. By Condition \ref{b.x}(ii),
$
{\rm I}(1,1) \leq Z \times \max_{i\in[p]}\sum_{j=1}^qI\{ |\hat\sigma_{i,j}^{(k)}|\geq \delta_1,|\sigma_{i,j}^{(k)}|\geq \delta_1 \} \leq Z\delta_1^{-\iota}s_1
$.
Applying Triangle inequality and Condition \ref{b.x}(ii) again, we have
\begin{align*}
{\rm I}(1,2)\leq &~	\max_{i\in[p]}\sum_{j=1}^q|\hat\sigma_{i,j}^{(k)}-\sigma_{i,j}^{(k)}|I\{ |\hat\sigma_{i,j}^{(k)}|\geq \delta_1,|\sigma_{i,j}^{(k)}|< \delta_1 \}+ \max_{i\in[p]}\sum_{j=1}^q|\sigma_{i,j}^{(k)}|I\{ |\sigma_{i,j}^{(k)}|< \delta_1 \}\\
\leq& ~\delta_1^{1-\iota}s_1+  \max_{i\in[p]}\sum_{j=1}^q|\hat\sigma_{i,j}^{(k)}-\sigma_{i,j}^{(k)}|I\{ |\hat\sigma_{i,j}^{(k)}|\geq \delta_1,|\sigma_{i,j}^{(k)}|< \delta_1 \}\,.
\end{align*}
 Taking $\theta\in(0,1)$, by Triangle inequality and Condition \ref{b.x}(ii), we have
\begin{align*}
&\max_{i\in[p]}\sum_{j=1}^q|\hat\sigma_{i,j}^{(k)}-\sigma_{i,j}^{(k)}|I\{ |\hat\sigma_{i,j}^{(k)}|\geq \delta_1,|\sigma_{i,j}^{(k)}|< \delta_1 \}\\
&~~~~~~~~~~~~~\leq \max_{i\in[p]}\sum_{j=1}^q|\hat\sigma_{i,j}^{(k)}-\sigma_{i,j}^{(k)}|I\{ |\hat\sigma_{i,j}^{(k)}|\geq \delta_1,|\sigma_{i,j}^{(k)}|\leq \theta \delta_1 \}\\
&~~~~~~~~~~~~~~~~+\max_{i\in[p]}\sum_{j=1}^q|\hat\sigma_{i,j}^{(k)}-\sigma_{i,j}^{(k)}|I\{ |\hat\sigma_{i,j}^{(k)}|\geq \delta_1,\theta \delta_1<|\sigma_{i,j}^{(k)}|< \delta_1 \}\\
&~~~~~~~~~~~~~\leq Z\times \max_{i\in[p]}\sum_{j=1}^qI\{|\hat\sigma_{i,j}^{(k)}-\sigma_{i,j}^{(k)}| \geq (1-\theta)\delta_1 \}+ Z\theta^{-\iota}\delta_1^{-\iota}s_1\,,
\end{align*}
which implies
$$
{\rm I}(1,2)\leq Z\times \max_{i\in[p]}\sum_{j=1}^qI\{|\hat\sigma_{i,j}^{(k)}-\sigma_{i,j}^{(k)}| \geq (1-\theta)\delta_1 \}+Z\theta^{-\iota}\delta_1^{-\iota}s_1+\delta_1^{1-\iota}s_1\,.
$$
Meanwhile, by Triangle inequality and Condition \ref{b.x}(ii), we have
\begin{align*}
{\rm I}(1,3)\leq  &~\max_{i\in[p]}\sum_{j=1}^q|\hat\sigma_{i,j}^{(k)}-\sigma_{i,j}^{(k)}|I\{ |\hat\sigma_{i,j}^{(k)}|< \delta_1,|\sigma_{i,j}^{(k)}|\geq \delta_1 \}\\
&+ \max_{i\in[p]}\sum_{j=1}^q|\hat\sigma_{i,j}^{(k)}|I\{ |\hat\sigma_{i,j}^{(k)}|< \delta_1,|\sigma_{i,j}^{(k)}|\geq \delta_1 \}\\
\leq&~  Z\delta_1^{-\iota}s_1+ \delta_1^{1-\iota}s_1\,.
\end{align*}
Hence, it holds that
$$
{\rm I}(1) \lesssim  Z\delta_1^{-\iota}s_1+ \delta_1^{1-\iota}s_1+Z\theta^{-\iota}\delta_1^{-\iota}s_1
+ Z\times \max_{i\in[p]}\sum_{j=1}^qI\{|\hat\sigma_{i,j}^{(k)}-\sigma_{i,j}^{(k)}| \geq (1-\theta)\delta_1 \}\,.$$
Selecting $\delta_1= C_*\{n^{-1}\log(pq)\}^{1/2}$ for some sufficiently large $C_*>0$,  by Markov inequality, we have
\begin{align*}
	&\mathbb{P}\bigg[\max_{i\in[p]}\sum_{j=1}^qI\{|\hat\sigma_{i,j}^{(k)}-\sigma_{i,j}^{(k)}| \geq (1-\theta)\delta_1 \}>\lambda \bigg]\\
	&~~~~~~~~~~~~~~\leq  \frac{1}{\lambda}\sum_{i=1}^p\sum_{j=1}^q\mathbb{P}\{|\hat\sigma_{i,j}^{(k)}-\sigma_{i,j}^{(k)}| \geq (1-\theta)\delta_1 \}\leq \frac{C}{\lambda}
\end{align*}
for any $\lambda>1$,
provided that $\log(pq)=o(n^c)$ for some constant $c\in(0,1)$ depending only on $r_1$ and $r_2$, which implies  $\max_{i\in[p]}\sum_{j=1}^qI\{|\hat\sigma_{i,j}^{(k)}-\sigma_{i,j}^{(k)}| \geq (1-\theta)\delta_1 \}=O_{\rm p}(1)$. Recall $Z=O_{\rm p}[\{n^{-1}\log(pq)\}^{1/2}]$. Therefore, ${\rm I}(1)=O_{\rm p}[s_1\{n^{-1}\log(pq)\}^{(1-\iota)/2}]$. Analogously, we also have ${\rm I}(2)=O_{\rm p}[s_2\{n^{-1}\log(pq)\}^{(1-\iota)/2}]$. By \eqref{div.y.i5}, we have
$
\|T_{\delta_1}\{\widehat{\bSigma}_{\bY,\xi}(k)\}-T_{\delta_1}\{{\bSigma}_{\bY,\xi}(k)\} \|_2=O_{\rm p}(\Pi_{1,n})$
for any $k\in[K]$. Together with $
\|T_{\delta_1}\{{\bSigma}_{\bY,\xi}(k)\}-{\bSigma}_{\bY,\xi}(k)  \|_2 \leq \delta_1^{1-\iota}(s_1s_2)^{1/2}=O(\Pi_{1,n})
$, we complete the proof of \eqref{eq:toshow}.

Due to $\widehat{\bM}_{1}=\sum_{k=1}^KT_{\delta_1}\{\widehat{\bSigma}_{\bY,\xi}(k)\}T_{\delta_1}\{\widehat{\bSigma}_{\bY,\xi}(k)^{\T}\}$ and ${\bM}_1=\sum_{k=1}^K{\bSigma}_{\bY,\xi}(k){\bSigma}_{\bY,\xi}(k)^{\T}$, by Triangle inequality, \eqref{eq:toshow} and Condition \ref{b.x}(i),
\begin{align*}
	\|\widehat{\bM}_{1}-\bM_1\|_2 \leq&~ \sum_{k=1}^{K}\|T_{\delta_1}\{\widehat{\bSigma}_{\bY,\xi}(k)\}-{\bSigma}_{\bY,\xi}(k)\|_2^2\\
	&+2\sum_{k=1}^{K}\|{\bSigma}_{\bY,\xi}(k)\|_2\|T_{\delta_1}\{\widehat{\bSigma}_{\bY,\xi}(k)\}-{\bSigma}_{\bY,\xi}(k)\|_2\\
	=&~O_{\rm p}(\Pi_{1,n})\,,
\end{align*}
provided that $\Pi_{1,n}=o(1)$ and $\log(pq)=o(n^{c})$ for some constant $c\in(0,1)$ depending only on $r_1$ and $r_2$.
Analogously, we also have $\|\widehat{\bM}_{2}-\bM_2\|_2=O_{\rm p}(\Pi_{1,n})$. $\hfill\Box$

\subsection{Proof of Lemma \ref{hatz.exp}}\label{sec:pflemma1}
By the same arguments for \eqref{eq:toshow}, if $\delta_2=C_{**}\{n^{-1}\log(pq)\}^{1/2}$ for some sufficiently large constant $C_{**}>0$, we have $
\|T_{\delta_2}\{\widehat{\bSigma}_{\check\bY}(k)\}-\bSigma_{\mathring{\bY}}(k)\|_2=O_{\rm p}(\Pi_{2,n})$. Write $\widetilde{\bP}=\widehat{\bP}\bE_1$ and $\widetilde{\bQ}=\widehat{\bQ}\bE_2$ for $(\bE_1,\bE_2)$ specified in Proposition \ref{hatp.exp}. Then $\bE_1^{\T}\widehat\bSigma_{\bZ,\eta}(k)\bE_2=\widetilde\bP^{\T}\widehat\bTheta^{\T}T_{\delta_2}\{\widehat{\bSigma}_{\check\bY}(k)\}\widetilde\bQ$.  Since
\begin{align*}
	\| \widehat\bTheta-\widetilde\bTheta \|_2=&~|\{\widehat\bQ\otimes\widehat\bP-(\bQ\bE_2^{\T})\otimes(\bP\bE_1^{\T})\}\bw|_2\\
	\lesssim&~ \|\widehat\bQ\otimes\widehat\bP-(\bQ\bE_2^{\T})\otimes(\bP\bE_1^{\T})\|_2\leq \|\widehat{\bQ}\bE_2-\bQ\|_2+\|\widehat{\bP}\bE_1-\bP\|_2\,,
\end{align*}
by Proposition \ref{hatp.exp}, we have $\|\widehat\bTheta-\widetilde\bTheta\|_2=O_{\rm p}(\Pi_{1,n})$. Note that $\|\widehat\bTheta\|_2=\|\widetilde{\bTheta}\|_2=|\bw|_2=O(1)$ and $\bSigma_{\bZ,\tilde{\eta}}(k)=\bP^{\T}\widetilde\bTheta^{\T}\bSigma_{\mathring{\bY}}(k)\bQ$.  Together with Condition \ref{sp2}(i), we have $\max_{k\in\{1,2\}}\|\bE_1^{\T}\widehat\bSigma_{\bZ,\eta}(k)\bE_2-\bSigma_{\bZ,\tilde{\eta}}(k)\|_2=O_{\rm p}(\Pi_{n})$ with $\Pi_n=\Pi_{1,n}+\Pi_{2,n}$. $\hfill\Box$

\subsection{Proof of Lemma \ref{s1.exp}}\label{sec:pfpn3}

 Recall $\bSigma_{\bZ,\tilde{\eta}}(k)=\bP^{\T}\widetilde\bTheta^{\T}\bSigma_{\mathring{\bY}}(k)\bQ$ with $\widetilde\bTheta=\bI_p\otimes \{(\bQ\otimes\bP)(\bE_2\otimes \bE_1)^{\T}\bw\}$, and $\|\widetilde\bTheta\|_2=|\bw|_2=O(1)$. By Condition \ref{sp2}(i), we have $\max_{k\in\{1,2\}}\|\bSigma_{\bZ,\tilde{\eta}}(k)\|_2\leq C$. Note that $\widetilde\bS_1 =\bSigma_{\bZ,\tilde{\eta}}(1)^{\T} \bSigma_{\bZ,\tilde{\eta}}(1)$ and $\widehat{\bS}_1=\widehat\bSigma_{\bZ,{\eta}}(1)^{\T} \widehat\bSigma_{\bZ,{\eta}}(1)$. By Triangle inequality and Lemma \ref{hatz.exp},
 \begin{align*}
 \|\bE_2^{\T}\widehat{\bS}_1\bE_2-\widetilde\bS_1\|_2 \leq&~\|\bE_1^{\T}\widehat{\bSigma}_{\bZ,\eta}(1)\bE_2-\bSigma_{\bZ,\tilde{\eta}}(1)\|_2^2+2\|\bSigma_{\bZ,\tilde{\eta}}(1)\|_2\|\bE_1^{\T}\widehat{\bSigma}_{\bZ,\eta}(1)\bE_2-\bSigma_{\bZ,\tilde{\eta}}(1)\|_2\\
 =&~O_{\rm p}(\Pi_{n})\,.
 \end{align*}
Analogously, we can also construct other results. $\hfill\Box$

\section{Additional simulation results}

\begin{table}[!tbp]
\renewcommand\arraystretch{1.2}
\centering
\begin{threeparttable}
\tiny
\caption{\textit{ Relative frequency estimates of $\mathbb{P}(\hat{d} = d)$ for $d=1$ in the refined estimation procedure based on $2000$ repetitions, where $\xi_t$ is determined by the PCA method and the random weight method. We set $(\delta_1,c_n)=(0,0)$ in \eqref{eq:ratio}, and consider the results based on $\widehat{\bM}_1$ and $\widehat{\bM}_2$, respectively. }}\label{SM:Dhat_d1}
\begin{tabular}{cc|ccc|ccc|ccc|ccc}
\hline\hline
        &     & \multicolumn{6}{c|}{PCA}                             & \multicolumn{6}{c}{Random weights}  \\ \cline{3-14}
        &     & \multicolumn{3}{c|}{$\hat{d}$ based on $\wh\bM_1$}   & \multicolumn{3}{c|}{$\hat{d}$ based on $\wh\bM_2$} & \multicolumn{3}{c|}{$\hat{d}$ based on $\wh\bM_1$}   & \multicolumn{3}{c}{$\hat{d}$ based on $\wh\bM_2$}   \\
$(p,\,q)$   & $n$   & $K=3$    & $K=5$    & $K=7$    & $K=3$    & $K=5$    & $K=7$  & $K=3$    & $K=5$    & $K=7$    & $K=3$    & $K=5$    & $K=7$   \\ \hline
(4,4)   & 300 & 100.00 & 100.00 & 100.00 & 100.00 & 100.00 & 100.00 & 99.80  & 100.00 & 100.00 & 99.80  & 99.90  & 100.00 \\
        & 600 & 100.00 & 100.00 & 100.00 & 100.00 & 100.00 & 100.00 & 99.95  & 100.00 & 99.95  & 99.95  & 100.00 & 99.95  \\
        & 900 & 100.00 & 100.00 & 100.00 & 100.00 & 100.00 & 100.00 & 100.00 & 100.00 & 100.00 & 99.90  & 100.00 & 99.95  \\
(8,8)   & 300 & 100.00 & 100.00 & 100.00 & 100.00 & 100.00 & 100.00 & 100.00 & 100.00 & 100.00 & 100.00 & 100.00 & 100.00 \\
        & 600 & 100.00 & 100.00 & 100.00 & 100.00 & 100.00 & 100.00 & 100.00 & 100.00 & 100.00 & 99.95  & 100.00 & 100.00 \\
        & 900 & 100.00 & 100.00 & 100.00 & 100.00 & 100.00 & 100.00 & 100.00 & 100.00 & 100.00 & 100.00 & 100.00 & 100.00 \\
(16,16) & 300 & 100.00 & 100.00 & 100.00 & 100.00 & 100.00 & 100.00 & 99.95  & 100.00 & 100.00 & 100.00 & 100.00 & 100.00 \\
        & 600 & 100.00 & 100.00 & 100.00 & 100.00 & 100.00 & 100.00 & 100.00 & 100.00 & 100.00 & 100.00 & 100.00 & 100.00 \\
        & 900 & 100.00 & 100.00 & 100.00 & 100.00 & 100.00 & 100.00 & 100.00 & 100.00 & 100.00 & 100.00 & 100.00 & 100.00 \\
(32,32) & 300 & 100.00 & 100.00 & 100.00 & 100.00 & 100.00 & 100.00 & 100.00 & 100.00 & 100.00 & 100.00 & 100.00 & 100.00 \\
        & 600 & 100.00 & 100.00 & 100.00 & 100.00 & 100.00 & 100.00 & 100.00 & 100.00 & 100.00 & 100.00 & 100.00 & 100.00 \\
        & 900 & 100.00 & 100.00 & 100.00 & 100.00 & 100.00 & 100.00 & 100.00 & 100.00 & 100.00 & 100.00 & 100.00 & 100.00 \\ \hline
(32,4)  & 300 & 100.00 & 100.00 & 100.00 & 100.00 & 100.00 & 100.00 & 99.95  & 100.00 & 100.00 & 99.95  & 100.00 & 100.00 \\
        & 600 & 100.00 & 100.00 & 100.00 & 100.00 & 100.00 & 100.00 & 100.00 & 100.00 & 100.00 & 100.00 & 100.00 & 100.00 \\
        & 900 & 100.00 & 100.00 & 100.00 & 100.00 & 100.00 & 100.00 & 100.00 & 100.00 & 100.00 & 100.00 & 100.00 & 100.00 \\
(64,4)  & 300 & 100.00 & 100.00 & 100.00 & 100.00 & 100.00 & 100.00 & 100.00 & 100.00 & 100.00 & 100.00 & 100.00 & 100.00 \\
        & 600 & 100.00 & 100.00 & 100.00 & 100.00 & 100.00 & 100.00 & 100.00 & 100.00 & 100.00 & 99.95  & 100.00 & 100.00 \\
        & 900 & 100.00 & 100.00 & 100.00 & 100.00 & 100.00 & 100.00 & 100.00 & 100.00 & 100.00 & 100.00 & 100.00 & 100.00 \\
(128,4) & 300 & 100.00 & 100.00 & 100.00 & 100.00 & 100.00 & 100.00 & 100.00 & 100.00 & 100.00 & 100.00 & 100.00 & 100.00 \\
        & 600 & 100.00 & 100.00 & 100.00 & 100.00 & 100.00 & 100.00 & 100.00 & 99.95  & 100.00 & 100.00 & 100.00 & 100.00 \\
        & 900 & 100.00 & 100.00 & 100.00 & 100.00 & 100.00 & 100.00 & 100.00 & 100.00 & 100.00 & 100.00 & 100.00 & 100.00 \\
(256,4) & 300 & 100.00 & 100.00 & 100.00 & 100.00 & 100.00 & 100.00 & 100.00 & 100.00 & 100.00 & 100.00 & 100.00 & 100.00 \\
        & 600 & 100.00 & 100.00 & 100.00 & 100.00 & 100.00 & 100.00 & 100.00 & 100.00 & 100.00 & 100.00 & 100.00 & 100.00 \\
        & 900 & 100.00 & 100.00 & 100.00 & 100.00 & 100.00 & 100.00 & 100.00 & 100.00 & 100.00 & 100.00 & 100.00 & 100.00 \\ \hline
(4,32)  & 300 & 100.00 & 100.00 & 100.00 & 100.00 & 100.00 & 100.00 & 100.00 & 100.00 & 100.00 & 100.00 & 100.00 & 100.00 \\
        & 600 & 100.00 & 100.00 & 100.00 & 100.00 & 100.00 & 100.00 & 100.00 & 100.00 & 100.00 & 100.00 & 100.00 & 100.00 \\
        & 900 & 100.00 & 100.00 & 100.00 & 100.00 & 100.00 & 100.00 & 100.00 & 100.00 & 100.00 & 100.00 & 100.00 & 100.00 \\
(4,64)  & 300 & 100.00 & 100.00 & 100.00 & 100.00 & 100.00 & 100.00 & 100.00 & 100.00 & 100.00 & 100.00 & 100.00 & 100.00 \\
        & 600 & 100.00 & 100.00 & 100.00 & 100.00 & 100.00 & 100.00 & 100.00 & 100.00 & 100.00 & 100.00 & 100.00 & 100.00 \\
        & 900 & 100.00 & 100.00 & 100.00 & 100.00 & 100.00 & 100.00 & 100.00 & 100.00 & 100.00 & 100.00 & 100.00 & 100.00 \\
(4,128) & 300 & 100.00 & 100.00 & 100.00 & 100.00 & 100.00 & 100.00 & 99.95  & 100.00 & 100.00 & 100.00 & 100.00 & 100.00 \\
        & 600 & 100.00 & 100.00 & 100.00 & 100.00 & 100.00 & 100.00 & 100.00 & 100.00 & 100.00 & 100.00 & 100.00 & 100.00 \\
        & 900 & 100.00 & 100.00 & 100.00 & 100.00 & 100.00 & 100.00 & 100.00 & 100.00 & 100.00 & 100.00 & 100.00 & 100.00 \\
(4,256) & 300 & 100.00 & 100.00 & 100.00 & 100.00 & 100.00 & 100.00 & 100.00 & 100.00 & 100.00 & 100.00 & 100.00 & 100.00 \\
        & 600 & 100.00 & 100.00 & 100.00 & 100.00 & 100.00 & 100.00 & 100.00 & 100.00 & 100.00 & 100.00 & 100.00 & 100.00 \\
        & 900 & 100.00 & 100.00 & 100.00 & 100.00 & 100.00 & 100.00 & 100.00 & 100.00 & 100.00 & 100.00 & 100.00 & 100.00 \\
\hline\hline
\end{tabular}
%\begin{tablenotes}
%  \item[*] ``New'' represents the proposed new method, and ``Original'' represents the proposed original method.
%\end{tablenotes}
\end{threeparttable}
\end{table}

\begin{table}[!tbp]
\renewcommand\arraystretch{1.2}
\centering
\begin{threeparttable}
\tiny
\caption{\textit{ Relative frequency estimates of $\mathbb{P}(\hat{d} = d)$ for $d=3$ in the refined estimation procedure based on $2000$ repetitions, where $\xi_t$ is determined by the PCA method and the random weight method. We set $(\delta_1,c_n)=(0,0)$ in \eqref{eq:ratio}, and consider the results based on $\widehat{\bM}_1$ and $\widehat{\bM}_2$, respectively. }}\label{SM:Dhat_d3}
\begin{tabular}{cc|ccc|ccc|ccc|ccc}
\hline\hline
        &     & \multicolumn{6}{c|}{PCA}                             & \multicolumn{6}{c}{Random weights}  \\ \cline{3-14}
        &     & \multicolumn{3}{c|}{$\hat{d}$ based on $\wh\bM_1$}   & \multicolumn{3}{c|}{$\hat{d}$ based on $\wh\bM_2$} & \multicolumn{3}{c|}{$\hat{d}$ based on $\wh\bM_1$}   & \multicolumn{3}{c}{$\hat{d}$ based on $\wh\bM_2$}   \\
$(p,\,q)$   & $n$   & $K=3$    & $K=5$    & $K=7$    & $K=3$    & $K=5$    & $K=7$  & $K=3$    & $K=5$    & $K=7$    & $K=3$    & $K=5$    & $K=7$   \\ \hline
$(8,\,8)$   & 300 & 78.85  & 79.85 & 80.35 & 79.00  & 79.75 & 80.20 & 68.90  & 70.75 & 72.00 & 68.60  & 70.55 & 72.10 \\
        & 600 & 82.45  & 82.15 & 81.65 & 82.50  & 82.40 & 82.30 & 70.75  & 71.30 & 71.95 & 70.65  & 70.90 & 71.65 \\
        & 900 & 85.55  & 85.05 & 84.50  & 85.55  & 84.60 & 84.70 & 73.20  & 72.40 & 72.80 & 72.55  & 72.20 & 72.60 \\
$(16,\,16)$ & 300 & 89.45  & 90.95 & 92.10  & 89.50  & 90.70 & 92.20 & 75.65  & 79.30 & 81.15 & 75.50  & 78.60 & 80.50 \\
        & 600 & 93.85  & 94.35 & 95.05 & 93.60  & 94.40 & 94.75 & 79.05  & 80.90 & 83.20 & 79.40  & 81.15 & 82.85 \\
        & 900 & 94.95  & 94.75 & 95.10  & 95.05  & 95.05 & 95.30 & 79.75  & 81.10 & 81.95 & 79.55  & 80.70 & 81.70 \\
$(32,\,32)$ & 300 & 94.30   & 96.25 & 96.45 & 94.50  & 96.15 & 96.35 & 84.00  & 86.15 & 87.80 & 83.60  & 86.20 & 88.00 \\
        & 600 & 96.20   & 96.95 & 97.60  & 96.20  & 97.15 & 97.50 & 84.60  & 86.15 & 87.30 & 84.55  & 86.05 & 87.35 \\
        & 900 & 97.20   & 97.80  & 98.35 & 97.10  & 97.55 & 98.30 & 84.70  & 86.75 & 87.50 & 84.80  & 86.35 & 87.25 \\
$(64,\,64)$ & 300 & 95.80   & 96.95 & 97.80  & 95.85  & 96.90 & 97.85 & 89.10  & 91.85 & 93.50 & 89.15  & 91.95 & 93.50 \\
        & 600 & 96.95  & 98.25 & 98.90  & 96.95  & 98.20 & 98.90 & 87.55  & 90.15 & 92.35 & 87.60  & 90.25 & 92.40 \\
        & 900 & 98.15  & 98.90  & 99.10  & 98.10  & 98.85 & 99.10 & 90.35  & 92.00 & 92.95 & 90.60  & 92.05 & 93.05 \\ \hline
$(32,\,8)$  & 300 & 88.65  & 90.20  & 91.55 & 85.85  & 86.25 & 87.60 & 74.60  & 77.60 & 80.00 & 70.85  & 73.50 & 75.25 \\
        & 600 & 92.95  & 93.65 & 94.50  & 91.05  & 91.10 & 91.30 & 77.95  & 78.50 & 80.95 & 74.85  & 75.25 & 76.30 \\
        & 900 & 94.45  & 95.25 & 96.00    & 92.90  & 93.05 & 93.00 & 80.75  & 81.90 & 82.85 & 78.45  & 77.95 & 78.45 \\
$(64,\,8)$  & 300 & 89.85  & 92.45 & 93.70  & 87.15  & 88.30 & 88.80 & 77.25  & 81.35 & 84.05 & 73.65  & 75.65 & 77.35 \\
        & 600 & 93.55  & 94.60  & 95.75 & 91.90  & 91.65 & 92.85 & 79.80  & 82.45 & 83.85 & 77.55  & 78.00 & 78.90 \\
        & 900 & 95.65  & 96.05 & 96.50  & 94.15  & 94.10 & 94.50 & 81.80  & 83.55 & 85.40 & 79.00  & 79.60 & 80.50 \\
$(128,\,8)$ & 300 & 91.40   & 93.55 & 94.90  & 89.20  & 89.95 & 90.20 & 79.75  & 82.20 & 84.50 & 74.75  & 75.90 & 77.55 \\
        & 600 & 95.05  & 95.65 & 96.55 & 93.65  & 93.65 & 93.45 & 80.40  & 83.25 & 85.75 & 77.10  & 77.65 & 78.80 \\
        & 900 & 96.45  & 97.05 & 97.35 & 95.30  & 95.30 & 95.55 & 81.80  & 83.85 & 85.50 & 78.20  & 78.55 & 79.10 \\
$(256,\,8)$ & 300 & 91.9   & 94.00    & 95.05 & 88.85  & 89.75 & 90.50 & 80.65  & 84.40 & 86.65 & 76.40  & 77.90 & 79.85 \\
        & 600 & 94.65  & 96.30  & 96.65 & 92.90  & 93.35 & 93.05 & 82.15  & 84.45 & 86.40 & 79.15  & 79.25 & 80.35 \\
        & 900 & 97.25  & 98.00    & 98.10  & 96.10  & 96.25 & 96.20 & 84.20  & 85.70 & 86.70 & 80.80  & 80.95 & 81.60 \\ \hline
$(8,\,32)$  & 300 & 84.30  & 85.80 & 86.55 & 87.10   & 89.65 & 91.20  & 73.45  & 75.05 & 76.70 & 76.80  & 80.05 & 82.15 \\
        & 600 & 90.40  & 89.90 & 89.55 & 91.95  & 93.00    & 93.35 & 76.95  & 76.85 & 77.95 & 80.00  & 81.55 & 83.25 \\
        & 900 & 93.40  & 93.35 & 93.35 & 94.60   & 95.30  & 96.00    & 77.15  & 77.75 & 77.10 & 80.00  & 80.70 & 82.10 \\
$(8,\,64)$  & 300 & 88.00  & 88.10 & 89.15 & 90.45  & 92.05 & 94.05 & 73.20  & 75.25 & 76.70 & 77.20  & 79.90 & 82.40 \\
        & 600 & 91.95  & 91.85 & 91.95 & 94.10   & 94.70  & 95.30  & 76.25  & 77.70 & 78.45 & 80.20  & 82.20 & 83.80 \\
        & 900 & 94.35  & 93.95 & 94.05 & 95.70   & 96.45 & 97.10  & 79.35  & 79.85 & 80.15 & 82.05  & 83.60 & 85.05 \\
$(8,\,128)$ & 300 & 87.50  & 88.50 & 88.90 & 91.05  & 92.75 & 93.70  & 74.50  & 77.05 & 78.55 & 79.05  & 82.40 & 85.20 \\
        & 600 & 93.10  & 93.55 & 93.30 & 94.45  & 95.55 & 96.40  & 78.45  & 79.30 & 80.50 & 82.00  & 83.85 & 85.80 \\
        & 900 & 94.45  & 94.55 & 94.65 & 96.20  & 97.00    & 97.60  & 78.75  & 79.55 & 79.70 & 81.65  & 83.90 & 85.45 \\
$(8,\,256)$ & 300 & 88.25  & 89.30 & 89.95 & 91.35  & 93.15 & 94.80  & 74.20  & 75.80 & 78.05 & 78.35  & 82.30 & 85.10 \\
        & 600 & 94.50  & 94.45 & 94.80 & 95.65  & 96.45 & 97.25 & 78.15  & 79.60 & 80.40 & 81.85  & 84.20 & 86.30 \\
        & 900 & 96.20  & 95.95 & 95.95 & 97.20   & 97.70  & 98.15 & 78.10  & 78.95 & 79.65 & 82.05  & 84.40 & 85.80 \\
\hline\hline
\end{tabular}
%\begin{tablenotes}
%  \item[*] ``New'' represents the proposed new method, and ``Original'' represents the proposed original method.
%\end{tablenotes}
\end{threeparttable}
\end{table}

\begin{table}[!tbp]
\renewcommand\arraystretch{1.2}
\centering
\begin{threeparttable}
\tiny
\caption{\textit{ Relative frequency estimates of $\mathbb{P}(\hat{d} = d)$ for $d=6$ in the refined estimation procedure based on $2000$ repetitions, where $\xi_t$ is determined by the PCA method and the random weight method. We set $(\delta_1,c_n)=(0,0)$ in \eqref{eq:ratio}, and consider the results based on $\widehat{\bM}_1$ and $\widehat{\bM}_2$, respectively. }}\label{SM:Dhat_d6}
\begin{tabular}{cc|ccc|ccc|ccc|ccc}
\hline\hline
        &     & \multicolumn{6}{c|}{PCA}                             & \multicolumn{6}{c}{Random weights}  \\ \cline{3-14}
        &     & \multicolumn{3}{c|}{$\hat{d}$ based on $\wh\bM_1$}   & \multicolumn{3}{c|}{$\hat{d}$ based on $\wh\bM_2$} & \multicolumn{3}{c|}{$\hat{d}$ based on $\wh\bM_1$}   & \multicolumn{3}{c}{$\hat{d}$ based on $\wh\bM_2$}   \\
$(p,\,q)$   & $n$   & $K=3$    & $K=5$    & $K=7$    & $K=3$    & $K=5$    & $K=7$  & $K=3$    & $K=5$    & $K=7$    & $K=3$    & $K=5$    & $K=7$   \\ \hline
$(12,\,12)$  & 300 & 73.35  & 78.15 & 81.80 & 73.05  & 78.60 & 81.80 & 65.80  & 71.15 & 75.40 & 64.85  & 70.55 & 75.20 \\
         & 600 & 77.85  & 81.50 & 84.85 & 76.85  & 81.45 & 84.85 & 66.95  & 71.45 & 75.40 & 66.85  & 71.80 & 75.15 \\
         & 900 & 80.15  & 82.90 & 85.10 & 81.25  & 84.05 & 85.70 & 70.25  & 73.80 & 76.20 & 70.40  & 74.45 & 76.15 \\
$(16,\,16)$  & 300 & 81.50  & 87.00 & 89.05 & 81.30  & 86.70 & 89.50 & 71.50  & 78.50 & 82.35 & 71.95  & 78.60 & 82.55 \\
         & 600 & 85.35  & 89.60 & 91.40 & 85.20  & 89.20 & 91.20 & 75.00  & 80.10 & 83.00 & 75.15  & 80.45 & 84.10 \\
         & 900 & 88.45  & 90.90 & 93.20 & 88.40  & 91.60 & 94.00 & 76.15  & 80.45 & 82.95 & 76.50  & 80.60 & 83.95 \\
$(32,\,32)$  & 300 & 90.65  & 94.90 & 96.40 & 90.70  & 94.50 & 96.20 & 84.50  & 90.30 & 92.85 & 84.05  & 90.15 & 92.75 \\
         & 600 & 92.50  & 96.40 & 97.80 & 92.65  & 96.15 & 97.85 & 85.40  & 90.70 & 93.20 & 85.45  & 90.45 & 92.70 \\
         & 900 & 93.40  & 96.00 & 97.55 & 93.65  & 95.80 & 97.55 & 85.40  & 90.15 & 92.45 & 85.40  & 90.65 & 92.30 \\
$(64,\,64)$  & 300 & 94.30  & 98.35 & 99.20 & 94.20  & 98.15 & 99.25 & 90.10  & 94.70 & 96.45 & 90.20  & 94.65 & 96.55 \\
         & 600 & 96.15  & 98.20 & 99.10 & 96.35  & 98.05 & 99.10 & 91.10  & 94.45 & 96.20 & 90.80  & 94.50 & 96.15 \\
         & 900 & 96.30  & 98.35 & 99.10 & 96.20  & 98.40 & 99.10 & 92.05  & 95.70 & 96.85 & 92.00  & 95.50 & 96.70 \\ \hline
$(32,\,12)$  & 300 & 85.25  & 90.85 & 94.55 & 78.05  & 82.70 & 86.20 & 77.80  & 83.85 & 87.30 & 69.25  & 74.00 & 77.85 \\
         & 600 & 89.55  & 93.75 & 95.00 & 84.60  & 87.90 & 89.70 & 77.80  & 84.10 & 87.10 & 70.70  & 76.05 & 79.15 \\
         & 900 & 90.65  & 93.50 & 95.75 & 86.95  & 90.05 & 91.50 & 77.95  & 84.05 & 87.60 & 71.40  & 76.25 & 79.20 \\
$(64,\,12)$  & 300 & 88.35  & 93.55 & 95.50 & 79.70  & 85.70 & 88.30 & 81.55  & 87.65 & 91.00 & 71.65  & 76.60 & 81.05 \\
         & 600 & 92.00  & 95.70 & 97.00 & 86.75  & 89.70 & 91.70 & 81.65  & 87.40 & 90.25 & 75.00  & 79.10 & 82.30 \\
         & 900 & 93.75  & 96.50 & 97.70 & 89.50  & 91.80 & 92.80 & 82.40  & 87.20 & 90.50 & 75.25  & 79.25 & 81.50 \\
$(128,\,12)$ & 300 & 90.80  & 95.10 & 96.80 & 83.10  & 87.05 & 89.95 & 82.40  & 89.00 & 92.15 & 73.30  & 79.00 & 82.55 \\
         & 600 & 94.05  & 96.45 & 97.60 & 89.65  & 92.35 & 92.95 & 82.65  & 88.30 & 91.25 & 74.35  & 79.50 & 81.90 \\
         & 900 & 94.60  & 96.85 & 98.40 & 91.25  & 93.00 & 94.70 & 85.15  & 90.10 & 92.80 & 78.05  & 81.35 & 83.55 \\
$(256,\,12)$ & 300 & 90.85  & 95.40 & 97.65 & 83.90  & 88.45 & 90.75 & 83.40  & 90.05 & 93.95 & 74.40  & 80.65 & 84.00 \\
         & 600 & 93.90  & 97.40 & 98.25 & 88.85  & 92.80 & 93.95 & 84.20  & 89.80 & 92.90 & 75.35  & 79.45 & 83.50 \\
         & 900 & 93.85  & 97.35 & 98.75 & 89.30  & 92.60 & 94.70 & 86.35  & 90.70 & 92.85 & 79.15  & 82.45 & 84.80 \\ \hline
$(12,\,32)$  & 300 & 77.95  & 82.50 & 86.15 & 84.85  & 90.60 & 93.25 & 68.50  & 74.75 & 78.00 & 76.50  & 83.50 & 87.50 \\
         & 600 & 83.25  & 87.55 & 89.70 & 88.55  & 93.35 & 95.90 & 71.05  & 76.35 & 79.65 & 77.50  & 84.95 & 88.25 \\
         & 900 & 87.15  & 89.80 & 91.00 & 90.60  & 94.15 & 95.90 & 71.05  & 76.65 & 80.10 & 78.20  & 84.15 & 87.25 \\
$(12,\,64)$  & 300 & 80.25  & 85.30 & 88.85 & 87.70  & 92.75 & 95.70 & 71.40  & 76.45 & 80.85 & 80.15  & 85.85 & 90.15 \\
         & 600 & 87.35  & 90.30 & 92.90 & 92.35  & 96.10 & 97.75 & 74.10  & 78.15 & 81.40 & 81.50  & 87.00 & 90.20 \\
         & 900 & 88.75  & 91.40 & 92.40 & 93.65  & 96.15 & 97.40 & 76.05  & 79.90 & 82.45 & 82.60  & 87.95 & 91.05 \\
$(12,\,128)$ & 300 & 82.80  & 88.20 & 90.90 & 90.55  & 95.10 & 97.70 & 73.40  & 78.35 & 82.15 & 82.90  & 88.90 & 91.45 \\
         & 600 & 88.45  & 91.20 & 93.45 & 93.70  & 96.75 & 98.45 & 75.45  & 79.25 & 81.95 & 83.75  & 88.50 & 90.95 \\
         & 900 & 89.25  & 92.15 & 94.10 & 94.40  & 97.40 & 98.15 & 75.30  & 79.70 & 82.50 & 83.55  & 89.00 & 91.70 \\
$(12,\,256)$ & 300 & 84.50  & 89.30 & 91.15 & 91.40  & 96.30 & 97.65 & 73.80  & 79.80 & 82.90 & 83.70  & 90.75 & 93.80 \\
         & 600 & 89.05  & 91.60 & 93.65 & 93.95  & 96.85 & 98.50 & 76.20  & 81.20 & 83.70 & 85.95  & 91.40 & 94.00 \\
         & 900 & 91.10  & 93.60 & 94.90 & 95.10  & 97.80 & 99.10 & 78.40  & 82.60 & 85.30 & 87.10  & 91.40 & 93.40 \\
\hline\hline
\end{tabular}
%\begin{tablenotes}
%  \item[*] ``New'' represents the proposed new method, and ``Original'' represents the proposed original method.
%\end{tablenotes}
\end{threeparttable}
\end{table}

\begin{table}[!tbp]
\renewcommand\arraystretch{1.2}
\centering
\begin{threeparttable}
\footnotesize
\caption{\textit{ Relative frequency estimates of $\mathbb{P}(\hat{d} = d)$ based on $2000$ repetitions with $\xi_t$ determined by the random weight method, where the direct estimation procedure and the refined estimation procedure are given in \eqref{eq:ratio1} and \eqref{eq:ratio}, respectively.}}\label{SM:originalDhat_random}
\begin{tabular}{ccccccccccccc}
\hline\hline
& & & \multicolumn{3}{c}{Refined} & Direct & & & \multicolumn{3}{c}{Refined} & Direct \\
& $(p,\,q)$  & $n$  & $K=3$    & $K=5$    & $K=7$  &   & $(p,\,q)$ & $n$   & $K=3$    & $K=5$    & $K=7$  & \\  \hline
\multirow{12}{*}{$d=1$} & $(4,\,4)$   & 300 & 99.80      & 100.00    & 100.00    & 99.15        & $(32,\,4)$   & 300 & 99.95      & 100.00    & 100.00    & 99.65        \\
                      &         & 600 & 99.95      & 100.00    & 99.95     & 99.65        &          & 600 & 100.00     & 100.00    & 100.00    & 99.85        \\
                      &         & 900 & 100.00     & 100.00    & 100.00    & 99.70        &          & 900 & 100.00     & 100.00    & 100.00    & 100.00       \\
                      & $(8,\,8)$   & 300 & 100.00     & 100.00    & 100.00    & 99.95        & $(64,\,4)$   & 300 & 100.00     & 100.00    & 100.00    & 99.85        \\
                      &         & 600 & 100.00     & 100.00    & 100.00    & 99.80        &          & 600 & 100.00     & 100.00    & 100.00    & 100.00       \\
                      &         & 900 & 100.00     & 100.00    & 100.00    & 99.80        &          & 900 & 100.00     & 100.00    & 100.00    & 99.85        \\
                      & $(16,\,16)$ & 300 & 99.95      & 100.00    & 100.00    & 99.90        & $(128,\,4)$  & 300 & 100.00     & 100.00    & 100.00    & 100.00       \\
                      &         & 600 & 100.00     & 100.00    & 100.00    & 99.90        &          & 600 & 100.00     & 99.95     & 100.00    & 99.90        \\
                      &         & 900 & 100.00     & 100.00    & 100.00    & 99.90        &          & 900 & 100.00     & 100.00    & 100.00    & 99.95        \\
                      & $(32,\,32)$ & 300 & 100.00     & 100.00    & 100.00    & 99.95        & $(256,\,4)$  & 300 & 100.00     & 100.00    & 100.00    & 99.95        \\
                      &         & 600 & 100.00     & 100.00    & 100.00    & 100.00       &          & 600 & 100.00     & 100.00    & 100.00    & 99.95        \\
                      &         & 900 & 100.00     & 100.00    & 100.00    & 99.90        &          & 900 & 100.00     & 100.00    & 100.00    & 99.90        \\ \hline
\multirow{12}{*}{$d=3$} & $(8,\,8)$   & 300 & 68.90      & 70.75     & 72.00     & 66.15        & $(32,\,8)$   & 300 & 74.60      & 77.60     & 80.00     & 70.00        \\
                      &         & 600 & 70.75      & 71.30     & 71.95     & 70.70        &          & 600 & 77.95      & 78.50     & 80.95     & 74.85        \\
                      &         & 900 & 73.20      & 72.40     & 72.80     & 73.65        &          & 900 & 80.75      & 81.90     & 82.85     & 77.55        \\
                      & $(16,\,16)$ & 300 & 75.65      & 79.30     & 81.15     & 72.70        & $(64,\,8)$   & 300 & 77.25      & 81.35     & 84.05     & 72.65        \\
                      &         & 600 & 79.05      & 80.90     & 83.20     & 77.85        &          & 600 & 79.80      & 82.45     & 83.85     & 76.30        \\
                      &         & 900 & 79.75      & 81.10     & 81.95     & 78.90        &          & 900 & 81.80      & 83.55     & 85.40     & 79.30        \\
                      & $(32,\,32)$ & 300 & 84.00      & 86.15     & 87.80     & 79.35        & $(128,\,8)$  & 300 & 79.75      & 82.20     & 84.50     & 72.40        \\
                      &         & 600 & 84.60      & 86.15     & 87.30     & 82.35        &          & 600 & 80.40      & 83.25     & 85.75     & 76.80        \\
                      &         & 900 & 84.70      & 86.75     & 87.50     & 82.60        &          & 900 & 81.80      & 83.85     & 85.50     & 78.75        \\
                      & $(64,\,64)$ & 300 & 89.10      & 91.85     & 93.50     & 83.90        & $(256,\,8)$  & 300 & 80.65      & 84.40     & 86.65     & 75.40        \\
                      &         & 600 & 87.55      & 90.15     & 92.35     & 84.20        &          & 600 & 82.15      & 84.45     & 86.40     & 79.40        \\
                      &         & 900 & 90.35      & 92.00     & 92.95     & 87.60        &          & 900 & 84.20      & 85.70     & 86.70     & 81.05        \\ \hline
\multirow{12}{*}{$d=6$} & $(12,\,12)$ & 300 & 65.80      & 71.15     & 75.40     & 51.20        & $(32,\,12)$  & 300 & 77.80      & 83.85     & 87.30     & 57.20        \\
                      &         & 600 & 66.95      & 71.45     & 75.40     & 56.70        &          & 600 & 77.80      & 84.10     & 87.10     & 61.35        \\
                      &         & 900 & 70.25      & 73.80     & 76.20     & 61.20        &          & 900 & 77.95      & 84.05     & 87.60     & 63.35        \\
                      & $(16,\,16)$ & 300 & 71.50      & 78.50     & 82.35     & 56.00        & $(64,\,12)$  & 300 & 81.55      & 87.65     & 91.00     & 59.30        \\
                      &         & 600 & 75.00      & 80.10     & 83.00     & 61.85        &          & 600 & 81.65      & 87.40     & 90.25     & 65.50        \\
                      &         & 900 & 76.15      & 80.45     & 82.95     & 65.25        &          & 900 & 82.40      & 87.20     & 90.50     & 68.00        \\
                      & $(32,\,32)$ & 300 & 84.50      & 90.30     & 92.85     & 65.30        & $(128,\,12)$ & 300 & 82.40      & 89.00     & 92.15     & 59.85        \\
                      &         & 600 & 85.40      & 90.70     & 93.20     & 72.00        &          & 600 & 82.65      & 88.30     & 91.25     & 65.85        \\
                      &         & 900 & 85.40      & 90.15     & 92.45     & 74.95        &          & 900 & 85.15      & 90.10     & 92.80     & 69.65        \\
                      & $(64,\,64)$ & 300 & 90.10      & 94.70     & 96.45     & 72.05        & $(256,\,12)$ & 300 & 83.40      & 90.05     & 93.95     & 59.70        \\
                      &         & 600 & 91.10      & 94.45     & 96.20     & 76.50        &          & 600 & 84.20      & 89.80     & 92.90     & 63.80        \\
                      &         & 900 & 92.05      & 95.70     & 96.85     & 80.55        &          & 900 & 86.35      & 90.70     & 92.85     & 71.80        \\
\hline\hline
\end{tabular}
%\begin{tablenotes}
%  \item[*] ``New'' represents the proposed new method, and ``Original'' represents the proposed original method.
%\end{tablenotes}
\end{threeparttable}
\end{table}

\begin{table}[!htbp]
\renewcommand\arraystretch{1.2}
\centering
\begin{threeparttable}
\tiny
\caption{\textit{ The averages and standard deviations (in parentheses) of $\rho^2(\bA, \wh \bA)$ and $\rho^2(\bB, \wh \bB)$ when $d=1$ based on $2000$ repetitions. All numbers in the table are the true numbers multiplied by $10^4$ for ease of presentation. }}\label{SM:tb1a}
\begin{tabular}{ccc|cc|cc|cc|cc}
\hline\hline
         &   &  &  \multicolumn{6}{c|}{The refined method} &    \multicolumn{2}{c}{The direct method} \\ \cline{4-9}
        &   &  & \multicolumn{2}{c|}{$K=3$}           & \multicolumn{2}{c|}{$K=5$}           & \multicolumn{2}{c|}{$K=7$}   &    \multicolumn{2}{c}{}            \\
$(p,q)$   & $n$ & $\xi_t$  & $\rho^2(\bA, \wh \bA)$& $\rho^2(\bB, \wh \bB)$ & $\rho^2(\bA, \wh \bA)$ &$\rho^2(\bB, \wh \bB)$ & $\rho^2(\bA, \wh \bA)$ & $\rho^2(\bB, \wh \bB)$  & $\rho^2(\bA, \wh \bA)$ & $\rho^2(\bB, \wh \bB)$  \\ \hline
$(4,\,4)$   & 300 & PCA    & 1.40(4.08)    & 1.57(5.01)    & 1.47(3.33)   & 1.70(5.36)    & 1.54(3.49)   & 1.78(5.75)  & 116.19(780.51)  & 2.89(34.79)     \\
        &  & Random & 12.80(142.16) & 19.53(277.13) & 7.49(68.53)  & 11.74(216.26) & 6.21(47.14)  & 5.45(36.49) & 286.68(1246.72) & 73.87(606.40)   \\
        & 600 & PCA    & 0.72(2.27)    & 0.66(1.85)    & 0.78(2.20)   & 0.71(1.83)    & 0.82(2.25)   & 0.73(1.73)  & 72.28(597.69)   & 1.41(24.11)     \\
        &  & Random & 7.05(57.12)   & 9.99(106.15)  & 4.23(27.67)  & 5.75(62.62)   & 3.38(20.48)  & 4.54(56.01) & 234.08(1140.36) & 49.29(535.16)   \\
        & 900 & PCA    & 0.49(1.48)    & 0.48(1.69)    & 0.51(1.49)   & 0.52(1.78)    & 0.53(1.52)   & 0.53(1.72)  & 45.94(419.11)   & 0.51(2.67)      \\
        &  & Random & 6.33(67.30)   & 9.02(109.43)  & 4.03(40.75)  & 4.26(50.37)   & 3.09(34.69)  & 3.72(50.18) & 200.23(1051.61) & 56.07(595.95)   \\ \hline
$(8,\,8)$   & 300 & PCA    & 0.51(0.72)    & 0.53(0.70)    & 0.56(0.79)   & 0.57(0.74)    & 0.58(0.79)   & 0.59(0.74)  & 123.71(839.74)  & 2.09(42.55)     \\
        &  & Random & 2.30(12.11)   & 2.47(19.20)   & 1.82(8.93)   & 1.87(11.25)   & 1.49(5.37)   & 1.49(5.82)  & 297.85(1275.58) & 46.21(516.24)   \\
        & 600 & PCA    & 0.25(0.30)    & 0.25(0.31)    & 0.28(0.33)   & 0.28(0.33)    & 0.29(0.33)   & 0.29(0.34)  & 63.73(606.12)   & 0.39(4.00)      \\
        &  & Random & 3.81(45.97)   & 3.68(41.39)   & 1.81(18.16)  & 1.50(9.78)    & 1.16(7.36)   & 1.06(5.39)  & 273.22(1169.78) & 35.67(411.25)   \\
        & 900 & PCA    & 0.16(0.21)    & 0.17(0.20)    & 0.18(0.23)   & 0.18(0.21)    & 0.19(0.23)   & 0.19(0.22)  & 54.95(509.13)   & 0.15(0.28)      \\
        &  & Random & 2.03(24.48)   & 2.12(26.34)   & 1.06(9.09)   & 1.10(7.79)    & 0.74(4.53)   & 0.80(5.63)  & 193.28(1057.94) & 47.96(575.80)   \\ \hline
$(16,\,16)$ & 300 & PCA    & 0.25(0.23)    & 0.24(0.24)    & 0.27(0.25)   & 0.27(0.26)    & 0.28(0.25)   & 0.28(0.26)  & 92.28(615.83)   & 0.23(0.38)      \\
        &  & Random & 3.11(91.16)   & 3.28(97.90)   & 0.78(3.37)   & 0.79(3.70)    & 0.71(3.05)   & 0.70(2.73)  & 345.86(1401.13) & 42.25(518.09)   \\
        & 600 & PCA    & 0.12(0.10)    & 0.12(0.11)    & 0.13(0.11)   & 0.13(0.12)    & 0.13(0.11)   & 0.14(0.12)  & 72.02(590.61)   & 0.11(0.27)      \\
        &  & Random & 0.78(8.96)    & 0.96(11.92)   & 0.43(1.94)   & 0.48(2.46)    & 0.35(1.20)   & 0.37(1.41)  & 213.02(1064.06) & 32.68(469.04)   \\
        & 900 & PCA    & 0.08(0.07)    & 0.08(0.07)    & 0.09(0.08)   & 0.09(0.08)    & 0.09(0.08)   & 0.09(0.08)  & 54.81(498.40)   & 0.08(0.30)      \\
        &  & Random & 0.68(6.08)    & 0.69(7.25)    & 0.41(2.53)   & 0.38(2.17)    & 0.32(1.76)   & 0.31(1.62)  & 198.95(1068.86) & 42.09(546.53)   \\ \hline
$(32,\,32)$ & 300 & PCA    & 0.12(0.11)    & 0.12(0.10)    & 0.13(0.11)   & 0.13(0.11)    & 0.14(0.11)   & 0.14(0.11)  & 90.75(656.49)   & 0.11(0.22)      \\
        &  & Random & 1.03(12.30)   & 1.03(13.02)   & 0.53(4.26)   & 0.56(5.12)    & 0.38(1.47)   & 0.38(1.67)  & 331.50(1402.45) & 46.96(579.20)   \\
        & 600 & PCA    & 0.06(0.05)    & 0.06(0.05)    & 0.06(0.05)   & 0.06(0.05)    & 0.07(0.05)   & 0.07(0.05)  & 66.48(493.96)   & 0.05(0.21)      \\
        &  & Random & 0.33(2.09)    & 0.36(2.52)    & 0.25(1.54)   & 0.26(1.85)    & 0.19(0.78)   & 0.20(0.91)  & 209.08(1078.56) & 24.29(415.27)   \\
        & 900 & PCA    & 0.04(0.03)    & 0.04(0.03)    & 0.04(0.03)   & 0.04(0.03)    & 0.04(0.03)   & 0.04(0.03)  & 65.96(576.73)   & 0.07(1.27)      \\
        &  & Random & 0.41(6.37)    & 0.34(3.72)    & 0.19(1.06)   & 0.20(1.16)    & 0.17(0.88)   & 0.17(0.88)  & 255.62(1232.73) & 39.15(557.05)   \\ \hline
$(32,\,4)$  & 300 & PCA    & 1.52(5.30)    & 0.15(0.62)    & 1.63(5.17)   & 0.16(0.59)    & 1.66(4.94)   & 0.16(0.60)  & 6.69(101.09)    & 0.15(0.91)      \\
        &  & Random & 14.49(182.43) & 1.80(30.72)   & 6.86(38.09)  & 0.70(5.96)    & 5.09(24.41)  & 0.50(3.60)  & 258.88(1250.20) & 86.82(819.03)   \\
        & 600 & PCA    & 0.70(1.72)    & 0.07(0.24)    & 0.76(1.78)   & 0.08(0.23)    & 0.79(1.76)   & 0.08(0.22)  & 6.19(168.04)    & 0.09(1.05)      \\
        &  & Random & 5.85(47.75)   & 0.69(7.29)    & 3.78(23.60)  & 0.46(4.78)    & 2.90(14.97)  & 0.29(1.64)  & 189.38(1094.05) & 45.36(535.51)   \\
        & 900 & PCA    & 0.46(1.22)    & 0.05(0.16)    & 0.50(1.34)   & 0.05(0.20)    & 0.53(1.40)   & 0.05(0.18)  & 5.73(187.57)    & 0.06(0.78)      \\
        &  & Random & 4.42(46.62)   & 0.88(15.17)   & 3.11(31.75)  & 0.38(4.92)    & 2.47(21.60)  & 0.25(2.66)  & 151.54(937.92)  & 30.34(417.63)   \\ \hline
$(64,\,4)$  & 300 & PCA    & 1.56(10.35)   & 0.07(0.40)    & 1.67(10.45)  & 0.08(0.50)    & 1.70(9.91)   & 0.08(0.35)  & 2.37(37.43)     & 0.09(1.48)      \\
        &  & Random & 10.61(115.46) & 0.95(25.12)   & 5.86(49.28)  & 0.29(3.29)    & 4.82(38.18)  & 0.28(3.48)  & 172.24(1024.22) & 49.87(579.15)   \\
        & 600 & PCA    & 0.73(4.33)    & 0.04(0.40)    & 0.80(4.56)   & 0.04(0.38)    & 0.82(4.48)   & 0.04(0.36)  & 1.38(33.07)     & 0.04(0.61)      \\
        &  & Random & 7.65(137.10)  & 4.02(167.92)  & 3.42(29.35)  & 0.18(1.95)    & 2.52(16.42)  & 0.13(0.97)  & 171.30(1054.63) & 42.43(569.89)   \\
        & 900 & PCA    & 0.48(2.60)    & 0.03(0.33)    & 0.51(2.39)   & 0.03(0.30)    & 0.53(2.35)   & 0.03(0.32)  & 1.83(44.50)     & 0.03(0.58)      \\
        &  & Random & 7.77(112.96)  & 0.94(21.85)   & 4.93(78.31)  & 0.54(14.86)   & 3.77(64.03)  & 0.40(12.27) & 138.40(918.24)  & 37.32(498.92)   \\ \hline
$(128,\,4)$ & 300 & PCA    & 1.51(4.92)    & 0.03(0.09)    & 1.62(4.75)   & 0.04(0.11)    & 1.67(4.68)   & 0.04(0.10)  & 1.55(7.75)      & 0.03(0.09)      \\
        &  & Random & 6.37(35.12)   & 0.15(1.05)    & 4.34(18.52)  & 0.10(0.51)    & 3.82(16.08)  & 0.09(0.43)  & 200.75(1108.93) & 52.39(617.07)   \\
        & 600 & PCA    & 0.70(1.65)    & 0.02(0.06)    & 0.76(1.71)   & 0.02(0.06)    & 0.79(1.73)   & 0.02(0.05)  & 0.80(6.26)      & 0.02(0.11)      \\
        &  & Random & 9.61(186.61)  & 1.86(76.41)   & 5.77(123.99) & 0.99(40.99)   & 5.48(140.16) & 1.42(60.95) & 171.28(1029.97) & 34.52(496.07)   \\
        & 900 & PCA    & 0.46(1.20)    & 0.01(0.02)    & 0.50(1.23)   & 0.01(0.02)    & 0.52(1.25)   & 0.01(0.02)  & 0.47(2.61)      & 0.01(0.02)      \\
        &  & Random & 5.19(52.51)   & 0.15(2.03)    & 2.94(27.35)  & 0.11(1.90)    & 2.41(19.41)  & 0.10(1.87)  & 120.52(844.39)  & 26.71(457.30)   \\ \hline
$(256,\,4)$ & 300 & PCA    & 1.42(2.39)    & 0.02(0.03)    & 1.54(2.47)   & 0.02(0.04)    & 1.59(2.52)   & 0.02(0.04)  & 1.36(4.25)      & 0.01(0.03)      \\
        &  & Random & 11.39(101.38) & 0.23(2.95)    & 5.92(32.24)  & 0.07(0.50)    & 4.56(20.26)  & 0.07(0.53)  & 207.73(1152.35) & 65.49(715.08)   \\
        & 600 & PCA    & 0.68(1.13)    & 0.01(0.02)    & 0.75(1.20)   & 0.01(0.02)    & 0.78(1.23)   & 0.01(0.02)  & 0.63(1.44)      & 0.01(0.02)      \\
        &  & Random & 4.54(30.47)   & 0.08(0.91)    & 3.15(18.73)  & 0.05(0.48)    & 2.46(10.29)  & 0.03(0.16)  & 116.60(766.12)  & 14.72(294.93)   \\
        & 900 & PCA    & 0.45(0.75)    & 0.00(0.01)    & 0.49(0.79)   & 0.01(0.01)    & 0.51(0.82)   & 0.01(0.01)  & 0.42(0.99)      & 0.00(0.01)      \\
        &  & Random & 5.25(83.46)   & 0.22(8.44)    & 2.37(14.58)  & 0.02(0.13)    & 1.94(10.41)  & 0.02(0.13)  & 139.62(936.96)  & 34.76(498.31)   \\ \hline
$(4,\,32)$  & 300 & PCA    & 0.15(0.77)    & 1.44(3.92)    & 0.16(0.73)   & 1.56(4.06)    & 0.16(0.76)   & 1.61(4.14)  & 0.22(4.03)      & 3.61(61.80)     \\
        &  & Random & 1.85(31.13)   & 11.29(138.14) & 0.58(5.19)   & 5.82(71.93)   & 0.45(2.86)   & 3.90(17.99) & 55.98(636.48)   & 212.75(1137.09) \\
        & 600 & PCA    & 0.07(0.17)    & 0.72(2.43)    & 0.07(0.16)   & 0.77(2.24)    & 0.07(0.16)   & 0.80(2.16)  & 0.07(0.45)      & 5.73(174.82)    \\
        &  & Random & 0.55(6.97)    & 4.04(30.94)   & 0.30(2.41)   & 2.89(20.11)   & 0.32(3.24)   & 2.46(16.42) & 30.20(451.62)   & 137.01(854.11)  \\
        & 900 & PCA    & 0.05(0.18)    & 0.46(1.30)    & 0.05(0.15)   & 0.51(1.43)    & 0.05(0.15)   & 0.52(1.39)  & 0.05(0.36)      & 2.23(56.31)     \\
        &  & Random & 0.44(6.36)    & 3.26(34.68)   & 0.28(3.13)   & 2.00(11.43)   & 0.16(0.98)   & 1.58(7.11)  & 21.08(394.75)   & 106.18(763.73)  \\ \hline
$(4,\,64)$  & 300 & PCA    & 0.08(0.57)    & 1.53(5.53)    & 0.08(0.57)   & 1.64(5.53)    & 0.09(0.58)   & 1.68(5.53)  & 0.08(0.57)      & 3.01(53.25)     \\
        &  & Random & 0.42(4.71)    & 7.09(49.33)   & 0.27(1.90)   & 5.28(25.39)   & 0.22(1.58)   & 4.05(16.39) & 66.66(688.85)   & 233.23(1203.09) \\
        & 600 & PCA    & 0.03(0.08)    & 0.75(3.57)    & 0.03(0.08)   & 0.81(3.43)    & 0.04(0.08)   & 0.84(3.50)  & 0.03(0.11)      & 0.91(9.53)      \\
        &  & Random & 0.45(12.25)   & 5.69(98.27)   & 0.16(1.57)   & 3.43(34.14)   & 0.11(0.51)   & 2.35(13.60) & 33.97(482.60)   & 140.10(911.49)  \\
        & 900 & PCA    & 0.02(0.21)    & 0.49(2.25)    & 0.03(0.22)   & 0.53(2.17)    & 0.03(0.24)   & 0.55(2.07)  & 0.03(0.38)      & 0.66(9.07)      \\
        &  & Random & 0.23(4.51)    & 4.03(41.95)   & 0.11(1.05)   & 2.11(12.89)   & 0.09(0.83)   & 1.61(7.15)  & 25.81(430.49)   & 133.21(874.30)  \\ \hline
$(4,\,128)$ & 300 & PCA    & 0.03(0.11)    & 1.38(2.83)    & 0.04(0.12)   & 1.49(2.88)    & 0.04(0.12)   & 1.54(2.94)  & 0.03(0.14)      & 1.35(4.02)      \\
        &  & Random & 1.00(35.99)   & 10.75(157.58) & 0.18(2.71)   & 5.95(51.45)   & 0.17(2.70)   & 4.80(35.85) & 50.44(606.51)   & 168.61(998.46)  \\
        & 600 & PCA    & 0.02(0.04)    & 0.65(1.08)    & 0.02(0.04)   & 0.71(1.14)    & 0.02(0.04)   & 0.74(1.18)  & 0.01(0.04)      & 0.61(1.30)      \\
        &  & Random & 0.17(1.81)    & 5.85(41.36)   & 0.09(0.75)   & 3.87(24.62)   & 0.07(0.50)   & 2.81(15.09) & 75.76(754.33)   & 189.43(1153.42) \\
        & 900 & PCA    & 0.01(0.03)    & 0.42(0.70)    & 0.01(0.02)   & 0.46(0.74)    & 0.01(0.02)   & 0.48(0.74)  & 0.01(0.03)      & 0.40(0.90)      \\
        &  & Random & 0.51(20.22)   & 6.51(158.36)  & 0.05(0.31)   & 2.44(15.15)   & 0.04(0.33)   & 2.00(10.96) & 21.10(370.42)   & 129.50(861.97)  \\ \hline
$(4,\,256)$ & 300 & PCA    & 0.02(0.04)    & 1.44(4.23)    & 0.02(0.04)   & 1.55(4.22)    & 0.02(0.04)   & 1.60(4.28)  & 0.02(0.28)      & 1.48(10.48)     \\
        &  & Random & 0.13(1.77)    & 8.17(59.57)   & 0.09(0.96)   & 5.84(41.18)   & 0.06(0.50)   & 4.71(31.75) & 63.87(698.11)   & 177.45(1079.03) \\
        & 600 & PCA    & 0.01(0.03)    & 0.70(2.33)    & 0.01(0.03)   & 0.75(2.27)    & 0.01(0.03)   & 0.78(2.28)  & 0.01(0.07)      & 0.74(5.44)      \\
        &  & Random & 3.67(155.60)  & 11.61(212.07) & 0.52(21.10)  & 5.71(101.41)  & 0.04(0.43)   & 3.44(35.35) & 32.48(448.41)   & 114.31(834.45)  \\
        & 900 & PCA    & 0.01(0.05)    & 0.46(1.71)    & 0.01(0.04)   & 0.50(1.66)    & 0.01(0.04)   & 0.52(1.67)  & 0.01(0.03)      & 0.49(3.57)      \\
        &  & Random & 0.20(4.55)    & 7.22(114.91)  & 0.08(1.69)   & 3.21(44.51)   & 0.03(0.54)   & 2.52(36.41) & 35.12(517.45)   & 106.52(844.96)  \\
        \hline\hline
\end{tabular}
%\begin{tablenotes}
%\item[1] The results for Test (1.1) are all updated.
%\item[2] For Test (1.2) and (1.3), the bold font means that the results are updated.
%\end{tablenotes}
\end{threeparttable}
\end{table}

\begin{table}[!htbp]
\renewcommand\arraystretch{1.2}
\centering
\begin{threeparttable}
\tiny
\caption{\textit{ The averages and standard deviations (in parentheses) of $\rho^2(\bA, \wh \bA)$ and $\rho^2(\bB, \wh \bB)$ when $d=3$ based on $2000$ repetitions. All numbers in the table are the true numbers multiplied by $10^2$ for ease of presentation. }}\label{SM:tb2}
\begin{tabular}{ccc|cc|cc|cc|cc}
\hline\hline
         &   &  &  \multicolumn{6}{c|}{The refined method} &    \multicolumn{2}{c}{The direct method} \\ \cline{4-9}
        &   &  & \multicolumn{2}{c|}{$K=3$}           & \multicolumn{2}{c|}{$K=5$}           & \multicolumn{2}{c|}{$K=7$}   &    \multicolumn{2}{c}{}            \\
$(p,q)$   & $n$ & $\xi_t$  & $\rho^2(\bA, \wh \bA)$& $\rho^2(\bB, \wh \bB)$ & $\rho^2(\bA, \wh \bA)$ &$\rho^2(\bB, \wh \bB)$ & $\rho^2(\bA, \wh \bA)$ & $\rho^2(\bB, \wh \bB)$  & $\rho^2(\bA, \wh \bA)$ & $\rho^2(\bB, \wh \bB)$  \\ \hline
$(8,\,8)$   & 300 & PCA    & 21.37(34.14) & 21.81(34.71) & 20.67(33.73) & 21.17(34.39) & 20.64(33.70) & 20.97(34.39) & 33.07(34.64) & 32.32(35.06) \\
        &  & Random & 30.47(39.56) & 31.04(40.17) & 28.98(38.97) & 29.58(39.87) & 27.73(38.70) & 28.53(39.49) & 42.45(37.89) & 41.78(38.44) \\
        & 600 & PCA    & 17.22(31.14) & 18.04(32.50) & 17.60(31.66) & 18.36(33.05) & 17.84(31.95) & 18.91(33.63) & 28.11(32.76) & 27.53(33.23) \\
        &  & Random & 28.75(39.20) & 28.85(39.39) & 27.92(39.01) & 28.16(39.31) & 27.39(38.78) & 28.01(39.35) & 38.54(38.05) & 37.90(38.00) \\
        & 900 & PCA    & 14.72(29.33) & 14.72(29.55) & 15.26(29.98) & 15.27(30.23) & 15.78(30.59) & 15.99(31.25) & 25.16(32.15) & 23.54(31.22) \\
        &  & Random & 26.08(38.12) & 26.35(38.24) & 26.96(38.65) & 27.40(39.03) & 26.83(38.61) & 26.97(39.03) & 35.87(37.71) & 34.73(37.68) \\ \hline
$(16,\,16)$ & 300 & PCA    & 11.22(28.07) & 11.47(28.32) & 9.89(26.62)  & 10.06(26.75) & 8.73(25.23)  & 9.01(25.44)  & 27.26(34.16) & 25.14(34.40) \\
        &  & Random & 24.43(39.87) & 24.60(40.23) & 21.23(38.06) & 21.45(38.52) & 19.82(37.04) & 19.85(37.34) & 41.17(39.71) & 39.37(40.76) \\
        & 600 & PCA    & 7.00(22.52)  & 6.98(22.40)  & 6.78(22.17)  & 6.84(22.10)  & 6.19(21.10)  & 6.27(21.17)  & 21.74(31.69) & 20.01(31.47) \\
        &  & Random & 21.57(38.14) & 21.60(38.44) & 19.93(37.03) & 19.85(37.28) & 17.70(35.46) & 17.73(35.75) & 34.72(39.19) & 32.71(39.37) \\
        & 900 & PCA    & 5.62(20.36)  & 5.60(20.03)  & 6.02(21.01)  & 6.14(21.13)  & 5.71(20.48)  & 5.85(20.76)  & 16.42(27.81) & 14.95(26.92) \\
        &  & Random & 20.28(37.66) & 20.40(37.80) & 19.22(36.95) & 19.28(36.93) & 18.51(36.54) & 18.53(36.57) & 32.23(38.78) & 31.02(39.30) \\ \hline
$(32,\,32)$ & 300 & PCA    & 6.31(22.88)  & 6.26(22.73)  & 4.50(19.14)  & 4.44(18.99)  & 4.30(18.64)  & 4.22(18.49)  & 24.32(34.08) & 22.48(34.19) \\
        &  & Random & 16.59(35.58) & 16.55(35.65) & 14.45(33.58) & 14.48(33.70) & 12.88(32.12) & 12.78(31.99) & 35.92(39.68) & 33.09(40.61) \\
        & 600 & PCA    & 4.11(18.47)  & 4.23(18.76)  & 3.44(16.83)  & 3.51(17.00)  & 2.87(15.32)  & 2.92(15.36)  & 16.86(29.32) & 15.16(29.26) \\
        &  & Random & 15.88(35.08) & 15.93(35.07) & 14.23(33.52) & 14.34(33.65) & 12.92(32.25) & 13.11(32.36) & 29.96(38.35) & 28.22(38.89) \\
        & 900 & PCA    & 3.22(16.22)  & 3.19(16.15)  & 2.68(14.57)  & 2.74(14.77)  & 2.13(12.70)  & 2.19(12.89)  & 14.38(27.12) & 12.95(26.42) \\
        &  & Random & 15.58(34.89) & 15.40(34.70) & 13.60(32.90) & 13.60(32.94) & 12.84(32.14) & 12.87(32.17) & 28.60(38.00) & 26.32(38.22) \\ \hline
$(64,\,64)$ & 300 & PCA    & 4.49(19.87)  & 4.46(19.87)  & 3.39(17.21)  & 3.38(17.24)  & 2.61(14.94)  & 2.61(14.97)  & 21.07(32.62) & 18.85(33.11) \\
        &  & Random & 11.34(30.88) & 11.28(30.87) & 8.76(27.30)  & 8.75(27.32)  & 7.01(24.58)  & 7.03(24.66)  & 31.51(38.40) & 28.35(39.53) \\
        & 600 & PCA    & 3.32(17.11)  & 3.29(17.11)  & 2.01(13.22)  & 2.01(13.30)  & 1.42(10.85)  & 1.42(10.94)  & 15.82(29.00) & 14.29(29.27) \\
        &  & Random & 12.73(32.56) & 12.66(32.51) & 10.35(29.61) & 10.27(29.60) & 8.01(26.45)  & 8.07(26.49)  & 28.14(38.21) & 25.59(38.61) \\
        & 900 & PCA    & 2.12(13.70)  & 2.12(13.61)  & 1.34(10.76)  & 1.31(10.55)  & 1.18(9.93)   & 1.16(9.82)   & 12.46(25.53) & 11.11(25.27) \\
        &  & Random & 9.87(29.18)  & 9.94(29.25)  & 8.22(26.81)  & 8.37(27.07)  & 7.28(25.42)  & 7.43(25.66)  & 24.91(36.42) & 22.80(37.00) \\ \hline
$(32,\,8)$  & 300 & PCA    & 13.11(31.02) & 11.27(27.45) & 11.84(29.47) & 10.38(26.44) & 10.46(27.79) & 9.31(25.34)  & 21.22(35.08) & 17.45(30.81) \\
        &  & Random & 26.88(41.89) & 23.76(38.18) & 23.99(40.16) & 21.66(37.19) & 21.50(38.77) & 19.38(35.92) & 34.94(43.13) & 29.66(39.05) \\
        & 600 & PCA    & 8.51(25.36)  & 7.29(22.10)  & 7.74(24.24)  & 6.65(21.19)  & 7.09(23.13)  & 6.18(20.41)  & 14.69(29.60) & 12.04(25.78) \\
        &  & Random & 23.18(39.95) & 20.28(35.85) & 23.07(39.78) & 20.18(35.73) & 20.78(38.24) & 18.20(34.45) & 29.50(41.44) & 25.15(37.14) \\
        & 900 & PCA    & 6.85(22.77)  & 5.60(19.28)  & 6.13(21.49)  & 5.00(17.96)  & 5.71(20.57)  & 4.53(16.86)  & 12.82(28.08) & 10.50(23.96) \\
        &  & Random & 20.62(38.32) & 18.16(34.47) & 19.49(37.43) & 17.50(34.31) & 18.51(36.81) & 16.70(33.69) & 26.92(40.38) & 23.10(36.28) \\ \hline
$(64,\,8)$  & 300 & PCA    & 11.37(30.03) & 9.45(25.66)  & 8.86(26.61)  & 7.63(23.20)  & 7.69(24.66)  & 6.55(21.40)  & 17.64(33.84) & 13.71(28.31) \\
        &  & Random & 23.90(41.03) & 21.16(37.17) & 20.50(38.63) & 18.29(35.19) & 17.90(36.38) & 15.84(33.26) & 30.71(43.12) & 26.16(38.67) \\
        & 600 & PCA    & 7.43(24.46)  & 6.22(20.84)  & 6.38(22.72)  & 5.44(19.43)  & 5.37(20.69)  & 4.56(17.64)  & 11.61(27.84) & 9.15(23.36)  \\
        &  & Random & 21.24(39.37) & 18.26(34.90) & 18.86(37.56) & 16.28(33.37) & 17.41(36.38) & 15.20(32.61) & 26.47(41.27) & 22.21(36.58) \\
        & 900 & PCA    & 5.30(20.64)  & 4.18(16.61)  & 4.86(19.73)  & 3.87(16.10)  & 4.56(18.92)  & 3.68(15.64)  & 9.28(25.13)  & 7.29(20.89)  \\
        &  & Random & 18.99(38.04) & 16.36(33.66) & 17.54(36.73) & 15.23(32.84) & 15.68(35.09) & 13.78(31.60) & 23.62(39.83) & 19.91(35.09) \\ \hline
$(128,\,8)$ & 300 & PCA    & 9.76(28.44)  & 8.05(24.26)  & 7.79(25.36)  & 6.43(21.76)  & 6.50(23.06)  & 5.45(19.97)  & 14.73(32.60) & 11.37(27.09) \\
        &  & Random & 21.58(39.99) & 18.22(35.19) & 18.70(38.00) & 16.35(34.12) & 16.47(36.07) & 14.56(32.56) & 29.89(43.76) & 25.08(38.48) \\
        & 600 & PCA    & 5.83(22.21)  & 4.63(18.05)  & 5.31(21.06)  & 4.31(17.37)  & 4.38(19.13)  & 3.66(16.22)  & 9.61(26.53)  & 7.28(21.68)  \\
        &  & Random & 20.48(39.36) & 17.32(34.43) & 17.73(37.00) & 15.36(32.92) & 15.18(34.80) & 13.38(31.31) & 25.19(41.68) & 20.95(36.16) \\
        & 900 & PCA    & 4.23(19.03)  & 3.45(15.88)  & 3.73(17.76)  & 3.09(15.07)  & 3.35(16.85)  & 2.79(14.32)  & 6.54(21.99)  & 4.95(17.82)  \\
        &  & Random & 18.84(38.24) & 16.07(33.57) & 16.89(36.52) & 14.76(32.43) & 15.19(34.97) & 13.41(31.37) & 23.17(40.51) & 19.35(35.06) \\ \hline
$(256,\,8)$ & 300 & PCA    & 9.03(27.70)  & 7.34(23.32)  & 6.97(24.38)  & 5.74(20.71)  & 5.90(22.42)  & 4.94(19.40)  & 13.83(32.23) & 10.58(26.53) \\
        &  & Random & 20.10(39.30) & 17.02(34.51) & 16.54(36.23) & 14.24(32.11) & 14.39(34.18) & 12.40(30.33) & 26.16(42.48) & 21.63(36.75) \\
        & 600 & PCA    & 6.12(22.96)  & 4.91(18.88)  & 4.39(19.38)  & 3.59(16.25)  & 4.03(18.61)  & 3.32(15.77)  & 8.05(25.16)  & 6.17(20.44)  \\
        &  & Random & 18.45(38.12) & 15.52(33.11) & 16.15(36.14) & 13.83(31.71) & 14.33(34.29) & 12.37(30.34) & 21.71(40.05) & 18.26(34.93) \\
        & 900 & PCA    & 3.31(16.83)  & 2.61(13.63)  & 2.46(14.36)  & 1.99(11.98)  & 2.43(14.25)  & 2.01(12.19)  & 5.44(20.52)  & 4.19(16.65)  \\
        &  & Random & 16.20(36.30) & 13.93(31.96) & 14.89(34.95) & 12.93(31.02) & 13.71(33.84) & 12.11(30.28) & 19.91(38.96) & 16.70(33.88) \\ \hline
$(8,\,32)$  & 300 & PCA    & 12.41(28.85) & 14.26(32.42) & 10.52(26.75) & 11.76(29.65) & 9.52(25.52)  & 10.53(28.00) & 18.76(31.95) & 22.98(36.59) \\
        &  & Random & 21.97(37.14) & 24.65(40.79) & 19.09(35.34) & 21.57(38.85) & 17.95(34.62) & 19.95(37.61) & 29.23(38.83) & 34.20(42.73) \\
        & 600 & PCA    & 7.89(22.63)  & 9.37(26.70)  & 7.25(21.88)  & 8.47(25.31)  & 7.13(22.00)  & 8.24(25.00)  & 11.97(25.83) & 15.04(30.51) \\
        &  & Random & 18.84(35.16) & 21.12(38.67) & 18.10(34.65) & 20.16(37.92) & 16.59(33.76) & 18.15(36.39) & 24.97(37.05) & 29.35(41.23) \\
        & 900 & PCA    & 5.86(19.66)  & 6.97(22.92)  & 5.46(19.06)  & 6.28(21.67)  & 4.96(18.15)  & 5.64(20.40)  & 10.04(23.92) & 12.32(27.81) \\
        &  & Random & 18.89(35.28) & 21.16(38.78) & 18.29(34.82) & 20.66(38.38) & 17.42(34.13) & 19.52(37.53) & 23.00(36.32) & 27.17(40.70) \\ \hline
$(8,\,64)$  & 300 & PCA    & 8.96(24.90)  & 10.96(29.30) & 7.76(23.23)  & 9.45(27.23)  & 6.20(20.71)  & 7.43(23.99)  & 14.11(28.76) & 17.91(34.06) \\
        &  & Random & 20.89(36.74) & 24.00(41.05) & 18.96(35.49) & 21.68(39.46) & 17.07(33.96) & 19.56(37.78) & 26.67(38.97) & 31.61(43.59) \\
        & 600 & PCA    & 5.75(19.79)  & 7.09(23.82)  & 5.25(18.99)  & 6.41(22.65)  & 4.85(18.41)  & 5.82(21.62)  & 10.12(24.71) & 12.69(29.52) \\
        &  & Random & 18.61(35.33) & 21.32(39.35) & 16.88(34.27) & 19.21(37.89) & 15.55(33.28) & 17.44(36.49) & 22.55(37.04) & 26.36(41.69) \\
        & 900 & PCA    & 4.40(17.40)  & 5.43(20.76)  & 3.85(16.22)  & 4.68(19.21)  & 3.58(15.74)  & 4.23(18.05)  & 7.02(20.39)  & 9.16(25.00)  \\
        &  & Random & 16.45(33.77) & 18.83(37.73) & 15.53(33.09) & 17.69(36.69) & 13.89(31.75) & 15.89(35.30) & 19.88(34.99) & 23.26(39.61) \\ \hline
$(8,\,128)$ & 300 & PCA    & 8.28(24.26)  & 10.09(28.77) & 6.91(22.40)  & 8.31(26.20)  & 6.28(21.55)  & 7.38(24.69)  & 12.02(27.82) & 15.31(33.31) \\
        &  & Random & 18.86(35.82) & 21.88(40.37) & 16.15(33.67) & 18.60(37.86) & 13.92(32.06) & 15.65(35.29) & 24.52(38.29) & 29.22(43.48) \\
        & 600 & PCA    & 5.13(19.02)  & 6.47(23.33)  & 4.49(18.08)  & 5.44(21.29)  & 3.80(16.47)  & 4.56(19.29)  & 7.58(22.14)  & 9.89(27.03)  \\
        &  & Random & 16.46(33.79) & 18.99(38.20) & 14.96(32.78) & 17.13(36.72) & 13.32(31.35) & 15.22(34.93) & 20.14(35.75) & 24.04(40.93) \\
        & 900 & PCA    & 3.62(15.88)  & 4.57(19.58)  & 3.00(14.19)  & 3.82(17.71)  & 2.67(13.35)  & 3.33(16.32)  & 5.13(17.94)  & 7.03(22.94)  \\
        &  & Random & 16.68(34.01) & 19.28(38.38) & 14.78(32.43) & 17.11(36.66) & 13.40(31.24) & 15.52(35.21) & 18.63(34.92) & 22.42(40.23) \\ \hline
$(8,\,256)$ & 300 & PCA    & 7.60(23.64)  & 9.48(28.26)  & 6.18(21.43)  & 7.70(25.55)  & 4.96(19.21)  & 6.10(22.65)  & 11.87(27.95) & 15.39(33.98) \\
        &  & Random & 19.25(36.08) & 22.46(40.96) & 16.22(33.89) & 18.59(38.06) & 13.67(31.91) & 15.64(35.59) & 24.23(38.40) & 29.11(43.87) \\
        & 600 & PCA    & 4.18(17.66)  & 5.03(20.85)  & 3.70(16.53)  & 4.37(19.22)  & 2.98(14.87)  & 3.56(17.27)  & 6.02(20.44)  & 7.72(24.58)  \\
        &  & Random & 16.05(33.63) & 18.85(38.39) & 14.32(32.11) & 16.58(36.48) & 12.43(30.20) & 14.50(34.42) & 19.80(35.58) & 23.80(41.20) \\
        & 900 & PCA    & 2.68(14.04)  & 3.37(17.01)  & 2.24(12.76)  & 2.81(15.48)  & 1.90(11.87)  & 2.25(13.90)  & 4.44(17.06)  & 5.85(21.08)  \\
        &  & Random & 15.87(33.56) & 18.48(38.29) & 14.04(32.06) & 16.32(36.30) & 12.91(30.97) & 14.99(35.06) & 18.52(34.99) & 22.35(40.61) \\
        \hline\hline
\end{tabular}
%\begin{tablenotes}
%\item[1] The results for Test (1.1) are all updated.
%\item[2] For Test (1.2) and (1.3), the bold font means that the results are updated.
%\end{tablenotes}
\end{threeparttable}
\end{table}

\begin{table}[!htbp]
\renewcommand\arraystretch{1.2}
\centering
\begin{threeparttable}
\tiny
\caption{\textit{ The averages and standard deviations (in parentheses) of $\rho^2(\bA, \wh \bA)$ and $\rho^2(\bB, \wh \bB)$ when $d=6$ based on $2000$ repetitions. All numbers in the table are the true numbers multiplied by $10^2$ for ease of presentation. }}\label{SM:tb3}
\begin{tabular}{ccc|cc|cc|cc|cc}
\hline\hline
         &   &  &  \multicolumn{6}{c|}{The refined method} &    \multicolumn{2}{c}{The direct method} \\ \cline{4-9}
        &   &  & \multicolumn{2}{c|}{$K=3$}           & \multicolumn{2}{c|}{$K=5$}           & \multicolumn{2}{c|}{$K=7$}   &    \multicolumn{2}{c}{}            \\
$(p,q)$   & $n$ & $\xi_t$  & $\rho^2(\bA, \wh \bA)$& $\rho^2(\bB, \wh \bB)$ & $\rho^2(\bA, \wh \bA)$ &$\rho^2(\bB, \wh \bB)$ & $\rho^2(\bA, \wh \bA)$ & $\rho^2(\bB, \wh \bB)$  & $\rho^2(\bA, \wh \bA)$ & $\rho^2(\bB, \wh \bB)$  \\ \hline
$(12,\,12)$  & 300 & PCA    & 40.56(35.31) & 41.12(35.42) & 38.48(34.34) & 39.09(35.01) & 36.42(34.27) & 36.74(34.60) & 67.46(25.17) & 67.74(25.66) \\
         &  & Random & 47.61(37.28) & 47.57(37.45) & 44.98(37.09) & 45.04(36.97) & 43.78(36.68) & 43.68(37.04) & 72.24(24.63) & 72.07(25.12) \\
         & 600 & PCA    & 36.22(34.25) & 36.87(34.59) & 34.80(33.92) & 35.84(34.29) & 33.00(33.37) & 33.13(33.60) & 63.80(26.43) & 63.11(27.00) \\
         &  & Random & 45.20(36.84) & 46.73(37.41) & 42.90(36.65) & 43.70(37.40) & 40.24(36.11) & 41.01(36.77) & 68.45(26.47) & 68.11(26.92) \\
         & 900 & PCA    & 33.26(32.99) & 33.37(33.59) & 32.16(32.99) & 32.74(33.59) & 30.40(32.77) & 31.46(33.49) & 59.33(27.44) & 59.71(27.76) \\
         &  & Random & 41.86(37.28) & 42.18(37.66) & 40.41(37.23) & 40.71(37.34) & 39.97(37.14) & 39.15(36.92) & 66.22(27.81) & 65.93(28.06) \\ \hline
$(16,\,16)$  & 300 & PCA    & 30.40(34.65) & 31.24(35.58) & 27.01(32.90) & 27.44(33.64) & 25.51(32.26) & 25.97(32.85) & 66.84(27.37) & 67.07(27.97) \\
         &  & Random & 39.74(39.02) & 39.56(39.32) & 34.64(37.64) & 34.64(37.97) & 31.83(36.46) & 31.71(36.97) & 71.15(27.30) & 70.43(28.05) \\
         & 600 & PCA    & 25.50(33.07) & 26.42(33.62) & 23.16(31.22) & 23.76(31.74) & 22.43(30.41) & 22.95(31.19) & 61.61(29.03) & 61.71(29.47) \\
         &  & Random & 35.90(38.32) & 36.02(38.78) & 31.98(37.11) & 31.92(37.41) & 30.64(36.25) & 30.70(36.69) & 67.05(28.86) & 66.81(29.54) \\
         & 900 & PCA    & 22.23(31.11) & 22.69(31.34) & 21.46(30.31) & 21.58(30.66) & 19.66(28.92) & 19.90(29.17) & 57.23(30.36) & 56.84(30.56) \\
         &  & Random & 33.22(37.51) & 33.82(38.00) & 30.59(36.51) & 30.92(37.18) & 29.09(36.01) & 29.41(36.24) & 65.23(29.61) & 64.67(30.52) \\ \hline
$(32,\,32)$  & 300 & PCA    & 15.92(30.33) & 16.03(30.34) & 12.86(26.19) & 13.07(26.32) & 11.87(24.82) & 11.65(24.59) & 62.84(32.14) & 62.05(33.62) \\
         &  & Random & 22.01(35.90) & 21.86(35.83) & 17.22(31.19) & 17.40(31.47) & 15.60(29.82) & 15.65(29.59) & 68.36(31.99) & 67.03(32.91) \\
         & 600 & PCA    & 13.36(27.88) & 13.42(28.11) & 10.55(23.87) & 10.47(23.72) & 9.26(21.62)  & 9.24(21.62)  & 56.17(33.77) & 55.85(34.63) \\
         &  & Random & 20.38(34.95) & 20.57(35.03) & 16.41(30.98) & 16.65(31.35) & 14.06(28.44) & 14.01(28.54) & 62.45(33.25) & 62.25(34.23) \\
         & 900 & PCA    & 12.08(26.48) & 11.96(26.38) & 9.76(23.22)  & 9.72(23.20)  & 8.29(20.89)  & 8.35(21.03)  & 51.79(34.20) & 51.72(35.06) \\
         &  & Random & 19.88(34.86) & 20.05(35.00) & 16.06(31.41) & 15.76(31.06) & 14.08(28.82) & 14.08(28.79) & 60.57(34.07) & 59.96(34.99) \\ \hline
$(64,\,64)$  & 300 & PCA    & 9.08(24.48)  & 8.95(24.60)  & 5.56(17.26)  & 5.31(17.22)  & 5.02(16.13)  & 4.92(16.09)  & 58.34(35.20) & 57.18(36.93) \\
         &  & Random & 13.87(30.92) & 13.87(31.06) & 9.54(25.16)  & 9.23(24.65)  & 7.64(22.09)  & 7.52(21.82)  & 63.40(34.63) & 62.35(36.63) \\
         & 600 & PCA    & 6.71(21.66)  & 6.71(21.44)  & 5.16(17.78)  & 5.26(17.86)  & 4.48(15.52)  & 4.60(15.64)  & 51.07(36.19) & 50.99(37.09) \\
         &  & Random & 12.35(29.57) & 12.33(29.36) & 8.77(24.70)  & 8.72(24.65)  & 7.05(21.52)  & 7.09(21.60)  & 59.08(35.56) & 57.75(37.34) \\
         & 900 & PCA    & 6.14(20.55)  & 6.22(20.45)  & 4.28(16.16)  & 4.21(15.99)  & 3.65(14.15)  & 3.87(14.81)  & 45.30(36.03) & 45.10(37.40) \\
         &  & Random & 10.82(27.62) & 11.23(28.10) & 8.05(23.49)  & 8.25(23.76)  & 6.33(20.36)  & 6.29(20.19)  & 56.43(35.89) & 56.24(37.54) \\ \hline
$(32,\,12)$  & 300 & PCA    & 27.61(35.49) & 24.97(32.38) & 22.83(32.06) & 20.94(29.96) & 21.27(29.99) & 19.59(28.30) & 58.78(36.34) & 52.08(33.75) \\
         &  & Random & 33.65(38.89) & 31.53(37.11) & 30.05(37.01) & 28.08(35.18) & 27.62(35.35) & 25.87(33.77) & 63.13(36.23) & 57.33(34.75) \\
         & 600 & PCA    & 22.51(32.50) & 19.84(29.13) & 19.30(29.71) & 17.32(26.77) & 17.47(28.25) & 15.93(25.97) & 52.22(36.48) & 46.39(33.68) \\
         &  & Random & 33.33(39.35) & 30.91(36.79) & 28.37(36.55) & 26.19(34.63) & 26.44(35.27) & 24.52(33.71) & 58.25(38.30) & 52.07(36.09) \\
         & 900 & PCA    & 20.04(31.31) & 17.55(27.46) & 17.68(28.72) & 16.17(26.16) & 15.29(26.39) & 14.27(24.39) & 45.36(37.06) & 41.01(33.89) \\
         &  & Random & 32.11(39.02) & 29.35(36.43) & 27.68(36.97) & 25.24(34.62) & 25.19(35.25) & 23.62(33.27) & 56.47(38.69) & 50.94(36.13) \\ \hline
$(64,\,12)$  & 300 & PCA    & 21.21(33.64) & 18.75(29.91) & 16.28(28.65) & 14.83(26.16) & 14.16(26.11) & 13.31(24.40) & 50.83(40.49) & 43.59(36.75) \\
         &  & Random & 28.22(38.74) & 24.98(35.10) & 23.28(35.30) & 20.97(32.82) & 20.64(32.51) & 18.97(30.39) & 56.09(41.49) & 48.91(37.99) \\
         & 600 & PCA    & 16.56(30.47) & 14.84(26.53) & 12.78(25.88) & 11.51(22.85) & 11.52(23.84) & 10.72(21.67) & 40.32(39.23) & 34.46(34.77) \\
         &  & Random & 26.47(38.38) & 23.59(34.74) & 22.19(35.26) & 20.19(32.55) & 18.31(32.37) & 16.89(30.30) & 48.90(42.36) & 42.70(38.34) \\
         & 900 & PCA    & 13.77(27.23) & 12.39(24.15) & 11.58(23.97) & 10.59(21.73) & 10.84(22.77) & 9.67(20.45)  & 34.55(37.47) & 30.12(33.22) \\
         &  & Random & 25.62(37.86) & 22.78(34.46) & 22.32(35.27) & 20.22(32.35) & 18.46(32.09) & 17.14(30.04) & 47.11(42.00) & 41.03(37.52) \\ \hline
$(128,\,12)$ & 300 & PCA    & 16.92(31.76) & 14.71(27.98) & 12.77(26.68) & 11.18(23.88) & 11.35(24.57) & 9.99(22.09)  & 43.14(42.41) & 36.16(37.13) \\
         &  & Random & 24.94(38.51) & 22.10(34.84) & 18.92(33.75) & 16.91(30.63) & 16.55(30.77) & 15.08(28.80) & 51.06(44.23) & 43.90(40.00) \\
         & 600 & PCA    & 12.31(27.14) & 10.41(22.98) & 9.96(23.48)  & 9.01(20.69)  & 9.02(21.68)  & 8.19(19.13)  & 31.45(38.67) & 26.84(33.69) \\
         &  & Random & 23.57(38.06) & 20.86(34.17) & 18.60(34.04) & 16.72(30.91) & 16.45(31.29) & 14.77(28.89) & 44.74(44.04) & 38.37(39.19) \\
         & 900 & PCA    & 11.53(26.13) & 9.78(22.13)  & 9.19(22.70)  & 8.08(20.14)  & 7.42(19.68)  & 6.84(17.88)  & 28.39(37.68) & 23.13(31.75) \\
         &  & Random & 21.63(36.56) & 19.18(32.94) & 17.34(32.58) & 15.50(29.69) & 13.60(28.72) & 12.23(26.54) & 40.10(43.70) & 34.23(38.57) \\ \hline
$(256,\,12)$ & 300 & PCA    & 14.40(30.74) & 12.15(26.13) & 10.18(24.60) & 8.84(21.68)  & 8.34(21.36)  & 7.44(19.04)  & 38.49(43.44) & 30.97(36.53) \\
         &  & Random & 22.34(37.63) & 19.18(33.56) & 15.93(31.87) & 14.03(29.01) & 12.40(27.57) & 11.07(25.32) & 47.96(45.82) & 40.03(40.09) \\
         & 600 & PCA    & 11.09(26.86) & 8.92(21.74)  & 7.37(20.73)  & 6.25(17.68)  & 6.98(19.85)  & 5.80(16.69)  & 28.09(39.05) & 22.57(32.35) \\
         &  & Random & 21.19(37.30) & 18.02(32.79) & 15.86(32.34) & 14.10(29.41) & 12.18(28.07) & 10.88(25.85) & 42.66(45.56) & 35.70(39.71) \\
         & 900 & PCA    & 10.58(26.54) & 8.46(21.19)  & 7.01(20.30)  & 6.03(17.29)  & 5.65(17.21)  & 4.70(14.56)  & 25.51(37.99) & 20.43(31.46) \\
         &  & Random & 18.27(35.08) & 16.03(31.27) & 14.53(31.33) & 12.89(28.31) & 11.87(28.13) & 10.87(25.86) & 34.93(43.44) & 29.42(37.85) \\ \hline
$(12,\,32)$  & 300 & PCA    & 24.16(32.73) & 26.76(35.47) & 20.63(29.86) & 21.98(31.56) & 19.84(28.58) & 21.40(30.51) & 52.06(34.42) & 57.19(36.79) \\
         &  & Random & 32.28(37.34) & 35.24(39.85) & 27.27(34.83) & 29.38(36.88) & 25.50(33.56) & 27.80(35.55) & 57.70(34.72) & 64.06(36.58) \\
         & 600 & PCA    & 20.81(29.79) & 22.90(32.94) & 17.30(26.80) & 19.17(29.56) & 16.39(25.61) & 18.03(27.89) & 45.45(33.40) & 50.55(36.58) \\
         &  & Random & 30.85(36.66) & 33.66(39.59) & 26.25(34.36) & 27.57(36.32) & 23.51(32.84) & 25.00(34.76) & 53.82(35.45) & 60.09(37.52) \\
         & 900 & PCA    & 17.88(27.47) & 20.20(31.08) & 16.55(26.24) & 17.88(28.65) & 15.11(24.79) & 16.08(26.68) & 41.88(33.62) & 45.86(36.83) \\
         &  & Random & 29.47(36.24) & 31.78(39.07) & 24.57(34.25) & 26.43(36.72) & 22.84(32.55) & 25.67(35.45) & 50.66(35.90) & 56.79(38.42) \\ \hline
$(12,\,64)$  & 300 & PCA    & 18.81(29.71) & 21.98(34.31) & 15.52(26.63) & 17.67(30.11) & 13.84(24.91) & 15.13(26.95) & 43.79(36.44) & 50.15(40.54) \\
         &  & Random & 26.25(35.96) & 29.13(39.23) & 22.14(33.90) & 24.08(36.10) & 19.66(31.53) & 21.11(33.47) & 48.71(38.47) & 56.17(41.99) \\
         & 600 & PCA    & 14.92(26.29) & 17.25(30.40) & 12.27(23.47) & 13.67(26.35) & 11.32(22.10) & 12.22(24.09) & 35.17(34.71) & 41.48(39.23) \\
         &  & Random & 23.70(34.47) & 26.38(37.98) & 20.39(32.53) & 22.63(35.64) & 17.98(30.49) & 19.96(33.26) & 42.35(37.78) & 48.83(42.36) \\
         & 900 & PCA    & 12.91(24.46) & 14.74(28.29) & 11.88(23.30) & 13.18(25.93) & 10.83(21.48) & 11.63(23.78) & 30.89(33.14) & 36.40(38.43) \\
         &  & Random & 23.34(34.36) & 25.92(38.06) & 18.78(31.40) & 21.00(34.47) & 16.82(29.49) & 18.85(32.30) & 40.96(37.86) & 47.13(41.95) \\ \hline
$(12,\,128)$ & 300 & PCA    & 14.20(26.97) & 16.33(31.15) & 11.81(23.54) & 13.09(26.70) & 10.04(20.93) & 10.69(23.11) & 35.81(36.77) & 43.20(42.56) \\
         &  & Random & 21.36(34.39) & 23.90(37.87) & 17.05(31.01) & 18.45(33.30) & 15.21(28.88) & 16.95(31.32) & 43.84(39.62) & 51.52(43.87) \\
         & 600 & PCA    & 11.18(23.76) & 12.89(27.77) & 8.66(20.34)  & 9.90(23.44)  & 7.17(17.67)  & 8.22(20.23)  & 27.61(33.79) & 33.40(39.93) \\
         &  & Random & 19.73(33.42) & 22.47(37.46) & 16.69(31.11) & 18.24(33.55) & 14.87(29.21) & 16.34(31.72) & 36.92(38.52) & 43.80(43.97) \\
         & 900 & PCA    & 9.75(22.14)  & 11.44(25.88) & 8.01(19.24)  & 8.66(21.17)  & 7.46(18.61)  & 7.98(20.31)  & 23.93(32.35) & 28.51(38.09) \\
         &  & Random & 19.70(33.50) & 22.09(37.31) & 15.45(30.12) & 17.43(33.02) & 14.11(28.56) & 15.22(30.52) & 35.95(38.50) & 42.22(43.82) \\ \hline
$(12,\,256)$ & 300 & PCA    & 12.49(26.66) & 14.63(30.54) & 8.70(21.43)  & 9.79(24.00)  & 7.65(19.85)  & 8.49(21.46)  & 32.32(37.07) & 39.74(43.65) \\
         &  & Random & 19.05(33.41) & 21.93(37.58) & 13.40(28.55) & 15.10(31.16) & 11.24(25.77) & 12.63(27.81) & 38.47(40.39) & 46.23(45.88) \\
         & 600 & PCA    & 9.03(22.23)  & 10.50(26.23) & 6.85(18.37)  & 7.81(21.48)  & 5.69(15.98)  & 6.55(18.65)  & 23.85(33.66) & 29.05(39.61) \\
         &  & Random & 16.65(31.98) & 19.00(35.57) & 12.66(27.96) & 13.54(30.08) & 10.44(25.29) & 11.34(27.09) & 34.51(39.31) & 40.99(45.01) \\
         & 900 & PCA    & 7.69(20.32)  & 9.27(24.13)  & 5.63(16.30)  & 6.65(19.02)  & 4.49(13.94)  & 5.06(15.43)  & 18.95(30.61) & 23.06(36.51) \\
         &  & Random & 15.01(30.29) & 17.27(34.30) & 12.10(27.14) & 13.73(30.27) & 10.72(25.66) & 11.79(28.00) & 31.11(38.39) & 36.90(44.20) \\ %\hline
        \hline\hline
\end{tabular}
%\begin{tablenotes}
%\item[1] The results for Test (1.1) are all updated.
%\item[2] For Test (1.2) and (1.3), the bold font means that the results are updated.
%\end{tablenotes}
\end{threeparttable}
\end{table}

\end{document}

%% file: QYalias.tex
 \renewcommand{\theequation}{\arabic{equation}}
\newtheorem{theorem}{Theorem}
\newtheorem{lemma}{Lemma} 
\newtheorem{proposition}{Proposition}

\def\singlespace{\def\baselinestretch{1}\@normalsize}

\def\wh{\widehat}

\newcommand{\cov}{{\rm Cov}}
\newcommand{\diag}{{\rm diag}}

\def\ga{\gamma}
\def\de{\delta}

\def\la{\lambda}

\newcommand{\ve}{{\varepsilon}}

\newcommand{\bA}{{\mathbf A}}
\newcommand{\bB}{{\mathbf B}}
\newcommand{\bF}{{\mathbf F}}
\newcommand{\bE}{{\mathbf E}}
\newcommand{\bG}{{\mathbf G}}
\newcommand{\bH}{{\mathbf H}}
\newcommand{\bI}{{\mathbf I}}
\newcommand{\bK}{{\mathbf K}}

\newcommand{\bM}{{\mathbf M}}
\newcommand{\bQ}{{\mathbf Q}}

\newcommand{\bP}{{\mathbf P}}
\newcommand{\bR}{{\mathbf R}}
\newcommand{\bS}{{\mathbf S}}
\newcommand{\bU}{{\mathbf U}}
\newcommand{\bV}{{\mathbf V}}
\newcommand{\bW}{{\mathbf W}}
\newcommand{\bX}{{\mathbf X}}
\newcommand{\bY}{{\mathbf Y}}
\newcommand{\bZ}{{\mathbf Z}}
\newcommand{\ba}{{\mathbf a}}
\newcommand{\bb}{{\mathbf b}}

\newcommand{\bh}{{\mathbf h}}

\newcommand{\br}{{\mathbf r}}

\newcommand{\bu}{{\mathbf u}}
\newcommand{\bv}{{\mathbf v}}
\newcommand{\bw}{{\mathbf w}}
\newcommand{\bx}{{\mathbf x}}

\newcommand{\bbeta}  {\boldsymbol{\beta}}

\newcommand{\bdelta} {\boldsymbol{\delta}}

\newcommand{\bSigma}{\mathbf \Sigma}
\newcommand{\bDelta}{\boldsymbol{\Delta}}
\newcommand{\bgamma}{\boldsymbol{\gamma}}

\newcommand{\bve}{\mbox{\boldmath$\varepsilon$}}

\newcommand{\bTheta} {\mathbf \Theta}

\newcommand{\bGamma} {\mathbf \Gamma}
\newcommand{\bLambda} {\boldsymbol{\Lambda}}
\newcommand{\bC}{{\mathbf C}}
\newcommand{\bD}{{\mathbf D}}
\newcommand{\bzero}{{\mathbf 0}}

\newcommand{\calR}{{\mathcal R}}

\def\6bullets{\bullet\bullet\bullet\bullet\bullet\bullet}

\DeclareMathAlphabet\EuScriptBF{U}{eus}{b}{n}

\newcommand{\eulbE}{\EuScriptBF E}

\newcommand{\eulbX}{\EuScriptBF X}
\newcommand{\eulbY}{\EuScriptBF Y}

\def\JASA{{\sl Journal of the American Statistical Association}}

\def\AS{{\sl The Annals of Statistics}}

\def\JE{{\sl Journal of Econometrics}}